\documentclass[onecolumn]{aastex62}
\usepackage{shortcuts-for-paper}

\graphicspath{{figures/}}

\begin{document}
\title{Angular Correlation Function Estimators Accounting for Contamination from Probabilistic Distance Measurements}

\author[0000-0003-2296-7717]{Humna Awan}
\affiliation{Department of Physics $\&$ Astronomy, Rutgers University, 136 Frelinghuysen Rd., Piscataway, NJ 08554}
\author{Eric Gawiser}
\affiliation{Department of Physics $\&$ Astronomy, Rutgers University, 136 Frelinghuysen Rd., Piscataway, NJ 08554}
\affiliation{Center for Computational Astrophysics, Flatiron Institute, 162 5th Ave., New York, NY 10010}

\correspondingauthor{Humna Awan}
\email{awan@physics.rutgers.edu}

\shorttitle{Weighted Galaxy Clustering}
\shortauthors{Awan $\&$ Gawiser}

\submitjournal{The Astrophysical Journal}
\begin{abstract}
With the advent of surveys containing millions to billions of galaxies, it is imperative to develop analysis techniques that utilize the available statistical power.\deleted{We focus on two-point angular auto- and cross-correlation function estimators that account for distance uncertainties in redshift-binned analyses.} \replaced{Since we are interested in the clustering of galaxies}{In galaxy clustering}, even small sample contamination arising from distance uncertainties can lead to large artifacts, which the standard estimator does not account for. \replaced{We illustrate how to use a matrix inversion to correct any correlation function estimator for contamination. We then introduce an estimator that assigns each galaxy a weight in each redshift bin based on its probability of being in that bin.}{We first introduce a formalism, termed decontamination, that corrects for sample contamination by utilizing the observed cross-correlations in the contaminated samples; this corrects any correlation function estimator for contamination. Using this formalism, we present a new estimator that uses the standard estimator to measure correlation functions in the contaminated samples but then corrects for contamination. We also introduce a weighted estimator that assigns each galaxy a weight in each redshift bin based on its probability of being in that bin.} We demonstrate that \replaced{, after decontamination, the new estimator is unbiased and leads to a framework for minimizing variance arising from sample contamination. The estimator}{these estimators effectively recover the true correlation functions and their covariance matrices. \replaced{We also note that o}{O}ur estimators can correct for sample contamination caused by misclassification between object types as well as photometric redshifts; they} should be particularly helpful for studies of galaxy evolution and baryonic acoustic oscillations, where forward-modeling \added{the clustering signal using} the contaminated redshift distribution is undesirable.
\end{abstract}

\section{Introduction\label{sec: intro}}
Various probes exist to study the cause of cosmic acceleration, one of which is the evolution of large-scale structure (LSS) as traced by clustering in the spatial distribution of galaxies \citep{Cooray+02}. The standard metric to quantify galaxy clustering is the two-point correlation function (CF) or its Fourier transform, the power spectrum. Galaxy clustering can be measured in 3D using spectroscopic surveys, where precise radial information is available, or by measuring the 2D correlations in tomographic redshift bins when only photometric data is available.

Several large astronomical surveys are coming online in the next decade, allowing access to an unprecedented amount of data and hence the ability to measure the evolution of LSS to high precision. These surveys include the \href{https://www.lsst.org/}{Large Synoptic Survey Telescope (LSST)} \citep{Abell+09}, \href{https://www.desi.lbl.gov/}{Dark Energy Spectroscopic Instrument} \citep{DESI}, \href{http://sci.esa.int/euclid/}{Euclid} \citep{Laureijs+11}, and \href{https://wfirst.gsfc.nasa.gov/}{WFIRST} \citep{Spergel+15}. The large datasets, however, present new challenges, among which are understanding, mitigating, and accounting for the impacts of systematic uncertainties that exceed the statistical uncertainties; these include uncertainties due to sample contamination, arising either due to photometric redshift uncertainties or spectroscopic line misidentification. Various studies have presented methods to mitigate these effects; e.g., \citet{Elsner+16} and \citet{Leistedt+16} present mode projection as a way to account for systematics, and \citet{Shafer+15} present methodology to handle multiplicative errors like photometric calibration errors.

Various estimators exist to measure the CFs, with the most widely used one introduced in \LSt\ (referred to as LS93 hereafter); see e.g., \citet{Kerscher+00} for a comparison of the various analog estimators, while \citet{Vargas+13} and  \citet{Bernstein1994} are example studies that consider involved optimizations of the estimators. These estimators can also be extended for various purposes using the overarching idea of `marked' statistics, which employ weights, or `marks', for different quantities: they can be used to account for additional dependencies in the correlation functions \citep[see e.g.,][]{Sheth+04, Harker+06, Skibba+06, White+09, Sheth+05, Robaina+12, Hernandez+18, White2016}, extract characteristic-dependent correlations  \citep[see e.g.,][]{Beisbart+00, Armijo+18}, or be used to account for different systematics or to extract target features. For instance, \citet{FKP} present a simple weighting that accounts for the signal-to-noise differences coming from each tomographic volume (which was applied e.g., when measuring the Baryonic Acoustic Oscillations (BAO) in \citealt{Eisenstein+2005}); \citet{Ross+17} extend the weights in \citet{FKP} to handle photometric redshift (\pz) uncertainties for BAO measurements while \citet{Peacock+04} extend them to account for luminosity-dependent clustering, which then are extended by \citet{Pearson+16} for minimal variance in cosmological parameters; \citet{Zhu+15} and \citet{Blake+19} use weights to optimize the BAO measurements; \citet{Bianchi+18} employ weights to account for spectroscopic fibre assignment; \citet{Ross+12} use them to handle systematics, as do \citet{Morrison+15}; while \citet{Bianchi+17} and \citet{Percival+17} employ them for 3D correlations to not only correct for missing observations but to improve clustering measurements. 

In this paper, we focus on the impacts of sample contamination on the angular correlation functions (ACF). As alluded to earlier, ACFs are especially relevant for photometric surveys, for which we can either measure the projected CFs (e.g., see \citealt{Zehavi+02, Zehavi+11}) or the ACFs in redshift bins (e.g., see \citealt{Crocce+16, Balaguera+18, Abbott+18}). Note that one can also measure the ACFs without the tomographic binning (e.g., as in \citealt{Connolly+02, Scranton+02}) but that disallows mapping the evolution of the galaxy clustering. Photo-$z$ uncertainties make measuring ACFs in tomographic bins more challenging as the uncertainties introduce spurious cross-correlations across the \replaced{various}{redshift} bins (e.g., see \citealt{Bailoni+17} for a study on the impacts of bin cross-correlations on cosmological parameters) and smear out valuable cosmological information, including the BAO (e.g., as in \citealt{Chaves+18}). \replaced{Now, s}{Since} the traditional ACF estimators do not account for contamination arising from \pz\ uncertainties, the standard tomographic clustering analysis entails estimating $N(z)$, i.e., the number of galaxies as a function of redshift, in each nominal redshift bin and forward modeling the contaminated ACFs using the $N(z)$ estimates (e.g., as in \citealt{Crocce+16, Balaguera+18, Abbott+18}); also see e.g., \citet{Newman08} for a discussion on estimating $N(z)$. While this method allows cosmological parameter estimation, it suffers some key limitations as \replaced{it does not measure the ACFs in independent redshift bins, which is necessary for galaxy evolution studies}{forward modeling is not commonly used outside of cosmology}. Furthermore, the variance on the cosmological parameters could potentially be reduced if sample contamination were accounted for directly, instead of being forward modeled, \replaced{e.g., when focusing on measuring the}{to yield a higher S/N} BAO signal from photometric samples.

We propose a method\deleted{ology} to measure the ACFs \textit{while} accounting for contamination and \deleted{more importantly, }without needing to forward model the $N(z)$. Specifically, \replaced{our estimator for the ACFs incorporates not just the \pz\ point estimates but each galaxy's entire \pz\ probability distribution function (PDF; of which \pz\ is only representative), by weighting each galaxy based on its \pz\ PDF.}{we first introduce a formalism that uses the observed cross correlations to account for sample contamination. Using this formalism, we propose our first estimator, which still uses the \pz\ point estimates and the standard CF estimator, but corrects for contamination. Then, we introduce a new estimator that incorporates not just the \pz\ point estimates but each galaxy's entire \pz\ probability distribution function (PDF; of which \pz\ is only representative), by weighting each galaxy based on its \pz\ PDF.} \replaced{While our method}{We note that while the second estimator} extends the idea of marked statistics, as discussed above, it differs from the applications in the literature on several fronts. In particular, it avoids the loss of information caused by placing galaxies in a single redshift bin based on their \pz s, thereby allowing us to counter the impacts of sample contamination \deleted{on the variance }with the statistical power of a large dataset, as well as \added{potentially} allowing low-variance measurements of the full correlation function\added{s}. We return to some of these \replaced{studies}{points} for a more thorough discussion of the various differences\added{ between our work and that in the literature}.

\replaced{In order to present the ideas in the most coherent way, t}{T}his paper is structured as follows: in Section~\ref{sec: 2pt}, we formally introduce the ACF and its \replaced{simplest}{standard} estimator. In Section~\ref{sec: estimators}, we introduce terminology to address sample contamination in the most general sense, followed by \replaced{a method way}{our first estimator} to correct for sample contamination\added{; we refer to this as the \ttt{Decontaminated} estimator}. In Section~\ref{sec: pweighted}, we introduce a weighted  estimator in which the weights can be chosen to track the probability of each \replaced{galaxies}{galaxy} lying in \replaced{a given}{each} redshift bin\added{; we refer to this as the \ttt{Weighted} estimator; it is followed by a \ttt{Decontaminated Weighted} estimator that estimates the true CFs}. We present our validation method in Section~\ref{sec: validation + results}, where we start with a toy example to illustrate the impacts of \pz\ uncertainties, followed by a realistic example of measuring the ACFs in three redshift bins, demonstrating the effectiveness of the estimators in recovering the true correlation functions \added{and their covariance matrices} in the presence of sample contamination. We discuss our results in Section~\ref{sec: discussion} and conclude in Section~\ref{sec: conclusions}.
\section{2D Two-Point Correlation Function\label{sec: 2pt}}
The most common statistic to study galaxy clustering is the two-point correlation function. The 2D angular correlation function $w_{\alpha\beta}(\theta)$ measures the excess probability of finding a galaxy of Type-$\alpha$ at an angular distance $\theta$ from a galaxy of Type-$\beta$, in comparison with a random distribution \citep{Peebles}:
\eq{
	dP_{\alpha\beta}(\theta) = \eta_{\alpha} \eta_{\beta}\bsqbr{1+w_{\alpha\beta}(\theta)} d\Omega_\alpha d\Omega_\beta
\label{eq: w}
}
where $dP_{\alpha\beta}(\theta)$ is the probability of finding a pair of galaxies of Type-$\alpha\beta$ at an angular distance $\theta$, $\eta_{\alpha}$ is the observed sky density of Type-$\alpha$ galaxies in the projected catalog, and $d\Omega$ is the solid angle element at separation $\theta$. An estimator for the correlation function can be constructed as the ratio of number of data-data pairs compared to the number of random-random pairs at a given angular separation:
\eq{
	w_{\alpha\beta}(\thetak) = \frac{(DD)_{\alpha\beta}(\thetak)}{(RR)_{\alpha\beta}(\thetak)} - 1
\label{eq: dd/rr}
}
where $(DD)_{\alpha\beta}(\thetak)$ is the normalized number of data-data pairs at angular separation $\thetak$, and $(RR)_{\alpha\beta}(\thetak)$ is that for the random-random pairs; the index $k$ emphasizes the binned nature of the estimator. We note that Equation~\ref{eq: dd/rr} leads to an auto-correlation function when $\alpha=\beta$ and cross-correlation otherwise; for the cross-correlation, we explicitly consider independent random catalogs for the two populations, accounting for the case when the two samples do not completely overlap in their angular range. We also note that each histogram can be written using the Heaviside step function, defined as
\eq{
	\Theta(x) = \begin{cases}
				0, & x<0 \\
				1, & x\geq 0
			\end{cases}
\label{eq: heaviside}
}
For instance, for the auto-correlation, we have
\eq{
	(DD)_{11}(\thetak)
	= \frac{
		\sum_i^{N_1} \sum_{j > i}^{N_1}
		\Theta(\theta_{ij}-\theta_{\mathrm{min}, k})
		[1-\Theta(\theta_{ij}-\theta_{\mathrm{max}, k})]
		}{
		\sum_i^{N_1} \sum_{j > i}^{N_1}
		}
	\equiv \frac{\sum_i^{N_1} \sum_{j > i}^{N_1} \heavisides{ij} }{\sum_i^{N_1} \sum_{j > i}^{N_1} }
	= \frac{\sum_i^{N_1} \sum_{j \neq i}^{N_1} \heavisides{ij} }{\sum_i^{N_1} \sum_{j\neq i}^{N_1} }
	\label{eq: dd auto}
}
where
\eq{
	\heavisides{ij} \equiv \Theta(\theta_{ij}-\theta_{\mathrm{min},k}) [1-\Theta(\theta_{ij}-\theta_{\mathrm{max},k})]
	\label{eq: heaviside short}
}
Here, $\theta_{{ij}}$ is the angular separation between the $i$th and $j$th galaxy in the data sample of ${N_1}$ galaxies, and we have explicitly written out the histogram: the $k$th bin counts the number of galaxy pairs at separations $\theta_{\mathrm{min}, k}\leq \theta_{ij}<\theta_{\mathrm{max}, k}$. Note that the normalized histograms can be calculated either by considering all unique pairs or with double counting, as long as the normalization accounts for the total pairs; the denominator in the case where we count only the unique pairs yields the familiar count of $N_1(N_1-1)/2$ pairs.

Similar to Equation~\ref{eq: dd auto}, we can write the histogram for the cross-correlation function as
\eq{
	(DD)_{12}(\thetak) = \frac{\sum_i^{N_1} \sum_{j}^{N_2} \heavisides{ij} }{\sum_i^{N_1} \sum_{j}^{N_2}}
	\label{eq: dd cross}
}
where sample $\alpha$ contains $N_\alpha$ galaxies.

We note here that the estimator in Equation~\ref{eq: dd/rr} differs only slightly from the estimator introduced in LS93 (referred to hereafter as the LS estimator). In the absence of sample contamination, the LS estimator is unbiased and has Poissonian variance but we choose to work with the simpler estimator since the LS estimator accounts for edge-effects that become subdominant to sample contamination when using large galaxy surveys. Specifically, we note that the $DD/RR$ estimator presented above is as (un)biased as the LS estimator (see Equation 48 in \LS) and its variance reduces to Poissonian variance in the limit of large $N$ (see Equations 42, 48 in \LS). \added{We refer to the $DD/RR$ estimator as the \ttt{Standard} estimator, when comparing with the new estimators.}

\section{Standard Estimator and Contaminants\label{sec: estimators}}
We start with the case of two galaxy types in the observed sample, Type-$A$  and Type-$B$; either one acts as a contaminant in relation to the other. We assume that we have some method that gives us the probability of each observed galaxy $i$ of being Type-$A$, $\ttt{q}_i^A$ or Type-$B$, $\ttt{q}_i^B$; example methods include, e.g., integration of a galaxy's \pz\ PDF in the target redshift bin or a Bayesian classifier as presented in \citet{Leung+15}. Assuming that our observed galaxy sample comprises only the two types of galaxies, we have $\ttt{q}_i^A + \ttt{q}_i^B = 1$, where $i$ runs over all the galaxies in the observed sample.

Now, assuming that the classifier is unbiased, we can use the classification probabilities to estimate the fraction of objects that are contaminants for a given target sample. For this purpose, however, we must divide the full observed sample into target subsamples, i.e., in the 2-sample case, the observed Type-$A$ and Type-$B$ galaxies.\footnote{A simple way to do this would be to assign all galaxies with $\ttt{q}_i^A>0.5$ to target sample $A$ and the rest to target sample $B$.} Then, our classifier provides the probability of each observed Type-$A$ galaxy $i$ to be truly of Type-$A$, $q_{i}^{AA}$, as well as the probability of each observed Type-$A$ galaxy to be truly of Type-$B$, $q_{i}^{AB}$. Hence, we have
\eq{
	q_{i}^{AA} +  q_{i}^{AB} = q_{j}^{BA} +  q_{j}^{BB} = 1
	\label{eq: q identity}
}
where $i$ runs over the observed Type-$A$ sample and $j$ runs over the observed Type-$B$ sample. We can then use the classification probabilities on the observed subsamples to estimate the contamination. That is, we have the fraction of observed Type-$A$ galaxies that are true Type-$A$ or Type-$B$ galaxies given by
\eq{
	\fAA = \ev{q_{i}^{AA}} \ ; \ \fAB = \ev{q_{i}^{AB}}
	\label{eq: fs}
}
where the average is over the observed Type-$A$ sample. Equation~\ref{eq: q identity} translates into the expected identities on the fractions:
\eq{
	 \fAA + \fAB = \fBA + \fBB = 1
	 \label{eq: f identity}
}

These ideas can be generalized to $M$ galaxy samples of Types $A_1, A_2, ..., A_M$, with the classification probabilities on the entire observed sample given by $\ttt{q}_{A_1}, \ttt{q}_{A_2}, ... , \ttt{q}_{A_M}$. Once the full observed catalog is divided into $M$ target subsamples, we have the probability of $i$th observed galaxy of Type-$A_j$ being of Type-$A_m$ given by  $q_{A_j A_m, i}$ and the fraction of observed Type-$A_j$ galaxies that are Type-$A_m$ galaxies given by $f_{A_j A_m}$.
\subsection{Decontamination\label{sec: direct}}
Using the standard ACF estimator, correlations from known contaminated samples can be corrected for by using the fractions $f_{\alpha\beta}$ as defined in Equation~\ref{eq: fs}; see e.g., \citet{Grasshorn+18}, \citet{Addison+18} for a similar approach. Formally, this is done by writing the observed correlation functions in terms of the true correlation functions by considering the type of galaxy that contributes to each data pair. Here we work with two target galaxy samples, Type-$A$ and Type-$B$; the generalized case is discussed in Appendix~\ref{sec: direct: general}.

Since we have two types of galaxies, we aim to calculate two auto-correlations and one cross-correlation from the contaminated sample: $\wAAtrue, \wABtrue, \wBBtrue$. However, if we calculate the correlations on the subsamples directly, we get $\wAAobs, \wABobs, \wBBobs$, which differ from the true correlations due to sample contamination. To construct the relation between the two, lets consider $\wABobs$  which gets its contributions from four types of pairs: 1) Observed Type-$A$ galaxies that are true Type-$A$, paired with observed Type-$B$ that are true Type-$A$, contributing $\fAA\fBA \wAAtrue$ to the observed correlation, 2) Observed Type-$A$ that are true Type-$A$, paired with observed Type-$B$ that are true Type-$B$, contributing $\fAA\fBB\wABtrue$, 3) Observed Type-$B$ that are true Type-$A$, paired with observed Type-$A$ that are true Type-$B$, contributing $\fAB\fBA\wABtrue$, and 4) Observed Type-$A$ that are true Type-$B$, paired with observed Type-$B$ that are true Type-$B$, contributing $\fAB\fBB\wBBtrue$. Therefore, we have
\eq{
	\wABobs= \fAA\fBA\wAAtrue+ \bcbr{\fAA\fBB+\fBA\fAB}\wABtrue+\fAB\fBB\wBBtrue
\label{eq: wABobs}
}
The auto correlations follow similarly, leading us to
\eq{
	\begin{bmatrix}
		\wAAobs \\ \wABobs \\ \wBBobs
	\end{bmatrix}
	=
	\begin{bmatrix}
		\fAA^2  & 2\fAA\fAB & \fAB^2\\
		\fAA\fBA & \fAA\fBB+\fAB\fBA & \fAB\fBB \\
		\fBA^2 & 2\fBB\fBA & \fBB^2
	\end{bmatrix}
	\begin{bmatrix}
		\wAAtrue \\ \wABtrue \\ \wBBtrue
	\end{bmatrix}
\label{eq: direct obs}
}
where we note that the contribution from the true cross correlation to the observed auto correlations simplifies (as opposed for that to the observed cross correlation). We also present a \added{formal} derivation of the result above using Equation~\ref{eq: w} in Appendix~\ref{sec: dP: standard}. Now, using these equations, we can construct the \added{\ttt{Decontaminated}} estimators $\wAAest, \wBBest, \wABest$ for the true correlation functions  $\wAAtrue, \wBBtrue, \wABtrue$ given by
\eq{
	\begin{bmatrix}
		\wAAest \ \ \wABest \ \ \wBBest
	\end{bmatrix}^T
	=
	[D_{\rm{S}}]^{-1}
	\begin{bmatrix}
		\wAAobs \ \ \wABobs \ \ \wBBobs
	\end{bmatrix}^T
\label{eq: direct}
}
where $[D_{\rm{S}}]$ is the square matrix in Equation~\ref{eq: direct obs}, which must be invertible\footnote{For the matrix to be non-invertible, its determinant must be zero, which, after many algebraic manipulations, simplifies to the constraint $(\fAA\fBB-\fAB\fBA)^3 = 0$. Given Equation~\ref{eq: f identity}, this leads to $\fAA=\fBA$ and $\fBB= \fAB$, implying that $\wAAobs = \wABobs = \wBBobs$, i.e., all the observed correlation functions are equal and hence disallow distinguishing the contributions from the true correlation functions. We do not expect the contamination rate to be high enough to enable this special case.}. Appendix~\ref{sec: direct: general} presents the \replaced{decontaminated standard}{\ttt{Decontaminated}} estimators for the generalized case of working with $M$ target subsamples. \added{We also note that this decontamination formalism could be easily applied to the LS estimator; the decontamination matrix $[D_{\rm{S}}]$ does not inherently depend on the usage of the $DD/RR$ estimator.}

Given their construction, the \replaced{decontaminated standard}{\ttt{Decontaminated}} estimators are unbiased (under the assumption that the contamination fractions are represented by the average classification probabilities); see Appendix~\ref{sec: direct: bias} for more details. As for the variance, the decontamination \deleted{essentially }leads to a quadrature sum of the variance of the standard estimators for each of the auto- and cross-correlations \added{in the absence of covariance between the observed correlations}; the closed form expression for the variance \added{as well as the general covariance} of the estimators is presented in Appendix~\ref{sec: direct: variance}. Note that this overarching idea of using contamination fractions is similar to that presented in \citet{Benjamin+10} but their focus is on estimating the contamination fractions \textit{from} the contaminated correlations, for which they resort to approximating the decontamination matrix as diagonal. Since we expect sufficiently strong correlations across the different target samples (e.g., between the neighboring \pz\ bins for a tomographic clustering analysis), the simplification of ignoring some contamination fractions becomes undesirable.

\section{A New, Weighted Estimator\label{sec: pweighted}}
Here, we present an estimator for the observed correlation function that accounts for pair weights, i.e., each pair of galaxies is weighted to account for its contribution to the target correlation function, e.g., by the classification probability of each contributing galaxy (alongside other parameters). This way, we consider the \textit{entire} observed catalog, containing $\Ntot$ galaxies of both Type-$A$ and Type-$B$, each with their respective classification probabilities. That is, we propose a \replaced{weighted}{\ttt{Weighted}} estimator for the observed correlation function:
\eq{
	\widetilde{w}_{\alpha\beta}^{\rm{obs}}(\thetak) = \frac{(\widetilde{DD})_{\alpha\beta}(\thetak)}{RR(\thetak)} - 1
	\label{eq: marked}
}
where $\alpha, \beta$ are the types, e.g., $\widetilde{w}_{AA}^{\rm{obs}}$ denotes the estimator for the observed Type-$A$ auto-correlation while $\widetilde{w}_{AB}^{\rm{obs}}$ denotes the cross-correlation. Here, we define weighted data-data pair counts as
\eq{
	(\widetilde{DD})_{\alpha\beta}(\thetak) = \frac{ \sum_i^{\Ntot} \sum_{j\neq i}^{\Ntot}
									\pw{ij} \heavisides{ij}
								}{
								\sum_i^{\Ntot} \sum_{j\neq i}^{\Ntot} \pw{ij}
								}
	\label{eq: marked dd}
}
where $\pw{ij}$ is the pair weight, with the pair comprised of the $i$th and $j$th galaxies, while the weighting is over all $\Ntot$ galaxies in the observed catalog. We note that the normalization is needed to match the normalization of unweighted correlation functions (Equations~\ref{eq: dd auto}, \ref{eq: dd cross}). Equation~\ref{eq: marked dd} therefore allows us to calculate the different weighted data-data pair counts, e.g., $(\widetilde{DD})_{AA}, (\widetilde{DD})_{AB}, (\widetilde{DD})_{BB}$. We also note that $RR(\thetak)$ is formally $(RR)_{\alpha\beta}(\thetak)$ since different galaxy samples can have different selection functions. However, since we consider all the galaxies in the observed sample, not just the target subsamples, we take $RR(\thetak)$ to trace the full survey geometry. \added{We also note that using the $DD/RR$ estimator allows us to introduce pair weights more naturally here; the LS estimator would make it difficult given the $DR$ term to account for.} We include some notes on the implementation of the \replaced{weighted}{\ttt{Weighted}} estimator in Appendix~\ref{eq: weighted notes}.

In the simplest scenario, the pair weight could be linearly dependent on the probabilities of $i$th and $j$th objects being of Type $\alpha, \beta$ respectively, i.e., $ \pw{ij} = \ttt{w}_{i}^{\alpha}\ttt{w}_{j}^{\beta} = \ttt{q}_i^{\alpha} \ttt{q}_j^{\beta} \label{eq: simple weights}$. Note that this approach does not require us to break the observed sample into target subsamples as long as intelligent weights are assigned to each galaxy pair. Explicitly, if we have two observed galaxy types in our observed catalog, as was discussed at the beginning of Section~\ref{sec: estimators}, $\ttt{w}_{i}^A= q_i^{AA}$ for observed Type-$A$ while $\ttt{w}_{i}^A= q_i^{BA}$ for observed Type-$B$ galaxies. Similarly, $\ttt{w}_{i}^B= q_i^{AB}$ for observed Type-$A$ while $\ttt{w}_{i}^B= q_i^{BB}$ for observed Type-$B$. Also note that $\Ntot= \NobsA+\NobsB= \NtrueA+\NtrueB$. Finally, we highlight that our \replaced{weighted}{\ttt{Weighted}} estimator reduces to the \replaced{standard}{\ttt{Standard}} estimator if $\ttt{w}_{i}^\alpha$ is set to 1 for observed Type-$A$ galaxies and to 0 for observed Type-$B$ galaxies, and $\ttt{w}_{i}^\beta$ is set to 0 for observed Type-$A$ galaxies and to 1 for observed Type-$B$.

\subsection{Estimator Bias and Variance\label{sec: pweighted bias+variance}}
The estimator in Equation~\ref{eq: marked} is biased, as it considers the entire sample, including contaminants with different correlation functions. In order to estimate the true correlations using unbiased estimators, $\widehat{w}$, we require that their expectation value approach the true correlations. That is, we have
\eq{
	\ev{
	\begin{bmatrix}
		\wAAest \\ \wABest \\ \wBBest
	\end{bmatrix}
	}
	=
	\ev{
	[D_W]
	\begin{bmatrix}
		\wAAestobs \\ \wABestobs \\ \wBBestobs
	\end{bmatrix}
	}
	=
	\begin{bmatrix}
		\wAAtrue \\ \wABtrue \\ \wBBtrue
	\end{bmatrix}
\label{eq: marked true}
}
where $[D_{\rm{W}}]$ is a decontamination matrix, designed to make \deleted{sure }the estimators \deleted{are }unbiased. It is analogous to the decontamination matrix $[D_{\rm{S}}]$ in Equation~\ref{eq: direct}. Here we explicitly work with the two-sample case, with only Type-$A$ and Type-$B$ galaxies present in our sample.

As done to decontaminate the \replaced{standard}{\ttt{Standard}} estimators in Section~\ref{sec: direct}, we calculate the contributions that are coming from each of the true correlation functions to any given weighted correlation function. That is, we have
\eqs{
	\ev{\widetilde{w}_{\alpha\beta}^{\rm{obs}}(\thetak)}
	&= \frac{
		\bbr{
			\sum\limits_i^{\Ntot} \sum\limits_{j\neq i}^{\Ntot}
			\ttt{w}_{ij}^{\alpha\beta} \ttt{q}_{i}^{A}\ttt{q}_{j}^{A} 
		}
		\wAAtrue
		+
		\bbr{
			\sum\limits_i^{\Ntot} \sum\limits_{j\neq i}^{\Ntot}
			\ttt{w}_{ij}^{\alpha\beta}
			\bcbr{\ttt{q}_{i}^{A}\ttt{q}_{j}^{B}+\ttt{q}_{i}^{B}\ttt{q}_{j}^{A}}
			}
		\wABtrue
		+
		\bbr{
			\sum\limits_i^{\Ntot} \sum\limits_{j\neq i}^{\Ntot}
			\ttt{w}_{ij}^{\alpha\beta} \ttt{q}_{i}^{B}\ttt{q}_{j}^{B}
		}
		\wBBtrue
		}{
			\sum\limits_i^{\Ntot}
			\sum\limits_{j\neq i}^{\Ntot}
			\ttt{w}_{ij}^{\alpha\beta}
		}
\label{eq: westobs}
}
We present the full derivation of Equation~\ref{eq: westobs} in Appendix~\ref{sec: dP: full sample to weighted}. Consolidating the terms as done in Equation~\ref{eq: direct obs}, we have
\eq{
	\begin{bmatrix}
		\ev{\wAAestobs} \\ \ev{\wABestobs} \\ \ev{\wBBestobs}
	\end{bmatrix}
	=
	\begin{bmatrix}
		\autofrac{A}{A}{A} & \crossfrac{A}{A}{A}{B} & \autofrac{A}{A}{B} \\
		\autofrac{A}{B}{A} & \crossfrac{A}{B}{A}{B}  &  \autofrac{A}{B}{B}  \\
		 \autofrac{B}{B}{A} & \crossfrac{B}{B}{A}{B}  &  \autofrac{B}{B}{B}  \\
	\end{bmatrix}
	\begin{bmatrix}
		\wAAtrue \\ \wABtrue \\ \wBBtrue
	\end{bmatrix}
\label{eq: pweighted obs}
}
Therefore, the \replaced{decontaminated weighted}{\ttt{Decontaminated Weighted}} estimators are given by
\eq{
	\begin{bmatrix}
		\wAAest \ \ \wABest \ \ \wBBest
	\end{bmatrix}^T
	=
	[D_{\rm{W}}]^{-1}
	\begin{bmatrix}
		\wAAestobs \ \ \wABestobs \ \ \wBBestobs
	\end{bmatrix}^T
\label{eq: pweighted}
}
where $[D_{\rm{W}}]$ is the square matrix in Equation~\ref{eq: pweighted obs}. We note that each row in Equation~\ref{eq: pweighted} corresponds to final, unbiased weights on each pair, comprised of a sum of three weights -- a fact that can be utilized when optimizing weights for minimum variance. We present an example optimization that decontaminates while estimating the correlation functions in Appendix~\ref{sec: pweighted: optimize}.

We have checked Equation~\ref{eq: pweighted} in various limiting cases \replaced{and confirmed}{to confirm} the validity of its form.  Specifically, we first divided the total observed sample into subsamples, and then applied the simplifications that reduce the \replaced{weighted}{\ttt{Decontaminated Weighted}} estimators to \replaced{decontaminated standard}{\ttt{Decontaminated}} estimators (i.e., setting the pair weights for the target subsample to unity and the rest to zero, and approximating the classification probabilities with their averages); we confirm that Equation~\ref{eq: pweighted} does indeed reduce to Equation~\ref{eq: direct}\added{, demonstrating that \ttt{Decontaminated Weighted} is the generalized estimator}. We then tested the two limiting cases of no contamination and 100\% contamination, working with just the observed subsamples and using pair weights that are a linear product of the respective classification probabilities; we confirm that the reduced estimator recovers the truth when there is no contamination while it is indeterminate when there is 100\% contamination. Finally, we considered the entire observed sample and tested the limiting cases of no contamination and 100\% contamination, with pair weights that are a linear product of the respective classification probabilities, and arrive at true correlations both when there is no contamination and when there is 100\% contamination -- an advantage of using the full sample. We also present the analytical form of the variance of the \replaced{weighted}{\ttt{Weighted}} estimator in Appendix~\ref{sec: pweighted variance}; since the variance is a function of a four-point sum and depends non-trivially on the pair weights, we choose to estimate the variance numerically\added{ using bootstrap as described in Section~\ref{sec: results 2bin}}. Finally, we present the generalized estimator, i.e., applicable to $M$ target samples, in Appendix~\ref{sec: pweighted: general}.

\section{Validation and Results\label{sec: validation + results}}
In order to test our estimators, we consider the simplest relevant application: tomographic clustering analysis, i.e., the measurement of the ACF for galaxies in different redshift bins. Then, in the context of our terminology in Sections~\ref{sec: estimators}-\ref{sec: pweighted}, the different `types' of galaxies are essentially the galaxies in the different redshift bins. For this purpose, we use the publicly available v0.4\_r1.4 of MICE-Grand Challenge Galaxy and Halo Light-cone Catalog. The catalog is generated by populating the dark matter halos in MICE, which is an $N$-body simulation covering an octant of the sky at $0 \leq z \leq 1.4$. Most importantly for our purposes, the catalog follows local observational constraints, e.g., galaxy clustering as a function of luminosity and color, and incorporates galaxy evolution for realistic high-$z$ clustering -- allowing for a robust test of the estimators. More details about the catalog can be found in MICE publications: \citet{Fosalba+15a, Crocce+15, Fosalba+15b, Carretero+15, Hoffmann+15}. We query the catalog using \href{https://cosmohub.pic.es/home}{CosmoHub} \citep{Carretero+17}.

In order to test our method, we must have \pz s that are realistic for upcoming surveys like \added{the} LSST. Since MICE catalog \pz s are biased and exhibit a large scatter, we simulate adhoc \pz s using the true redshifts and assuming $\sigma_z = 0.03(1+z)$, the upper limit on the scatter mentioned in the LSST Science Requirements Document\footnote{\url{ https://docushare.lsstcorp.org/docushare/dsweb/Get/LPM-17 }; see also \citet{Abell+09}.}. Specifically, we model the \pz\ probability distribution function (PDF) for each galaxy as a Gaussian with its true redshift as the mean and $\sigma_z$ as the standard deviation. Then, we randomly draw from the PDF and assign the draw as the \pz\ of the galaxy; the ``observed PDF" is then a Gaussian with the random draw as the mean and $\sigma_z$ as the standard deviation. This method generates unbiased \pz s in a simple way.

\begin{figure}[t!]
	\hspace*{-0em}
	\begin{minipage}{0.25\paperwidth}
		\vspace{1em}
		\includegraphics[trim={5 5 160 25}, clip=false, width=0.3\paperwidth]{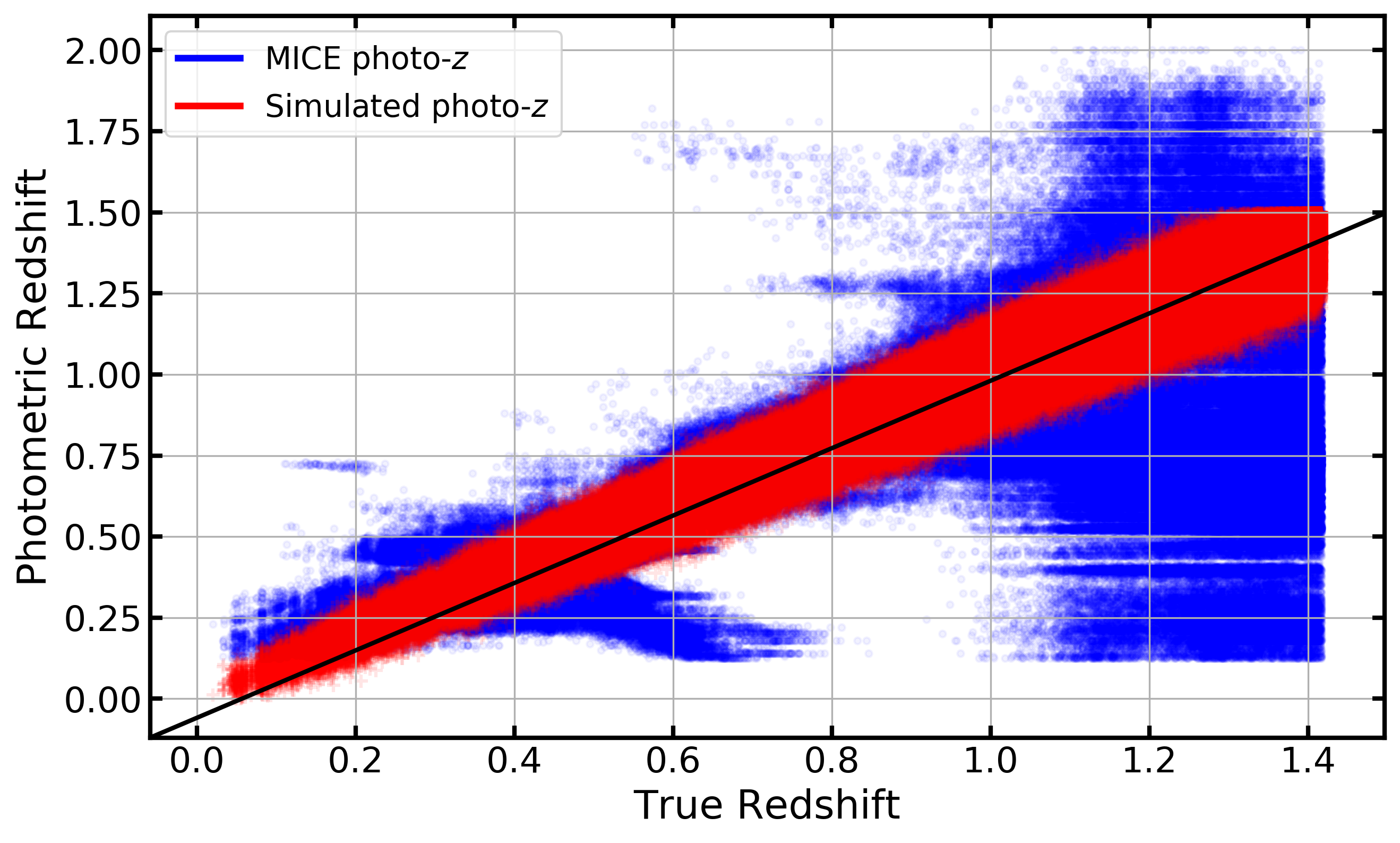}
	\end{minipage}\
	\hspace*{10em}
	\begin{minipage}{0.3\paperwidth}
		\includegraphics[trim={5 10 10 25}, clip=false, width=0.38\paperwidth]{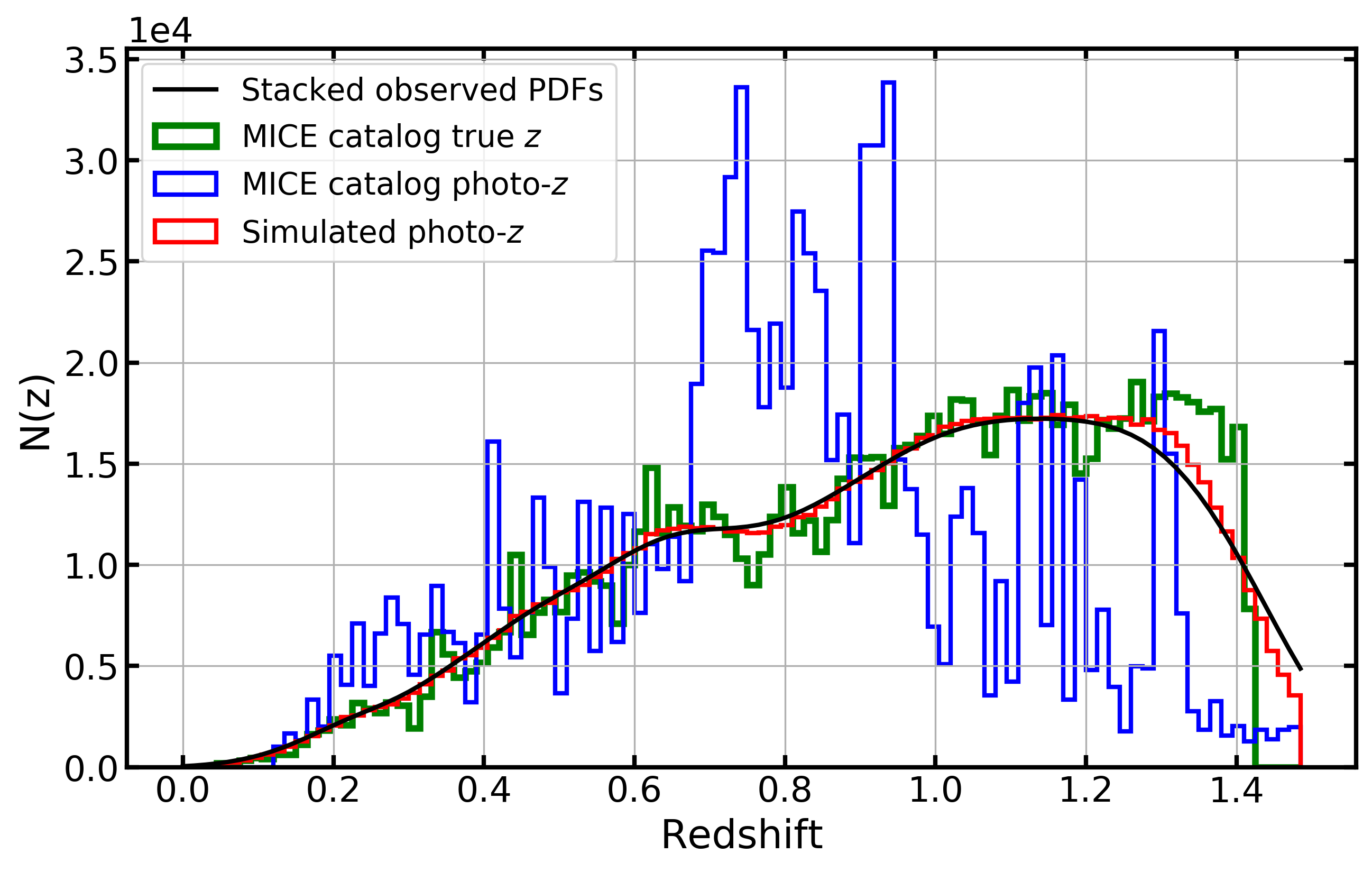}
	\end{minipage}\
	\caption{Illustration of the simulated \pz s. \textit{Left}: Comparison between true redshift  and MICE catalog \pz s \added{(blue)} vs. those simulated here \added{(red)}. \textit{Right}: Comparison between the different $N(z)$ distributions: true $N(z)$; those based on MICE catalog \pz s  vs. those simulated assuming Gaussian PDFs with $\sigma_z = 0.03(1+z)$.  The red, blue, green are $N(z)$ estimates from binning the respective redshifts, while the black \replaced{curves are}{curve is} based on \replaced{PDF stacking}{stacking the observed \pz\ PDFs}.  We see that our simulated \pz s are well-behaved and are able to recover the true $N(z)$ effectively. These plots are created using only the galaxies with $0 \leq $ RA $\leq 5$ deg, $0 \leq $ Dec $\leq 5$ deg, yielding 994,863 galaxies at $0 \leq z \leq 1.4$.
}
	\label{fig: photoz}
\end{figure}

Figure~\ref{fig: photoz} illustrates our simulated \pz s: the left panel compares the MICE catalog \pz s and the simulated \pz s with the true redshifts, while the right panel shows $N(z)$, the number of galaxies as a function of redshift, as estimated by binning the redshifts as well as by stacking the \pz\ PDFs. We see that our simulated \pz\ PDFs and the consequent \pz s effectively recover the overall true galaxy number distribution. Also note that the $N(z)$ from simulated \pz\ (solid red) and observed (solid black) PDFs are very similar, indicating that our simulated \replaced{``observed PDFs''}{observed \pz\ PDFs} are \replaced{overall}{nearly} unbiased. 

Now, the true catalog essentially consists of the location of the galaxies on the sky (RA, Dec) and the true redshift, while the observed catalog consists of the RA, Dec and \pz s. In order to test the effects of contamination, we must work with observed subsamples, i.e., galaxies with \pz s in the target redshift bin; these differ from the true subsamples, which are galaxies with their true redshifts in the target redshift bins. Note that this subsampling is not necessary for the \replaced{weighted}{\ttt{Weighted}} estimator, introduced in Section~\ref{sec: pweighted}, which only needs the \pz\ PDFs for all the observed galaxies. We use \ttt{TreeCorr} \citep{Jarvis+04} to calculate the correlation functions.

\begin{figure}[t!]
	\centering
	\includegraphics[width=.5\linewidth, trim={5 5 5 5}, clip=true]{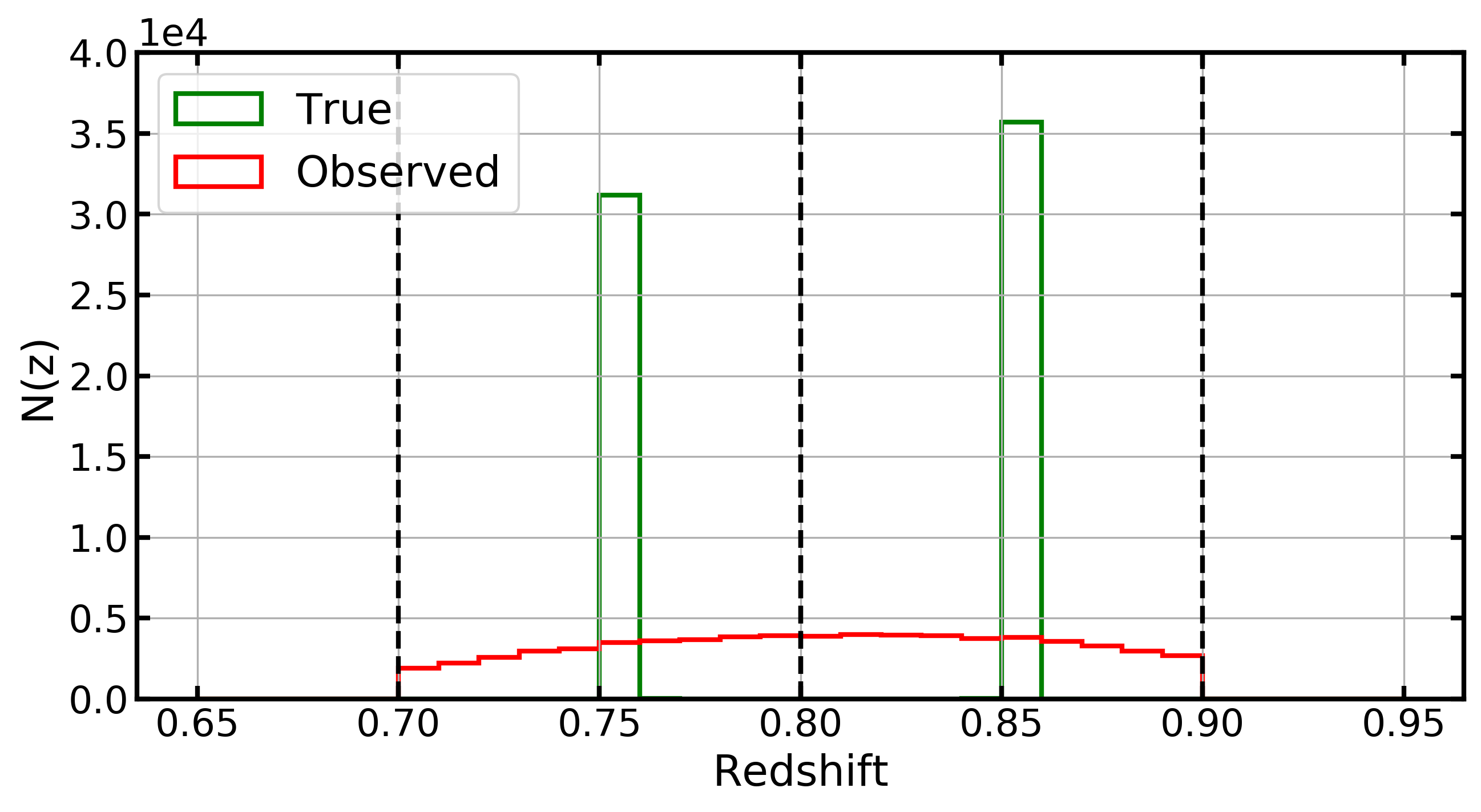}
	\caption{True and observed redshift histograms for the idealized galaxy sample of Section~\ref{sec: results 2bin}, with redshift bin edges shown using the vertical dashed lines. \replaced{P}{We see that p}hoto-$z$ uncertainties lead to a smearing of the redshift information.}
	\label{fig: mock zhists}
\end{figure}

\replaced{
\subsection*{Mock Example\label{sec: results 2bin}}
}{
\subsection{Toy Example\label{sec: results 2bin}}
}
In order to illustrate the impacts of \pz s, we consider a toy example: a clustering analysis using only two tomographic bins ($0.7 \leq z < 0.8$, $0.8 \leq z < 0.9$) with the true galaxy sample having galaxies only at $0.75 \leq z \leq 0.76$, $0.85 \leq z \leq 0.86$, but with the \pz\ scatter as mentioned before, i.e., $\sigma_z = 0.03(1+z)$. We query the true galaxies in \replaced{ten 5x5}{nine 10x10} deg$^2$ patches along Dec = 0; all patches have a similar number of galaxies \replaced{(13K-20K)}{(66K-78K)} and face similar \pz\ contamination rates (\replaced{21-27\%, 16-22\%}{22-25\%, 18-21\%} in the two tomographic bins, respectively). \added{To make explicit the impacts of redshift binning based on photo-$z$ point estimates, we show the true and observed positions of the galaxies in the two redshift bins in Figure~\ref{fig: ra-dec}, where we can see that the two distributions are different, with \pz\ uncertainties mixing the LSS between the two bins.} Figure~\ref{fig: mock zhists} shows the distributions of the true and photometric redshifts using one of the patches (with \replaced{13,599 galaxies, and 24\% and 20\%}{66,927 galaxies, and 23\% and 20\%} contamination in the two tomographic bins, respectively).

\begin{figure}[t!]
	\centering
	\includegraphics[trim={5 5 5 5}, clip=true, width=.5\paperwidth]{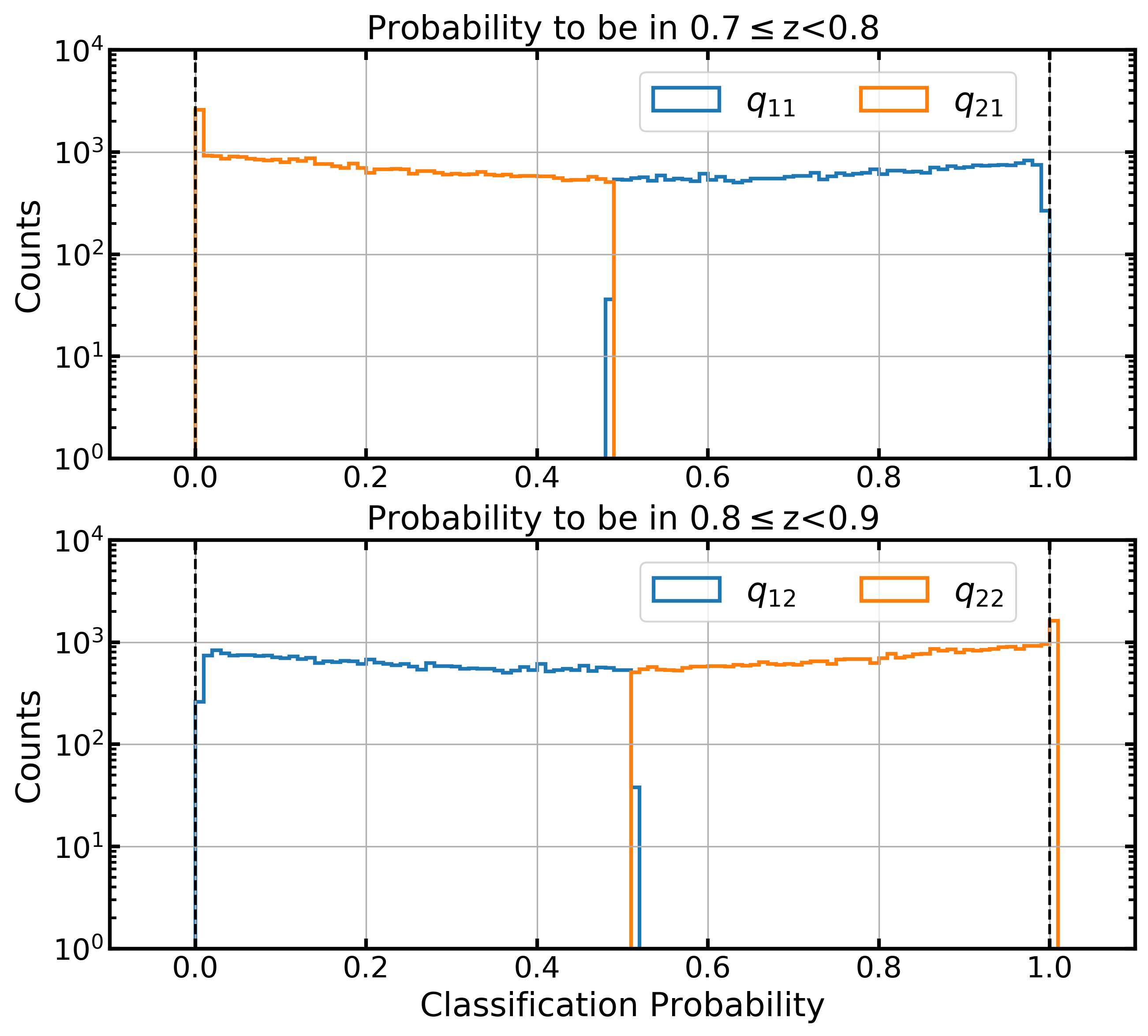}
	\caption{Distribution of the classification probabilities to be in bin 1 (upper panel) or bin 2 (lower panel) for the \replaced{mock}{toy} galaxy sample of Section~\ref{sec: results 2bin}. As introduced in Section~\ref{sec: estimators}, \replaced{$q^{\alpha\beta}$}{$q_{\alpha\beta}$} is the probability of the observed Type-$\alpha$ galaxy to be a true Type-$\beta$ galaxy. We see that given the \pz\ uncertainties, the probability to be in a given target tomographic bin has a broad range. \added{Note that the two panels are mirror images of one another, as dictated by the identity in Equation~\ref{eq: q identity}.}
	}
	\label{fig: mock qs}
\end{figure}

\begin{figure}[]
	\centering
	\includegraphics[width=\linewidth, trim={1 1 1 50}, clip=true]{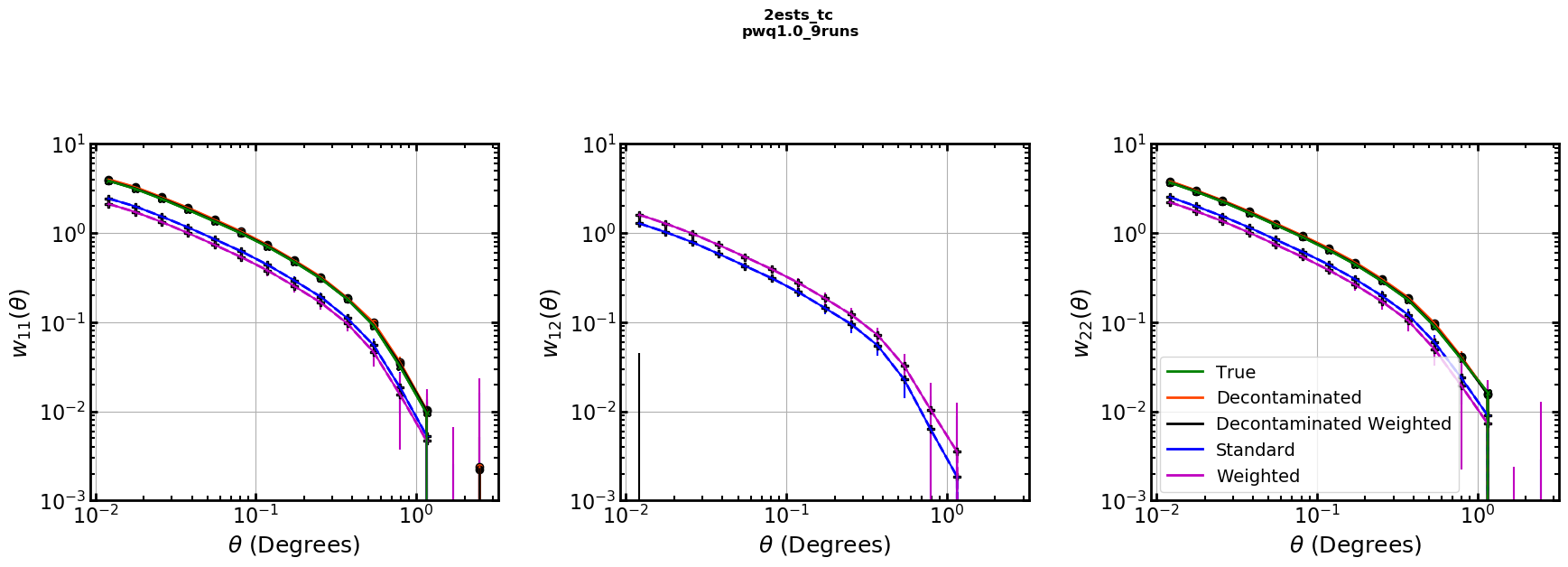}
	\caption{Correlation functions \added{estimates} and the \added{estimator} variance in the \replaced{mock}{toy} galaxy sample with only two redshift bins (presented in Section~\ref{sec: results 2bin}). We see that just as \replaced{standard decontamination}{\ttt{Decontamination}} (red) recovers the truth (green) using the correlations on the contaminated subsamples (blue), the \replaced{decontaminated weighted}{\ttt{Decontaminated Weighted}} estimator (black) recovers the truth from the \replaced{weighted}{\ttt{Weighted}} correlations on the entire observed sample (magenta), without needing to divide the observed sample into subsamples. \added{We also note that the decontaminated estimators reduce the variance on the CF estimates, as indicated by the error bars here.}}
	\label{fig: mock results}
\end{figure}

\begin{figure}[t!]
	\centering
	\includegraphics[width=\linewidth, trim={1 1 1 50}, clip=true]{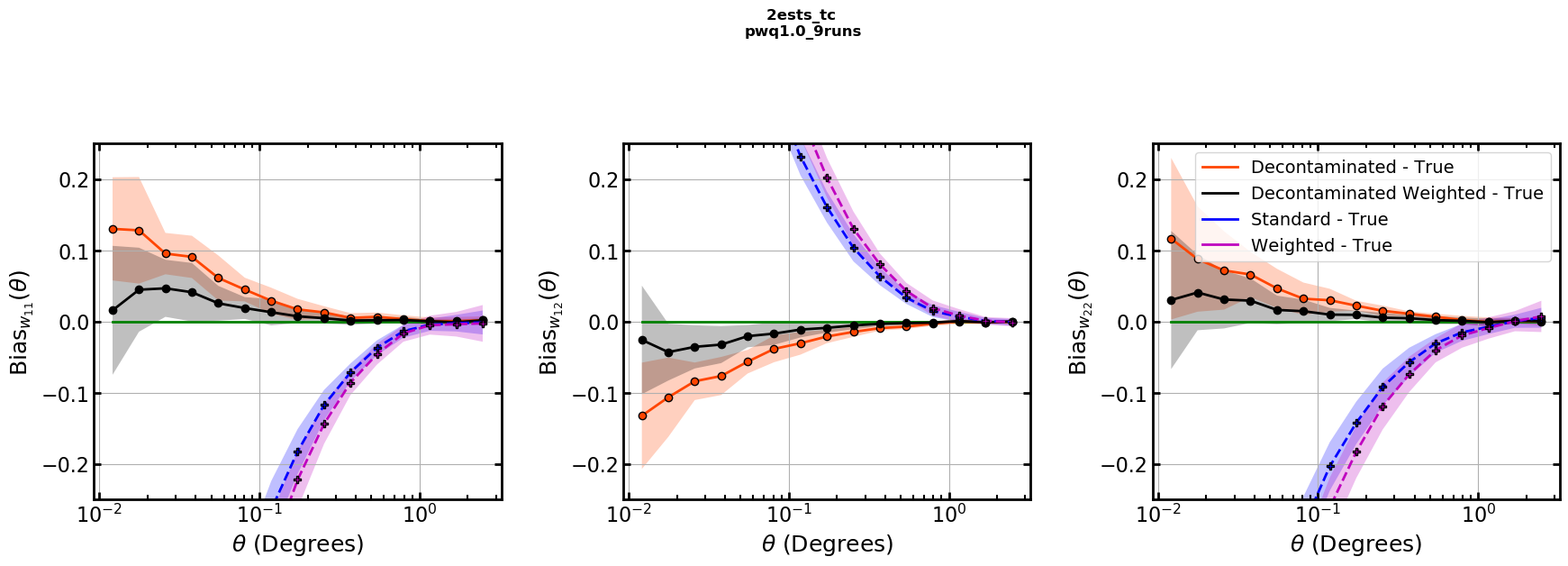}
	\caption{\replaced{Residuals for}{Bias in} correlation functions for the \replaced{mock}{toy} galaxy sample of Section~\ref{sec: results 2bin}, with 1$\sigma$ \replaced{errors}{uncertainties in each estimator} indicated with the shaded regions. We see that the \replaced{weighted}{\ttt{Decontaminated Weighted}} estimator (black) leads to \replaced{residuals}{a bias} smaller than \replaced{those}{that} from the \replaced{standard}{\ttt{Decontaminated}} estimator (red)\added{; the green line indicates zero bias.}}
	\label{fig: mock residuals}
\end{figure}

\added{
\begin{figure}[t!]
	\centering
	\includegraphics[width=0.75\linewidth, trim={5 5 5 5}, clip=true]{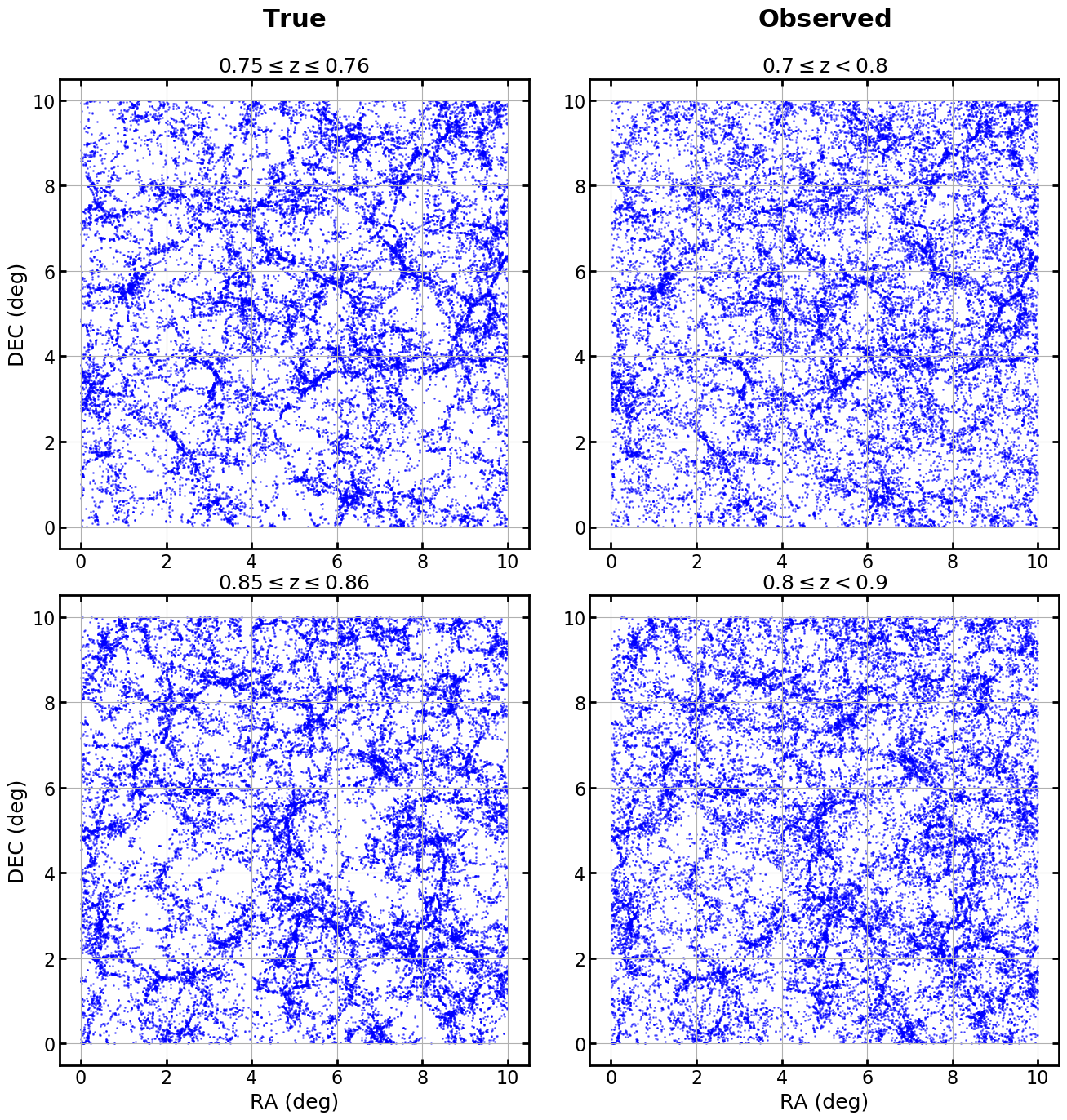}
	\caption{True and observed positions of galaxies for the idealized galaxy sample of Section~\ref{sec: results 2bin}, where all the true galaxies lie at $0.75 \leq z \leq 0.76$, $0.85 \leq z \leq 0.86$. We see that redshift binning of galaxies based on photo-$z$ point estimates modifies the LSS due to the redshift contamination.}
	\label{fig: ra-dec}
\end{figure}
}
Then, using the observed \added{\pz\ }PDFs, we calculate the classification probabilities as the integral of the PDFs within the target redshift bin. Note that since we are simulating only two bins, we use Gaussian PDFs truncated at $z=0.7$ and $z=0.9$ to ensure that we conserve the number of true and observed galaxies; this yields a slight bias in the PDF integrations, which we correct to make the overall classification probabilities unbiased, i.e., $\ev{q_i^{AB}} = f_{AB}$, where the average is checked over redshift intervals with $\Delta z = 0.02$\added{, while ensuring the de-biased probabilities remain in the range 0-1. For real data, this debiasing should be possible utilizing a limited set of spectroscopic redshifts}. Figure~\ref{fig: mock qs} shows the distribution of the final classification probabilities for all the galaxies in our observed sample.
In order to estimate the various correlation functions (two auto, one cross) and their variance, we consider the \replaced{10}{9} patches: the mean across the \replaced{ten}{nine} samples gives us the mean estimate of the respective correlation function while \replaced{the standard deviation across them describes the variance}{we calculate the estimator variance as $\ev{\bcbr{ \wabest{i}{} - \wabtrue{i}{}}^2}$ where $i$ runs over all the correlations (both auto and cross) and the expectation value is over all the realizations; note that this variance is not sensitive to the sample variance but only a measure of the estimator variance, which we can calculate explicitly given that we have access to the true CFs in each of the nine patches}. Note that for each of the patches, we calculate five types of the three correlation functions: those in the true subsamples; those using the \replaced{standard}{\ttt{Standard}} estimator on the contaminated observed subsamples, followed by those from the \replaced{decontaminated standard}{\ttt{Decontaminated}} estimators; and those using the \replaced{weighted}{\ttt{Weighted}} estimator, followed by the \replaced{decontaminated weighted}{\ttt{Decontaminated Weighted}} ones. Also, we use a random catalog that is 5x the size of the data catalog, and restrict \added{CF calculation} to \replaced{0.001-1deg}{0.01-3deg} scales. Figure~\ref{fig: mock results} shows our results, with both the correlation functions and their variance. As expected, the cross correlations with\replaced{out de}{ }contamination are non-negligible, taking signal away from the two auto-correlations. \added{Decontamination lowers the amplitude of the cross-correlations, and we}\deleted{We} find that both estimators \deleted{effectively }correct for the contamination and \replaced{recover the truth}{reduce the bias, leading to estimates closer to the truth}. This is more apparent in Figure~\ref{fig: mock residuals}, where we show the \replaced{residuals on}{bias in} the correlation functions (i.e., difference from the truth\added{ calculated as $\ev{ \wabest{i}{} - \wabtrue{i}{} }$ where $i$ runs over all the correlations (both auto and cross) and the expectation value is over all the realizations}). We note that the \replaced{weighted}{\ttt{Decontaminated Weighted}} estimator is unbiased after decontamination -- a reassuring result. \added{We also note that our decontaminated estimators reduce the variance on the CF estimates, as indicated by the error bars in Figure~\ref{fig: mock results}. } \deleted{Finally, Figure~\ref{fig: mock sigmas} shows the variance on the correlation functions where we see a lower variance in contaminated estimates given that they draw LSS from a larger volume, and that while correcting for contamination increases the variance, the weighted estimator leads to a variance similar to the standard one, demonstrating that using the full sample without subsampling does not lead to any loss of information.}  

\subsection{Realistic Example: Optimistic Case\label{sec: results 3bin}}
Now we consider a more realistic scenario: a true galaxy sample with $0.7\leq z \leq 1.0$, with three redshift bins ($0.7 \leq z < 0.8$, $0.8 \leq z < 0.9$, $0.9 \leq z < 1.0$) for the tomographic clustering analysis. As before, we query the galaxies in \replaced{ten 5x5}{nine 10x10} deg$^2$ patches along Dec = 0\added{, and model their \pz s\ assuming Gaussian PDFs for all the galaxies with $\sigma_z = 0.03(1+z)$ as discussed at the beginning of Section~\ref{sec: validation + results}}; all patches have \added{a} similar number of galaxies (\replaced{260K-300K}{1080K-1147K}) and face similar contamination (\replaced{23-27\%, 41-47\%, 19-25\%}{23-26\%, 44-46\%, 19-23\%} in the three tomographic bins, respectively). Note that our chosen bins\deleted{ and their width} are realistic, as a tomographic analysis for 10 redshift bins with $\Delta z = 0.1$ is currently planned for dark energy science studies with LSST \citep{DESC-SRD2018}\added{; our treatment of \pz s, however, is optimistic in the assumption of Gaussian \pz\ PDFs}.

\begin{figure}[t!]
	\centering
	\includegraphics[width=.5\linewidth]{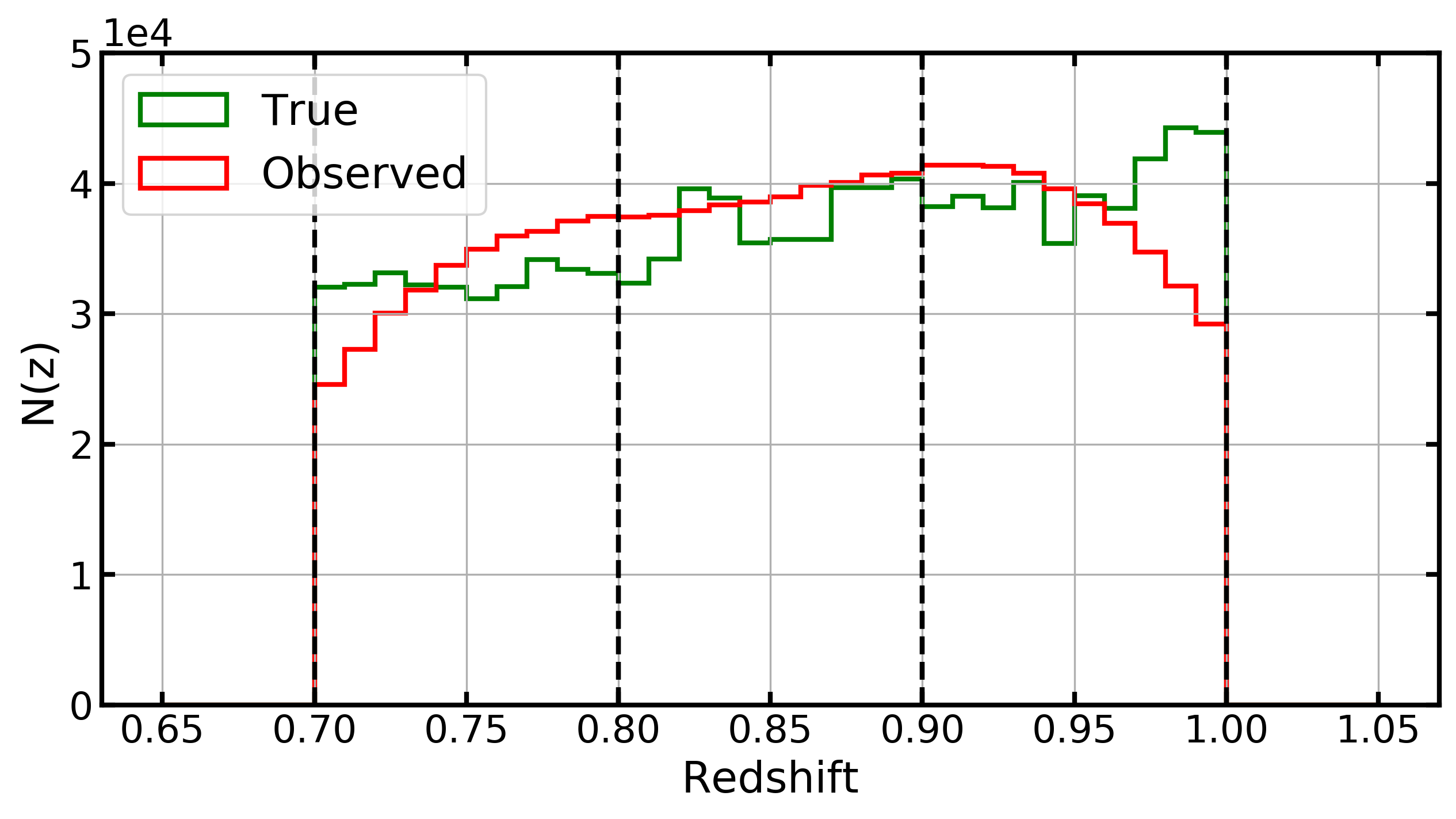}
	\caption{True and observed redshift histograms for the \added{mock }galaxy sample of Section~\ref{sec: results 3bin}, with bin edges shown using the vertical dashed lines. We see that the \pz\ uncertainties lead to a smearing of the redshift information, while the truncation \deleted{at }of the edge-bins makes the $N(z)$ biased near the outermost edges.}
	\label{fig: 3bin zhists}
\end{figure}

Figure~\ref{fig: 3bin zhists} shows the distributions of the true redshifts and the \pz s using one of the patches (with \replaced{263,836 galaxies, and 24\%, 47\% and 20\%}{1,095,404 galaxies, and 24\%, 45\% and 22\%} contamination in the three redshift bins, respectively). We note that the middle bin sees the largest and most realistic contamination -- the case that will be true for most of the LSST bins, hence making this example a relevant one. Note that the bin edges see the impacts of artificially having contamination from only one side.
\begin{figure}[t!]
	\centering
	\includegraphics[width=.5\linewidth]{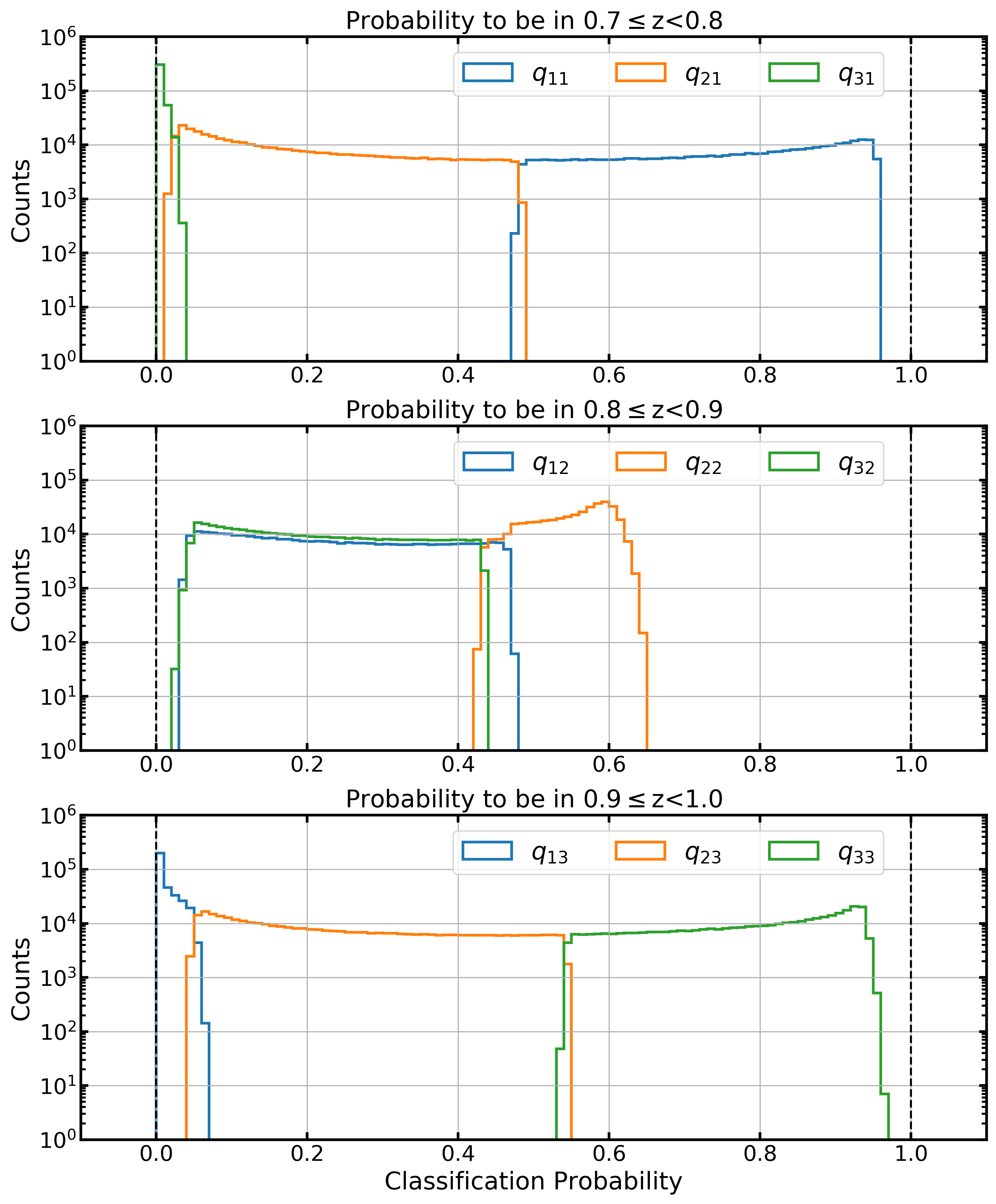}
	\caption{Distribution of the classification probabilities to be in the three target redshift bins for the \added{mock }galaxy sample of Section~\ref{sec: results 3bin}. The middle bin sees the largest contamination and therefore has no objects that have a very high probability to be in any target bin.}
	\label{fig: 3bin qs}
\end{figure}
\begin{figure}[t!]
	\centering
	\includegraphics[width=\linewidth, trim={1 1 1 50}, clip=true]{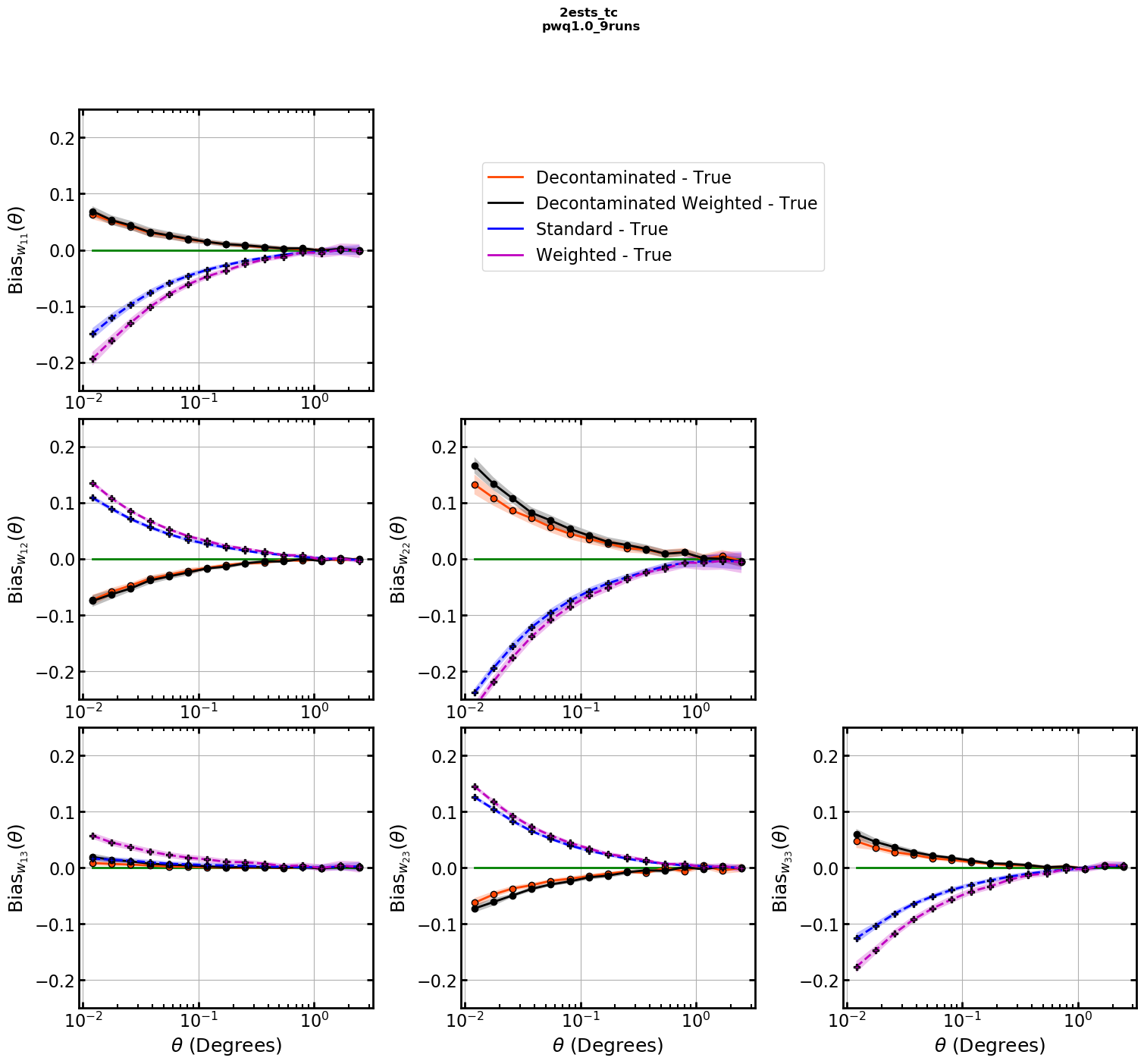}
	\caption{\replaced{Residuals}{Bias} in the correlation functions in the three sample case of Section~\ref{sec: results 3bin}, with 1$\sigma$ \replaced{errors}{uncertainties in each estimator} indicated with the shaded regions. We see that as in the \replaced{mock case}{toy example in Section~\ref{sec: results 2bin}}, just as \replaced{standard decontamination}{\ttt{Decontamination}} (red) \replaced{recovers the truth (green)}{reduces the bias} using the correlations on the contaminated subsamples (blue), the \replaced{decontaminated weighted}{\ttt{Decontaminated Weighted}} estimator (black) \replaced{recovers the truth}{reduces the bias} from the \replaced{weighted}{\ttt{Weighted}} correlations on the entire observed sample (magenta), without needing to divide the observed sample into subsamples\added{; the green line indicates zero bias}.}
	\label{fig: 3bin residuals}
\end{figure}

Figure~\ref{fig: 3bin qs} shows the distribution of the classification probabilities for all the galaxies. Again we note that given the large contamination rates for the middle bin, the classification probabilities are far from unity, indicating that no observed galaxy has a very high probability to be in any target bin. As before, we calculate the various correlations for each of the \replaced{10}{nine} patches and estimate the mean and the variance across the calculations. Figure~\ref{fig: 3bin residuals} \replaced{shows}{illustrates} our results, showing only the \replaced{residuals}{estimator bias} for brevity, where we see that \replaced{our weighted}{the \ttt{Decontaminated Weighted}} estimator leads to \replaced{residuals}{a bias} that \replaced{are}{is} comparable to \replaced{those}{that} using the \replaced{standard estimators}{\ttt{Decontaminated} estimator}\added{, both of which are smaller than from those without decontamination}. We note that the \replaced{standard}{\ttt{Decontaminated}} estimator performs similar to \replaced{weighted}{\ttt{Decontaminated Weighted}}, potentially due to the correlation functions in the three redshift bins being similar\added{. We also note that there is a weak residual bias in the decontaminated estimates, which is likely caused by our simple debiasing of the classification probabilities}. \deleted{Finally, Figure~\ref{fig: 3bin sigmas} shows the variance on the correlation functions where we see that while accounting for contamination increases the variance, the weighted estimator leads to a variance similar to that from the standard estimator.}
\replaced{ }{

As a more comprehensive metric for comparing the various estimators, we consider the covariances in correlation functions across the three redshift bins for an example $\theta$-bin. Specifically, given that we have access to the truth here, we first calculate the covariances in the estimators without accounting for the LSS sample variance -- this we term as the ``estimator covariance" and calculate as $\ev{\bcbr{ \wabest{i}{} - \wabtrue{i}{}}\bcbr{\wabest{j}{} - \wabtrue{j}{} }}$ where $i, j$ run over all the correlations (both auto and cross) and the expectation value is over all the realizations\footnote{We calculate covariances using the \ttt{numpy.cov} function, which automatically subtracts off the mean for each variable (which, in this case, is the residual bias for each estimator); the default parameters of the function also account for the lost degree-of-freedom (i.e., using $N-1$ when calculating the average, where $N$ is the number of realizations).}; note here that the diagonal of this covariance matrix is the estimator variance used to generate uncertainties shown in Figures~\ref{fig: mock results}-\ref{fig: mock residuals} and Figure~\ref{fig: 3bin residuals}. We show the estimator covariances for the \added{mock }galaxy sample considered here in Figure~\ref{fig: 3bin est cov}, where we see that without decontamination, the covariances are large, as expected given the strong mixing of the samples. Both decontaminated estimators effectively reduce the covariances, with \ttt{Decontaminated Weighted} outperforming \ttt{Decontaminated}.

Then we consider the covariances accounting for the LSS sample variance -- this we term as the ``full covariance" and calculate as $\ev{\bcbr{ \wabest{i}{} - \ev{\wabest{i}{}} }\bcbr{\wabest{j}{} - \ev{ \wabest{j}{}} }}$ where $i, j$ again run over all the correlations and the expectation value is over all the realizations; these are shown in Figure~\ref{fig: 3bin full cov}. We see that without decontamination, the clustering information is smeared across the CF-space and is much in contrast from the true covariances. However, both of our decontaminated estimators are able to approximate the true covariances effectively, hence achieving their purpose of correcting for sample contamination. We also note here that decontamination does not simply diagonalize the covariance matrices but instead reduces off-diagonal elements appropriately; diagonalization would not account for true covariances that exist between auto- and cross- CFs for neighboring bins due to shared LSS. Finally, comparing with Figure~\ref{fig: 3bin est cov}, we note that LSS sample variance largely dominates over the estimator variance for the 10x10 patches considered here -- a reassuring result; a comparison between the two sources of variance for larger effective survey area is left for future work.}

\added{
\begin{figure}[t!]
	\centering
	\includegraphics[width=0.7\linewidth, trim={365 1 1 20}, clip=true]{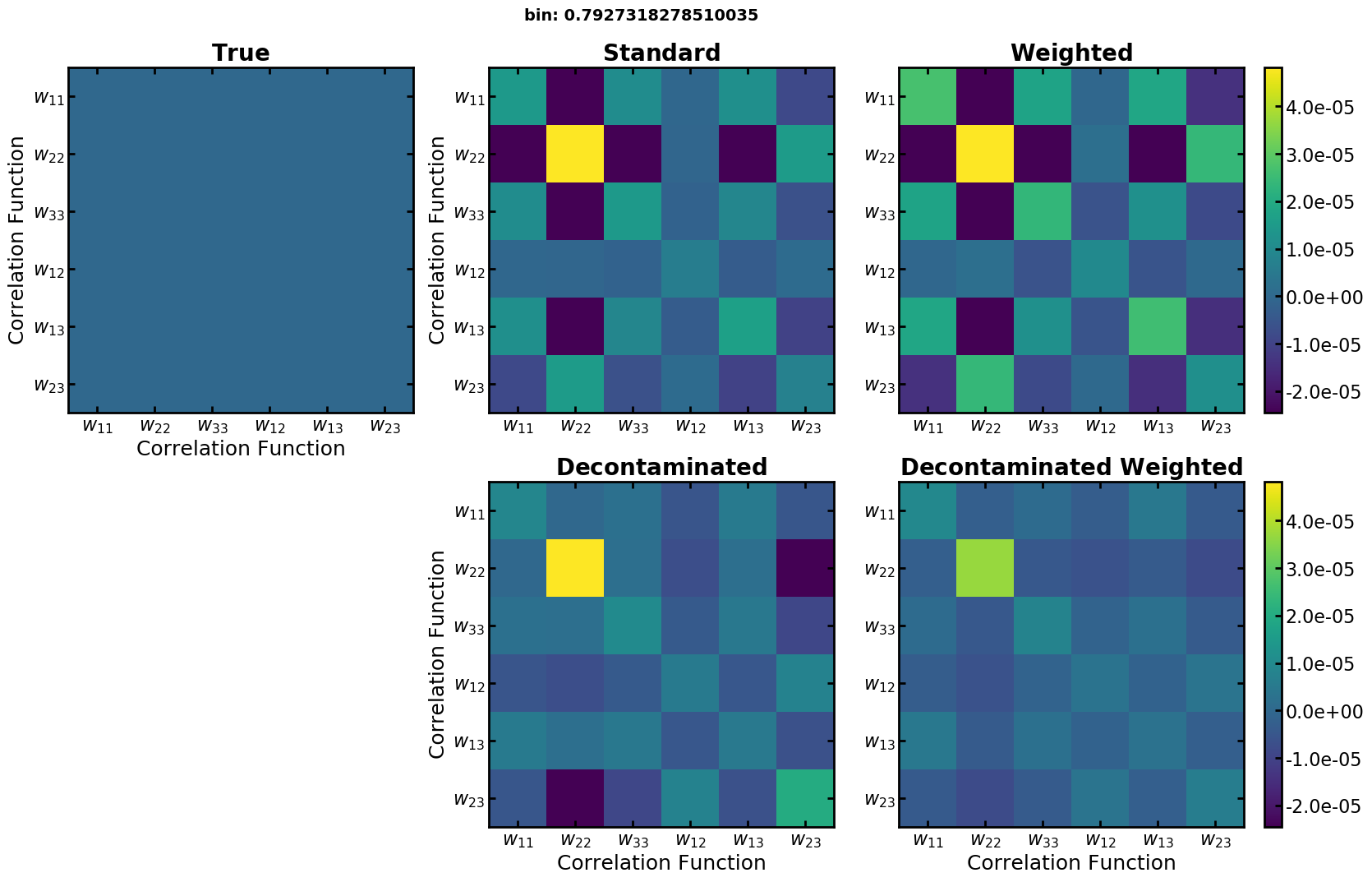}
	\caption{Estimator covariances across redshift bins for the case with three target redshift bins of Section~\ref{sec: results 3bin} for an example theta-bin (with $\theta=0.79$ degrees as nominal center of the bin in $\mathrm{log}(\theta)$); these probe the covariances in the estimators without accounting for LSS sample variance. Here, $w_{\alpha\beta}$ refers to the CF between galaxies in redshift bins $\alpha$ and $\beta$, and as noted in the text, we estimate the estimator covariance as $\ev{\bcbr{ \wabest{i}{} - \wabtrue{i}{}}\bcbr{\wabest{j}{} - \wabtrue{j}{} }}$ for each estimator, where $i, j$ run over all the correlations (both auto and cross) and the expectation value is over all the realizations. Note that this is not sensitive to sample variance since the true CF for each realization is subtracted from the observed CF for that realization. The left column shows estimator covariances in contaminated samples constructed using \pz\ point estimates before (top) and after (bottom) decontamination, while the right column shows the estimator covariances in CF estimates using our \ttt{Weighted} estimator before (top) and after (left) decontamination. We see that our new decontaminated estimators reduce the covariances, with \ttt{Decontaminated Weighted} outperforming \ttt{Decontaminated}.}
	\label{fig: 3bin est cov}
\end{figure}
\begin{figure}[t!]
	\centering
	\includegraphics[width=\linewidth, trim={1 1 1 20}, clip=true]{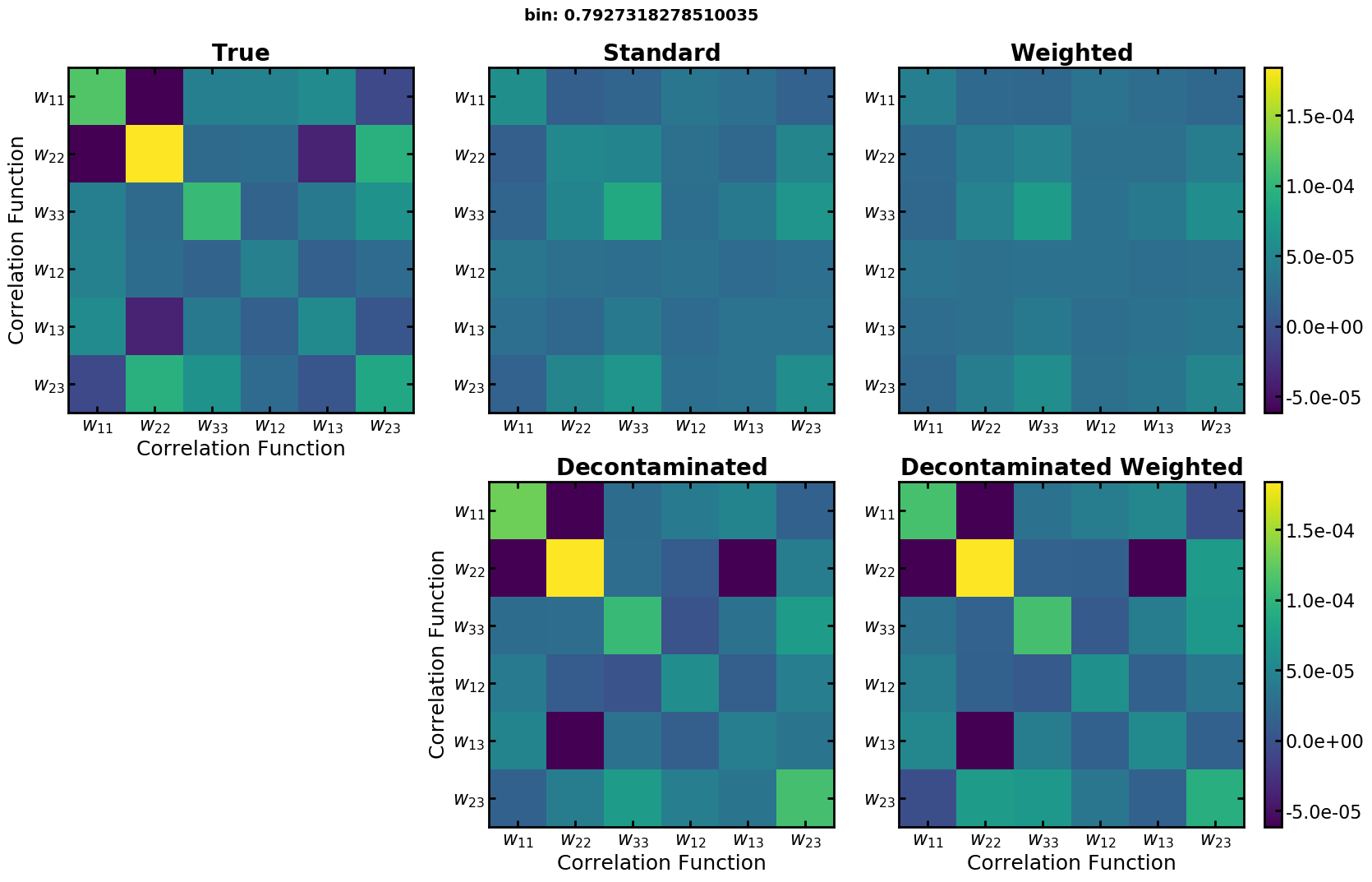}
	\caption{Full covariances across redshift bins for the case with three target redshift bins of Section~\ref{sec: results 3bin} for an example theta-bin (with $\theta=0.79$ degrees as nominal center of the bin in $\mathrm{log}(\theta)$); these probe the covariances in the estimators while accounting for LSS sample variance. Here, $w_{\alpha\beta}$ refers to the CF between galaxies in redshift bins $\alpha$ and $\beta$, and e.g., $w_{11}$ and $w_{12}$ are correlated since LSS at the boundary of the two bins makes $w_{12}$ non-zero and contributes to $w_{11}$. As noted in the text, we calculate these full covariances as $\ev{\bcbr{ \wabest{i}{} - \ev{\wabest{i}{}} }\bcbr{\wabest{j}{} - \ev{ \wabest{j}{}} }}$ for each estimator, where $i, j$ again run over all the correlations and the expectation value is over all the realizations. The top left panel shows the true covariances across multiple realizations of the LSS, the middle column shows covariances in contaminated samples constructed using \pz\ point estimates before (top) and after (bottom) decontamination, while the rightmost column shows the covariances in CF estimates using our \ttt{Weighted} estimator before (top) and after (left) decontamination. We see that our new decontaminated estimators approximate the true covariances, successfully accounting for sample contamination arising from \pz\ uncertainties.}
	\label{fig: 3bin full cov}
\end{figure}
}

\added{
\subsection{Realistic Example: Pessimistic Case\label{sec: results 3bin bimodal}}
Now we consider a more pessimistic scenario for the true galaxy sample of Section~\ref{sec: results 3bin}: instead of having all the galaxies with well-behaved Gaussian \pz\ PDFs, we assign half of the galaxies bimodal \pz\ PDFs -- a scenario where standard $N(z)$ forward modeling might be problematic. Specifically, the Gaussian \pz\ PDFs are constructed as described above: by drawing a random number from a Gaussian of width $\sigma = 0.03(1+z_{\mathrm{true}})$, with the observed \pz\ PDF being a Gaussian centered at $z_{\mathrm{draw}}$ and with width $\sigma = 0.03(1+z_{\mathrm{draw}})$. In contrast, the bimodal \pz\ PDF are constructed with one mode at the true redshift and another randomly chosen to be $\pm$ 0.13 away (while ensuring the second mode remains in the redshift range of 0.7-1.0); 0.13 separation mimics a degeneracy arising from Balmer vs. 4000\AA\ decrement at $\sim$7\% separations in $1+z$. This treatment leads to slightly higher contamination rates: 39-42\%, 54-57\%, 33-36\% in the three tomographic bins, respectively. To illustrate the difference between the two cases more explicitly, Figure~\ref{fig: pdfs} shows an example set of PDFs for the case of all-Gaussian PDFs vs. half-bimodal ones.

Figure~\ref{fig: 3bin zhists bimodal} shows the distributions of the true redshifts and the \pz s using one of the patches (with 1,095,404 galaxies as before, but now with 40\%, 55\% and 35\% contamination in the three redshift bins, respectively). Comparing it to Figure~\ref{fig: 3bin zhists}, we see that the distribution is slightly more biased, although the middle redshift bin sees a comparable observed redshift distribution; and as before, the bin edges see the impacts of artificially having contamination from only one side.

Figure~\ref{fig: 3bin qs bimodal} shows the classification probabilities for all the galaxies here; comparing it to Figure~\ref{fig: 3bin qs}, we see that the classification probabilities are now more varied, with more objects in the edge-bins with larger classification probabilities due to the bimodality in some of the \pz\ PDFs. As before, we calculate the various correlations for each of the nine patches and estimate the mean CFs and the covariances. Figure~\ref{fig: 3bin residuals bimodal} shows the residuals in the CF estimates, and we see that the decontaminated estimators are able to reduce the bias significantly. Figure~\ref{fig: 3bin est cov bimodal} shows the estimator covariance matrices where we see that as in the all-Gaussian case, our decontaminated estimators lead to lower estimator covariances, with \ttt{Decontaminated Weighted} outperforming \ttt{Decontaminated} slightly more strongly than in Figure~\ref{fig: 3bin est cov}. Finally, Figure~\ref{fig: 3bin full cov bimodal} shows the full covariance matrices. Here too, we see that as in Figure~\ref{fig: 3bin full cov} for the all-Gaussian case, our decontaminated estimators approximate the true covariances more effectively with those without decontamination.

\begin{figure}[t!]
	\hspace*{5em}
	\begin{minipage}{0.4\paperwidth}
		\includegraphics[trim={440 310 230 5}, clip=true, width=0.3\paperwidth]{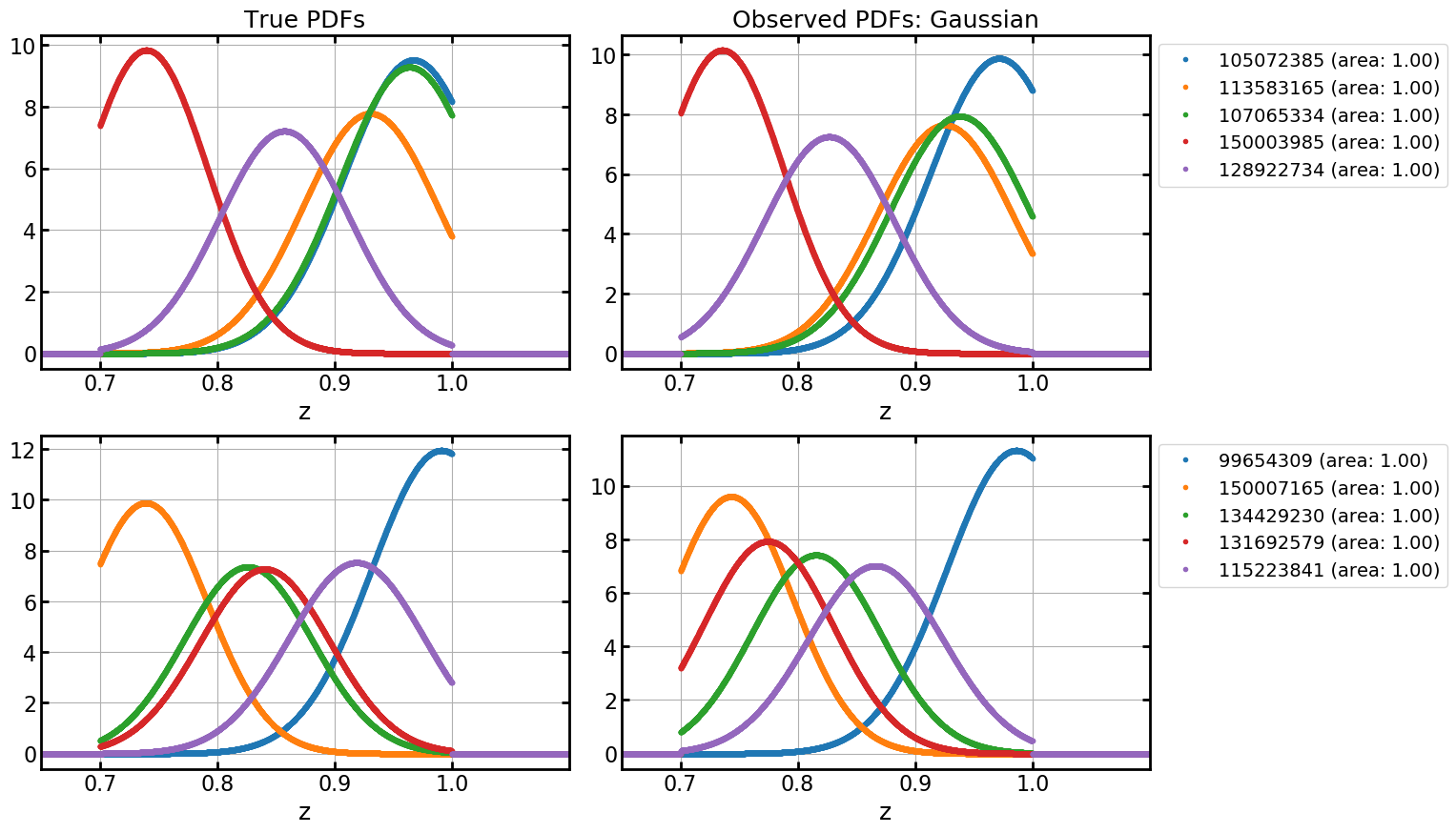}
	\end{minipage}\
	\hspace*{-2em}
	\begin{minipage}{0.4\paperwidth}
		\includegraphics[trim={440 310 230 5}, clip=true, width=0.3\paperwidth]{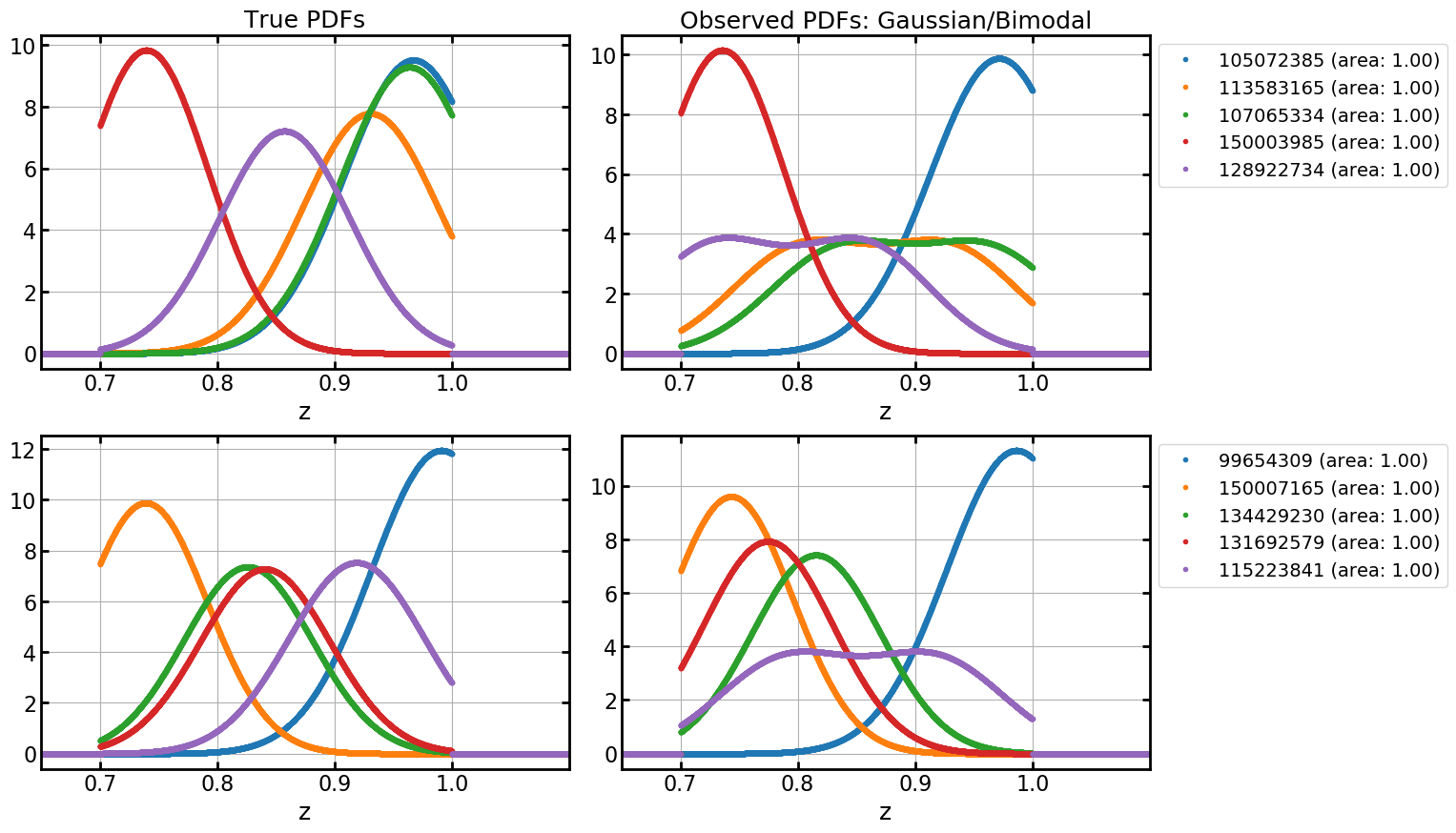}
	\end{minipage}\
	\caption{An example set of PDFs to compare the case of all-Gaussian PDFs of Section~\ref{sec: results 3bin} vs. the case presented in Section~\ref{sec: results 3bin bimodal} where half of the galaxies have bimodal PDFs. The left panel shows the observed \pz\ PDFs for the case of all-Gaussian PDFs while the right panel shows them for the case where half of the galaxies have bimodal PDFs. The colors correspond to the same objects across the panels.
	}
	\label{fig: pdfs}
\end{figure}
\begin{figure}[t!]
	\centering
	\includegraphics[width=.5\linewidth]{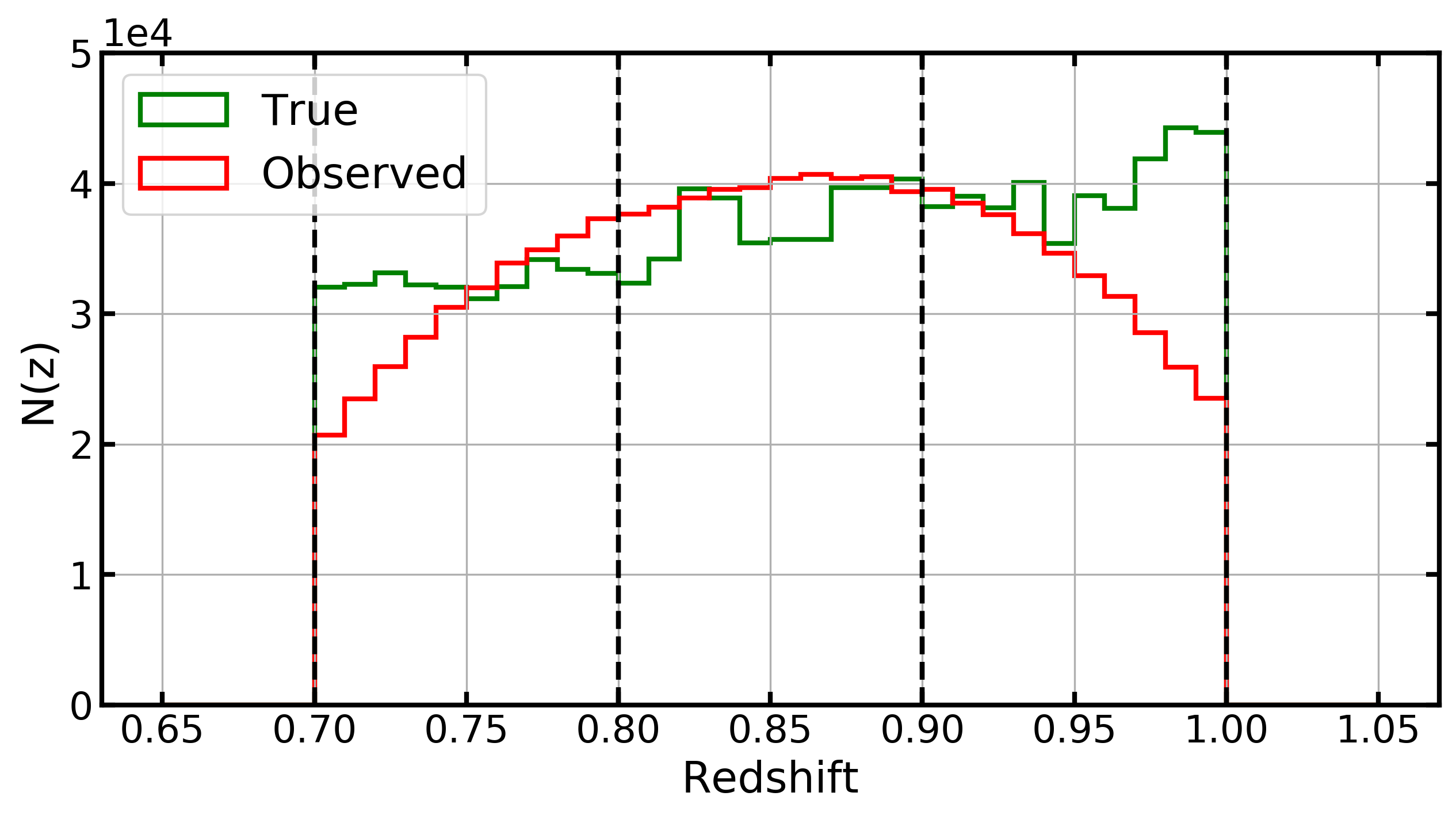}
	\caption{True and observed redshift histograms for the \added{mock }galaxy sample of Section~\ref{sec: results 3bin bimodal}. As in Figure~\ref{fig: 3bin zhists}, the bin edges shown using the vertical dashed lines. We see that as in Figure~\ref{fig: 3bin zhists}, the \pz\ uncertainties lead to a smearing of the redshift information, while the truncation of the edge-bins makes the $N(z)$ biased near the outermost edges.}
	\label{fig: 3bin zhists bimodal}
\end{figure}
\begin{figure}[t!]
	\centering
	\includegraphics[width=.5\linewidth]{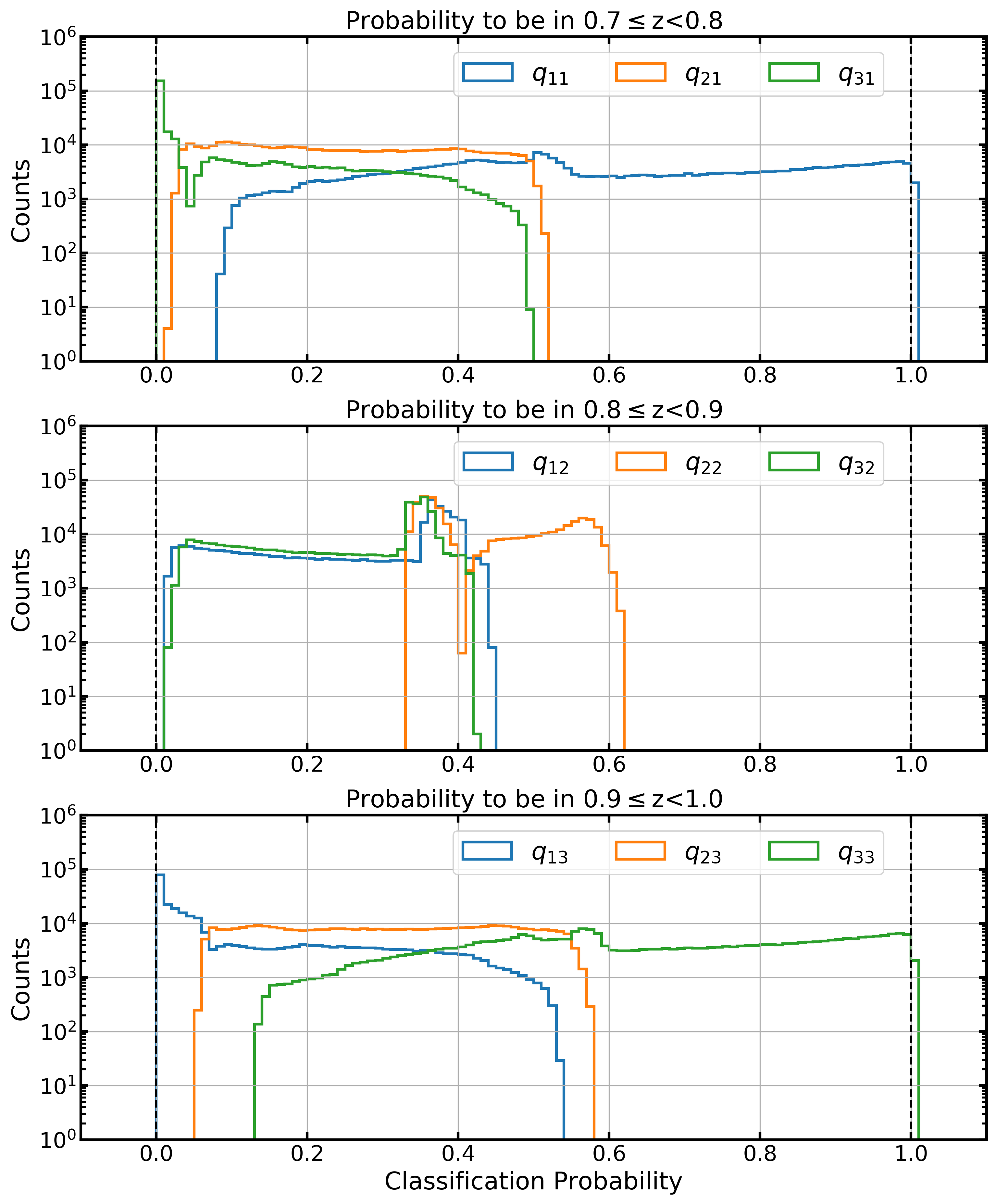}
	\caption{Distribution of the classification probabilities to be in the three target redshift bins for the \added{mock }galaxy sample of Section~\ref{sec: results 3bin bimodal}. As in Figure~\ref{fig: 3bin qs}, the middle bin sees the largest contamination and therefore has no objects that have a very high probability to be in any target bin.}
	\label{fig: 3bin qs bimodal}
\end{figure}
\begin{figure}[t!]
	\centering
	\includegraphics[width=\linewidth, trim={1 1 1 50}, clip=true]{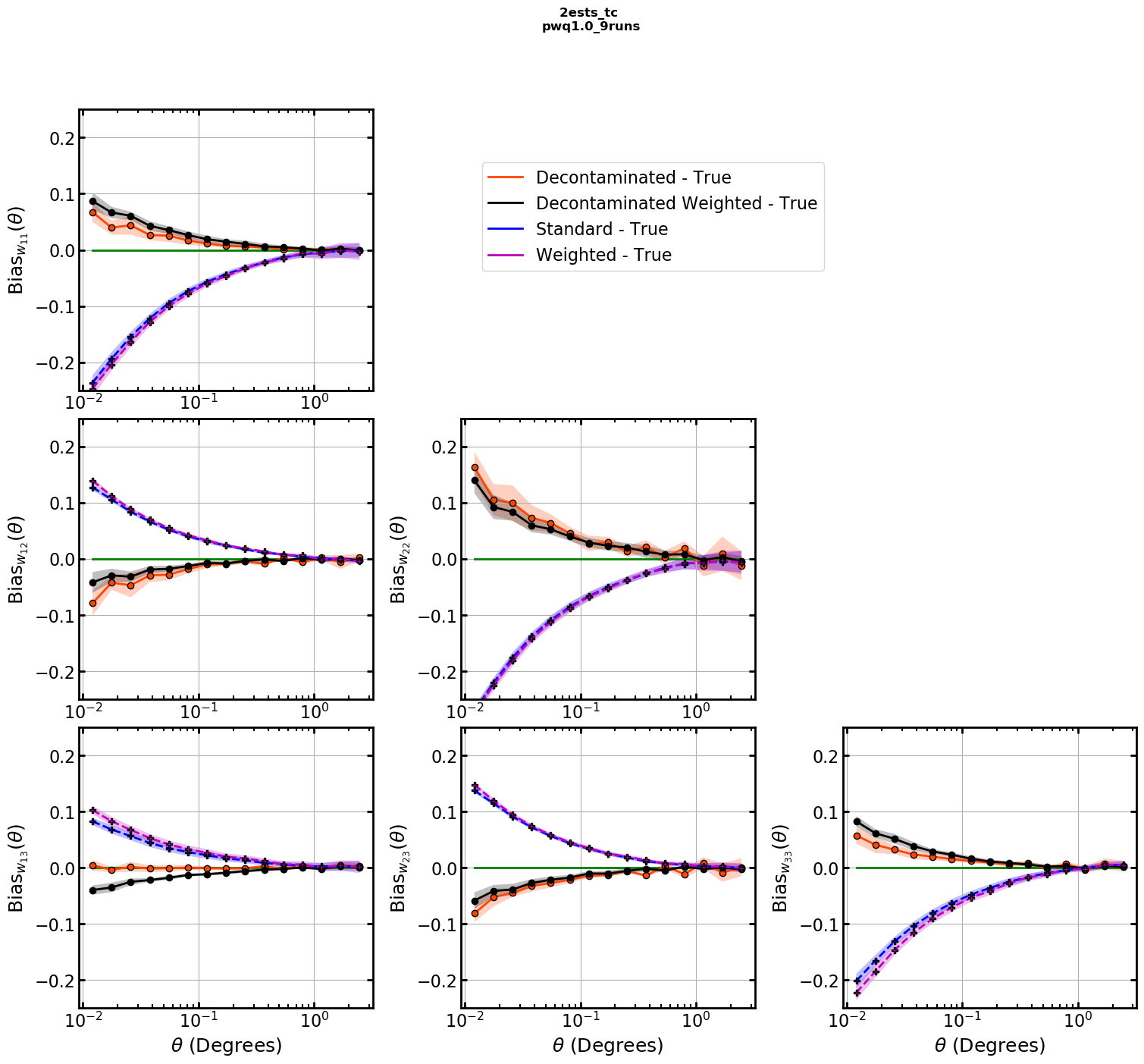}
	\caption{Bias in the correlation functions in the three sample case of Section~\ref{sec: results 3bin bimodal}. As in Figure~\ref{fig: 3bin residuals}, the 1$\sigma$ uncertainties in each estimator are indicated with the shaded regions. We see that as for the all-Gaussian \pz\ PDFs case, both decontaminated estimators significantly reduce the bias and lead to estimates closer to the truth.}
	\label{fig: 3bin residuals bimodal}
\end{figure}
\begin{figure}[t!]
	\centering
	\includegraphics[width=0.7\linewidth, trim={365 1 1 20}, clip=true]{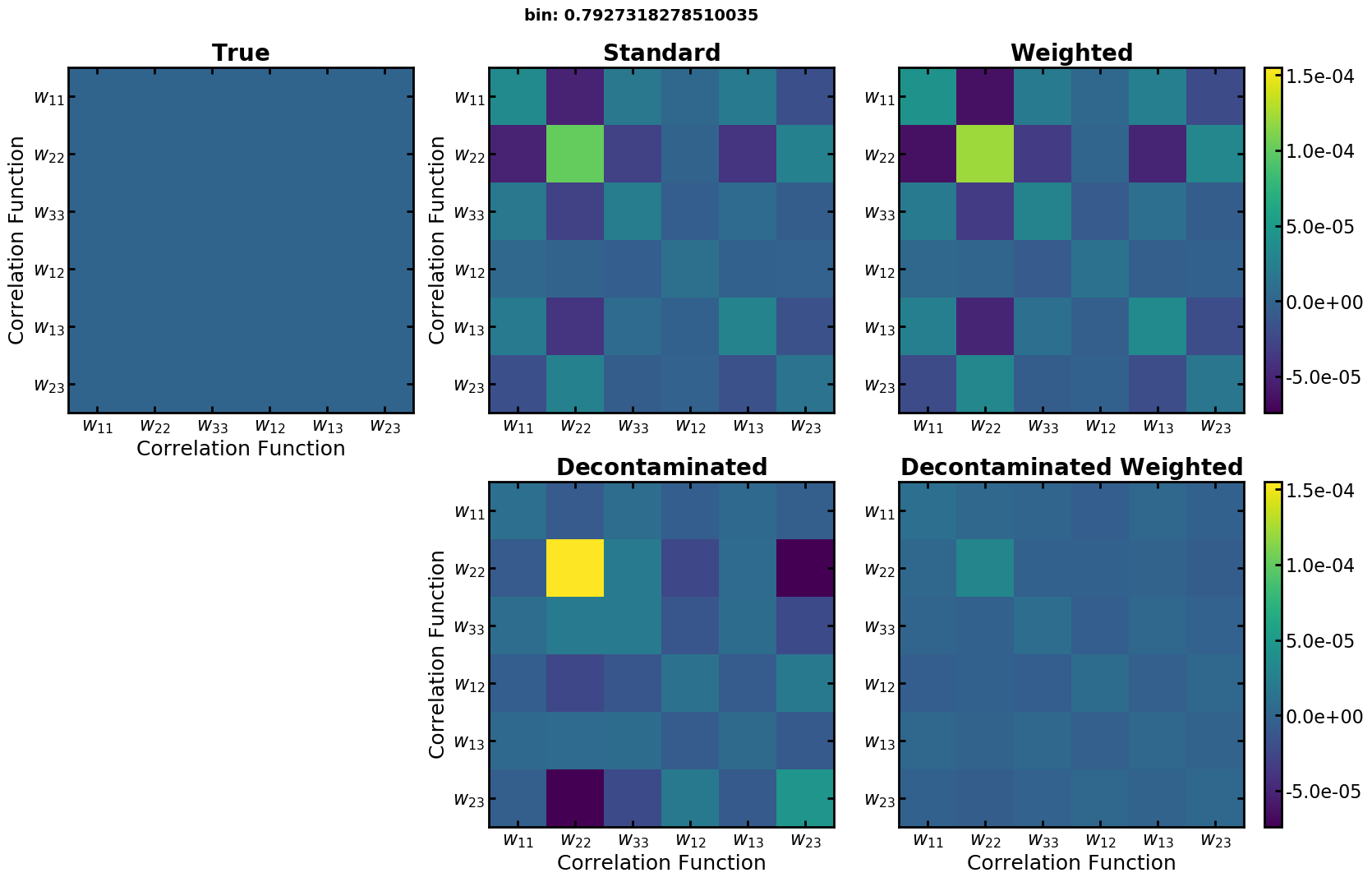}
	\caption{Estimator covariances across redshift bins for the case of Section~\ref{sec: results 3bin bimodal} for the same example theta-bin as in Figure~\ref{fig: 3bin est cov}. As in Figure~\ref{fig: 3bin est cov}, the left column shows estimator covariances in contaminated samples constructed using \pz\ point estimates before (top) and after (bottom) decontamination, while the right column shows the estimator covariances in CF estimates using our \ttt{Weighted} estimator before (top) and after (left) decontamination. We see that our new decontaminated estimators reduce the covariances, with \ttt{Decontaminated Weighted} outperforming \ttt{Decontaminated}.}
	\label{fig: 3bin est cov bimodal}
\end{figure}
\begin{figure}[t!]
	\centering
	\includegraphics[width=\linewidth, trim={1 1 1 20}, clip=true]{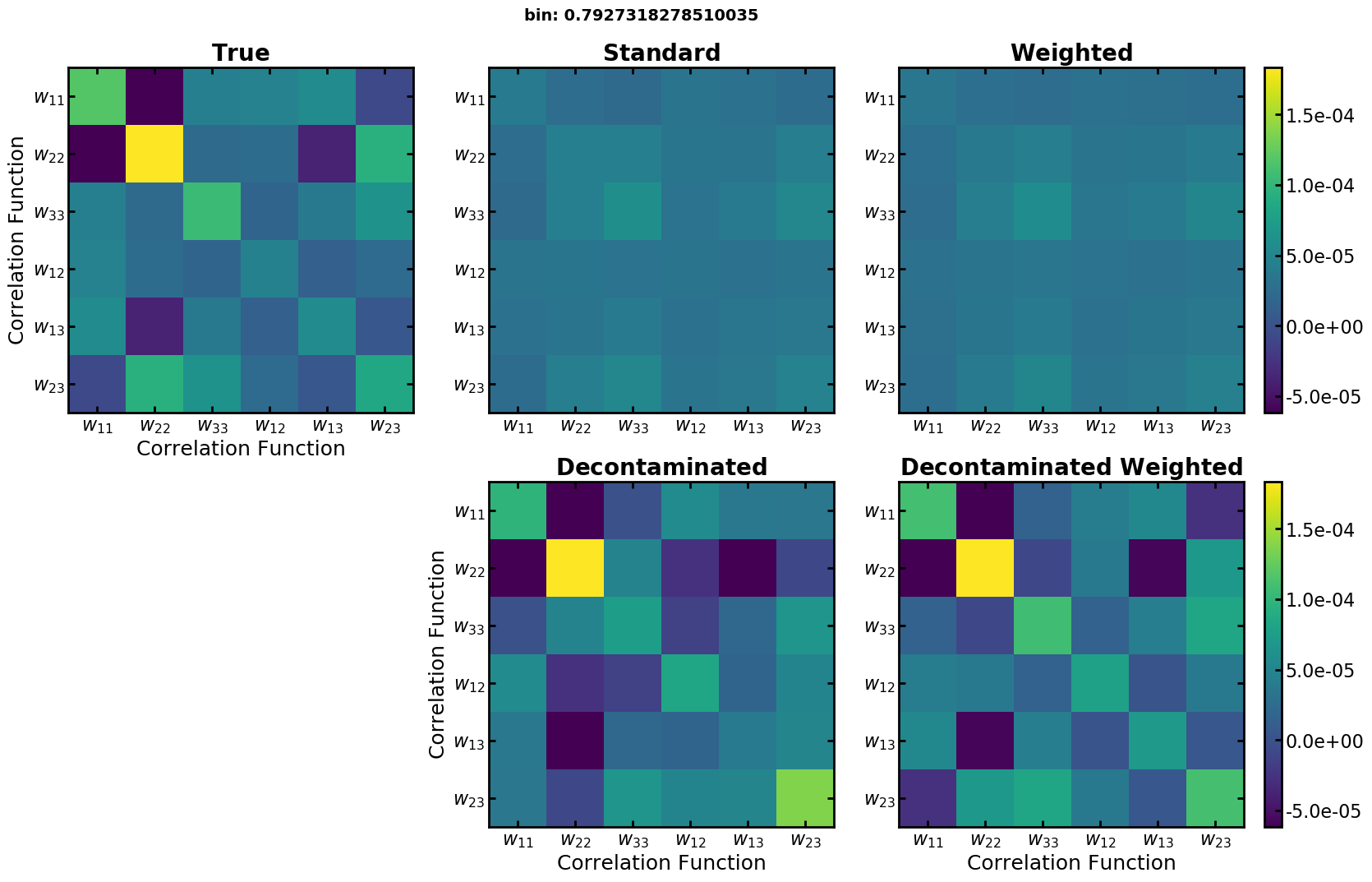}
	\caption{Full covariances across redshift bins for the case of Section~\ref{sec: results 3bin bimodal} for the same example theta-bin as in Figure~\ref{fig: 3bin full cov}. As in Figure~\ref{fig: 3bin full cov}, the top left panel shows the true covariances across multiple realizations of the LSS, the middle column shows covariances in contaminated samples constructed using \pz\ point estimates before (top) and after (bottom) decontamination, while the rightmost column shows the covariances in CF estimates using our \ttt{Weighted} estimator before (top) and after (left) decontamination. We see that our new decontaminated estimators approximate the true covariances, successfully accounting for sample contamination arising from \pz\ uncertainties.}
	\label{fig: 3bin full cov bimodal}
\end{figure}

}

This completes the demonstration of our new \replaced{estimator: it provides for a way to decontaminate correlations, using the full photo-$z$ PDFs and full observed samples, in a framework that can be extended e.g., to minimize variance.}{estimators: they provide for a way to decontaminate correlations, while the \replaced{weighted}{\ttt{Weighted}} estimator specifically allows using the full photo-$z$ PDFs and full observed samples, in a framework that can be extended e.g., to minimize variance.}

\section{Discussion\label{sec: discussion}}
We have presented a \replaced{method}{formalism} to estimate the ACFs in the presence of sample contamination arising from \pz\ uncertainties. We achieve this by \added{a two-fold process: using the information in the contaminated correlations and} utilizing the probabilistic information available via each galaxy's \pz\ PDF\added{ in each target redshift bin}. As mentioned in Section~\ref{sec: intro}, our method avoids forward modeling the contaminated ACFs based on estimated $N(z)$, which is the standard way to handle the \pz\  contamination \added{for cosmological analyses}. We note, however, that forward modeling is effective if \deleted{the goal is a low variance estimation of the cosmological parameters and}the contamination can be modeled effectively; a full investigation of measurements using our method vs. those using forward modeling is left for future work. We also note that the BAO signal is \deleted{generally }washed out by projection and hence its measurement \replaced{may}{should} benefit from our approach.

\replaced{We also note that our work is}{Our estimators are} distinct from \replaced{the various}{previous} work\deleted{s in the literature} employing weighted correlation functions, specifically on three accounts: \replaced{1) to our knowledge, there is no estimator in the literature that uses weights that are dependent on a galaxy's \pz\ PDF to handle \pz\ contamination for galaxy clustering analyses, 2) our proposed estimator considers the galaxies in the entire observed sample and not subsamples based on \pz\ bins}{1) our weighted estimator considers all galaxies in the entire observed sample as a part of every \pz\ bin, 2) to our knowledge, there is no literature on the usage of a decontamination matrix to correct for correlation function contamination, and our \ttt{Decontaminated Weighted} estimator presents a novel way to decontaminate marked correlation functions}, and 3) we weight only the data, and not the randoms. \added{As far as we are aware, the only other estimator in the literature that uses weights that are dependent on a galaxy's \pz\ PDF in a galaxy clustering analysis is \citet{asorey+16} but they employ a threshold to determine whether a galaxy contributes to a given redshift bin and do not allow contributions from a single galaxy to more than one bin.} \replaced{F}{In a further comparison with our work, f}or instance, \citet{Ross+17} employ weights to account for \pz\ uncertainty by weighting both the data and random galaxies in the target subsamples by inverse-variance weights. \citet{Blake+19} also weight both the data and random galaxies to increase the precision with which they can measure the BAO by accounting for the dependency on the environment of the measured signal\deleted{, alongside \citet{Bianchi+17} where both the data and random galaxies are weighted to account for missing information}. In somewhat of a contrast, \citet{Zhu+15} use both weighted data and random pairs, and unweighted random pairs for optimized BAO measurements, while \citet{Morrison+15} employ weighted randoms to account for mitigating survey systematics. \citet{Percival+17}, on the other hand, upweight only their data (data-data, data-random pairs, but not the random-random pairs) for 3D BAO measurements when the spectroscopic data is available only for a subset of the angular sample\added{ while \citet{Bianchi+17} employ a similar weighting to account for missing information}. \deleted{We highlight that none of these studies employs our concept of using the entire observed sample or uses the probabilistic information (available via e.g., PDFs) in the weights when it is available.We are \added{also} not aware of the usage of a decontamination matrix to correct for correlation function contamination in the literature, and our weighted estimator presents a novel way to decontaminate marked correlation functions.}

Since this work introduces a new estimator, we note various avenues for further development. For the 2D case, we can optimize the estimator to be minimum variance by introducing an additional parameter for each pair of galaxies, i.e., $\ttt{w}_{ij, \rm{opt}}^{\alpha\beta}$= $\Upsilon_{ij}(q, k) \ttt{w}_{ij}^{\alpha\beta}$, where $\Upsilon_{ij}(q, k)$ are the optimization parameters that minimize the variance of the estimator for each bin $k$. We note \added{again} that the \replaced{standard decontaminated}{\ttt{Decontaminated}} estimator presented in the text is in fact a special case of the \replaced{weighted}{\ttt{Decontaminated Weighted}} estimator, with the weights set to 1 when the probability is high enough to place an object in a given subsample and 0 otherwise\added{ and then with average contamination fractions used to decontaminate instead of the classification probabilities}.  It is \added{indeed} surprising that the \replaced{standard}{\ttt{Decontaminated}} estimator \replaced{appears to perform}{performs nearly} as well as our \replaced{simple probability-weighted}{\ttt{Decontaminated} probability-\ttt{Weighted}} estimator; this implies either a broad range of optimal weights or, more likely, that the optimal weights lie somewhere between these two simplistic approaches.  Optimization of the weights will be an important aspect of applying the new estimator. Furthermore, since we have introduced general pair weights, we can incorporate Bayesian priors on the correlation functions, based on current measurements, or when measuring correlation functions for different galaxy types, as then, we can incorporate priors that are dependent on the separations, e.g., accounting for one galaxy sample clustering strongly on smaller scales. This will call for an in-depth analysis of the covariance matrices for the various correlation functions. Also, we can extend the weighting scheme to harmonic space, where it will be relevant for a tomographic analysis for LSST (Awan et al., in prep).

We also note that our method can handle other kinds of contamination, e.g., star-galaxy contamination, where probabilistic models for whether an object is a star or a galaxy can inform the weights for each object in our observed sample\added{; this is possible since neither decontamination nor the pair weights have an explicit redshift dependence, hence allowing decontaminating and weighting any types} . Finally, we can also extend the 2D formulation to 3D, where it will be relevant for \href{http://hetdex.org/}{HETDEX} \citep{Hill+08}, Euclid and WFIRST, as they face emission line contaminants, as well as LSST where the projected correlation function will be measurable (without tomographic binning). Note that for the 3D case in real space, we must treat the random catalogs more carefully than in 2D; in the \replaced{latter}{2D} case\added{ considered here}, we have not made a distinction between random \replaced{samples}{catalogs} for the different samples as \replaced{the samples}{they} are spatially overlapping with the same selection function -- a case that does not hold for 3D.

\section{Conclusions\label{sec: conclusions}}
Cosmology is entering a data-driven era, with several upcoming galaxy surveys opening gateways for huge galaxy catalogs. Given the increased statistical power of our datasets, we face imminent challenges, including the need to account for systematic uncertainties that dominate the uncertainty budget on our measurements. In this paper, we have\deleted{ specifically} studied the treatment of \replaced{sample contaminants}{contamination} arising from \pz\ uncertainties when measuring the two-point angular correlation functions\deleted{, with a special emphasis on countering the increased variance in the measurement of angular correlation functions arising due to the contamination, with the statistical power of the large datasets}. We first \replaced{considered a simple approach}{introduced a simple formalism}: decontamination that uses the correlations in contaminated subsamples to estimate the true correlations. We then introduced a new estimator that accounts for the full \pz\ PDF of each galaxy to estimate the true correlations, \replaced{without requiring a subsampling of the observed galaxy sample using the measured \pz s}{allowing each galaxy to contribute to all bins (or samples) based on their probabilities}. We demonstrated the effectiveness of our method \added{in recovering true CFs and covariance matrix }on both a toy example and a realistic scenario that is scaleable for surveys like LSST. \added{We also note that our estimator can correct for contamination when measuring correlation functions of multiple galaxy populations, rather than \pz\ bins, alongside other kinds of contamination.}

We emphasize the need for more data-driven tools in order to truly utilize the statistical power of the large datasets. Here we have presented an estimator that incorporates the available probabilistic information to reduce the \added{bias and }variance in the measured correlation functions; this represents a step in the direction of reducing biases and uncertainties in the measurement of cosmological parameters from upcoming surveys.

\section*{Acknowledgements}
We thank David Alonso\replaced{ and Nelson Padilla}{, Nelson Padilla and Javier S\'{a}nchez} for their helpful feedback. H. Awan also thanks Kartheik Iyer and Willow Kion-Crosby for insightful discussions through the various stages of this work. H. Awan has been supported by the Rutgers Discovery Informatics Institute (RDI$^2$) Fellowship of Excellence in Computational and Data Science\added{(AY 2017-2020) and Rutgers University \& Bevier Dissertation Completion Fellowship (AY 2019-2020)}. This work has used resources from RDI$^2$, which are supported by Rutgers and the State of New Jersey; specifically, \replaced{the correlation function calculations}{our} analysis used the \ttt{Caliburn} supercomputer \citep{RDI2}. \added{The authors also acknowledge \href{http://oarc.rutgers.edu}{the Office of Advanced Research Computing (OARC)} at Rutgers, The State University of New Jersey for providing access to the \ttt{Amarel} cluster and associated research computing resources that have contributed to our work.} H. Awan also thanks the LSSTC Data Science Fellowship Program, which is funded by LSSTC, NSF Cybertraining Grant \#1829740, the Brinson Foundation, and the Moore Foundation, as participation in the program has benefited this work. This research was also supported by the Department of Energy (grants DE-SC0011636 and DE-SC0010008).
\bibliography{references}

\appendix
\replaced{
\section*{Standard Estimator: Decontamination, Bias and Variance\label{sec: direct bias + variance}}
}{\section{\ttt{Decontaminated} Estimator: Decontamination, Bias and Variance\label{sec: direct bias + variance}}}
\subsection{Decontamination Derivation\label{sec: dP: standard}}
Here, we re-derive the \deleted{standard} decontamination equation\deleted{s} (Equation~\ref{eq: direct obs}) using the definition of angular correlation function. We start with Equation~\ref{eq: w}, rewriting it as
\eqs{
	dP_{\alpha\beta}(\thetak) &= \eta_{\alpha\beta}^{\rm{pair}} \bsqbr{1+w_{\alpha\beta}(\thetak)} d\Omega_\alpha d\Omega_\beta
						=  \mathcal{N}_{\alpha\beta} \bsqbr{1+w_{\alpha\beta}(\thetak)} \frac{d\Omega_\alpha}{V_\alpha} \frac{d\Omega_\beta}{V_\beta}
	\label{eq: dp}
}
where $\eta_{\alpha\beta}^{\rm{pair}}$ is the observed sky density of Type-$\alpha\beta$ pairs of galaxies while $\mathcal{N}_{\alpha\beta}$ is the observed number of type-$\alpha\beta$ pairs. Assuming that we work with large surveys such that the integral constraint is nearly zero, $\mathcal{N}_{\alpha\beta} \rightarrow \ev{ \mathcal{N}_{\alpha\beta} }$, hence the simplification in the last line in the equation above. Since we consider samples in the same volume, $V_\alpha=V_\beta=V$ and $d\Omega_\alpha = d\Omega_\beta = d\Omega$. Therefore, for \replaced{standard}{the \ttt{Standard}} estimator, \added{for the case} where we have the correlations measured in the contamination subsamples, we have
\eqs{
	dP_{\alpha\beta}(\thetak)
	=  \mathcal{N}_{\alpha\beta, \rm{obs}} \bsqbr{1+w_{\alpha\beta}^{\rm{obs}}(\thetak)} \frac{d\Omega}{V} \frac{d\Omega}{V}
	= \sum_{\gamma,\delta} \mathcal{N}_{\alpha\beta, \rm{obs}}^{\gamma\delta, \rm{true}}  [ 1 + w_{\gamma,\delta}^{\rm{true}} (\thetak) ] \frac{d\Omega}{V} \frac{d\Omega}{V}
	\label{eq: dp 2}
}
where $w_{\alpha\beta}^{\rm{obs}}(\thetak)$ is the biased correlation function, measured using contaminated samples. Expanding the sum on the right hand side, we have
\eqs{
	\mathcal{N}_{\alpha\beta, \rm{obs}}^{\rm{tot}} \bsqbr{1+w_{\alpha\beta}^{\rm{obs}}(\thetak)}
	&= \mathcal{N}_{\alpha\beta, \rm{obs}}^{11, \rm{true}} \bsqbr{1+w_{11}^{\rm{true}}(\thetak)}
		+
		\mathcal{N}_{\alpha\beta, \rm{obs}}^{12, \rm{true}} \bsqbr{1+w_{12}^{\rm{true}}(\thetak)}
		\\
		&\hspace*{2em}
		+
		\mathcal{N}_{\alpha\beta, \rm{obs}}^{21, \rm{true}} \bsqbr{1+w_{21}^{\rm{true}}(\thetak)}
		+
		\mathcal{N}_{\alpha\beta, \rm{obs}}^{22, \rm{true}} \bsqbr{1+w_{22}^{\rm{true}}(\thetak)}
}
Since we have
\eq{
	\frac{ \mathcal{N}_{\alpha\beta, \rm{obs}}^{\gamma\delta, \rm{true}} }{ \mathcal{N}_{\alpha\beta, \rm{obs}}^{\rm{tot}} } = f_{\alpha\gamma}f_{\beta\delta}
}
\eq{
\Rightarrow
	\bsqbr{ 1+w_{\alpha\beta}^{\rm{obs}}(\thetak) }
	= f_{\alpha1} f_{\beta1} \bsqbr{1+w_{11}^{\rm{true}}(\thetak)}
		+
		\bcbr{ f_{\alpha1} f_{\beta2} + f_{\alpha2} f_{\beta1}   } \bsqbr{1+w_{12}^{\rm{true}}(\thetak)}
		+
		f_{\alpha2} f_{\beta2} \bsqbr{1+w_{22}^{\rm{true}}(\thetak)}
	\label{eq: dP: standard: gen}
}
Therefore, for $\alpha, \beta = 1, 2$, Equation~\ref{eq: dP: standard: gen} becomes
\eq{
	\bsqbr{ 1+w_{12}^{\rm{obs}}(\thetak) }
	= f_{11} f_{21} \bsqbr{1+w_{11}^{\rm{true}}(\thetak)}
		+
		\bcbr{ f_{11} f_{22} + f_{12} f_{21} } \bsqbr{1+w_{12}^{\rm{true}}(\thetak)}
		+
		f_{12} f_{22} \bsqbr{1+w_{22}^{\rm{true}}(\thetak)}
}
Now, since
\eqs{
	f_{11} f_{21} + \bcbr{ f_{11} f_{22} + f_{12} f_{21} } + f_{12} f_{22}
	= 	f_{11} \bsqbr{f_{21} + f_{22} } + f_{12} \bsqbr{f_{21} + f_{2} } = 1,
}
we have
\eqs{
	w_{12}^{\rm{obs}}(\thetak)
	&= f_{11} f_{21} w_{11}^{\rm{true}}(\thetak)
		+
		\bcbr{ f_{11} f_{22} + f_{12} f_{21} }
			w_{12}^{\rm{true}}(\thetak)
		+
		f_{12} f_{22} w_{22}^{\rm{true}}(\thetak)
	\label{eq: dp: w12 reduce}
}
which agrees with Equation~\ref{eq: direct obs}. Similar results follow for $(\alpha, \beta)$ = (1,1), (2, 2).

\subsection{Estimator Bias\label{sec: direct: bias}}
We expect that the \replaced{decontaminated standard}{\ttt{Decontaminated}} estimators are unbiased\deleted{,} given their construction (i.e., Equation~\ref{eq: wABobs}). However, for brevity, we formally show that they are indeed unbiased. By definition, an unbiased estimator is such that
\eq{
	\ev{\widehat{w}}= w_{\rm{true}}
\label{eq: unbiased estimator}
}
where the expectation value is over many realizations of the survey. Then, using Equations~\ref{eq: direct obs} and \ref{eq: direct}, we have
\eqs{
	\ev{
	\begin{bmatrix}
		\wAAest \ \ \wABest \ \ \wBBest
	\end{bmatrix}^T
	}
&=
	\ev{
	[D_{\rm{S}}]^{-1}
	\begin{bmatrix}
		\wAAobs \ \ \wABobs \ \ \wBBobs
	\end{bmatrix}^T
	}
\\%
&=
	[D_{\rm{S}}]^{-1}
	[D_{\rm{S}}]
	\begin{bmatrix}
		\wAAtrue \ \ \wABtrue \ \ \wBBtrue
	\end{bmatrix}^T
	=
	\begin{bmatrix}
		\wAAtrue \ \ \wABtrue \ \ \wBBtrue
	\end{bmatrix}^T
\label{eq: direct proof}
}
where the second equality follows by substituting Equation~\ref{eq: direct obs}. Hence, the \replaced{decontaminated standard}{\ttt{Decontaminated}} estimators are unbiased. We note here that $[D_{\rm{S}}]$ in Equation~\ref{eq: direct} is effectively a decontamination matrix: it removes the contamination from the biased estimates, $w_{\alpha \beta}^{\rm{obs}}$, in the presence of sample contamination. A similar argument follows for the case where we have $M$ target samples, using Equation~\ref{eq: direct general}. We also note that Equation~\ref{eq: direct proof} is valid only when $f_{\alpha\beta}$ are accurate averages of the classification probabilities.

\subsection{Estimator Variance\label{sec: direct: variance}}
As for the variance of the \replaced{decontaminated standard}{\ttt{Decontaminated}} estimator\added{s}, we can calculate it by using the variance in our observed correlations. That is, given Equation~\ref{eq: direct}, we have
\replaced{
\eq{
	\begin{bmatrix}
		\var{\wsabest{A}{A}}(\thetak) \ \ \var{\wsabest{A}{B}}(\thetak) \ \ \var{\wsabest{B}{B}}(\thetak)
	\end{bmatrix}^T
	=
	[D_{\rm{S}}]^{-2}
	\begin{bmatrix}
		\var{\wsabobs{A}{A}}(\thetak) \ \ \var{\wsabobs{A}{B}}(\thetak) \ \ \var{\wsabobs{B}{B}}(\thetak)
	\end{bmatrix}^T
}
}{
\eq{
	\begin{bmatrix}
		\var{\wsabest{A}{A}}(\thetak) \ \ \var{\wsabest{A}{B}}(\thetak) \ \ \var{\wsabest{B}{B}}(\thetak)
	\end{bmatrix}^T
	=
	\bcbr{ [D_{\rm{S}}]^{-1} }_{ij}^2
	\begin{bmatrix}
		\var{\wsabobs{A}{A}}(\thetak) \ \ \var{\wsabobs{A}{B}}(\thetak) \ \ \var{\wsabobs{B}{B}}(\thetak)
	\end{bmatrix}^T
\label{eq: direct variance}
}
}
\added{where $\bcbr{ [D_{\rm{S}}]^{-1} }_{ij}^2$ denotes that matrix resulting from squaring each individual coefficient in the matrix $[D_{\rm{S}}]^{-1}$. We also note that the above derivation assumes no covariance between the observed correlations (i.e., $\wsabobs{\alpha}{\beta}$), which is incorrect for the case of neighboring redshift bin given the shared LSS between them; this is discussed in more detail when we discuss the covariance matrices in Section~\ref{sec: results 3bin}. To consider the covariance matrix for the \ttt{Decontaminated} estimators, we start with Equation~\ref{eq: direct}, which is reproduced here:
 \eq{
	\begin{bmatrix}
		\wAAest \ \ \wABest \ \ \wBBest
	\end{bmatrix}^T
	=
	[D_{\rm{S}}]^{-1}
	\begin{bmatrix}
		\wAAobs \ \ \wABobs \ \ \wBBobs
	\end{bmatrix}^T
}
Given Equation~\ref{eq: direct proof}, we therefore have
 \eq{
 	\ev{
	\begin{bmatrix}
		\wAAest \ \ \wABest \ \ \wBBest
	\end{bmatrix}^T
	}
	=
	[D_{\rm{S}}]^{-1}
	\ev{
	\begin{bmatrix}
		\wAAobs \ \ \wABobs \ \ \wBBobs
	\end{bmatrix}^T
	}
}
where we assume that $[D_{\rm{S}}]$ is constant across the samples over which the expectation value is calculated. Now, using the above equations, we can write the variations in the estimators from their expectation value ($\equiv \Delta w \equiv w - \ev{w}$) as
 \eq{
	\begin{bmatrix}
		\Delta\wAAest \ \ \Delta\wABest \ \ \Delta\wBBest
	\end{bmatrix}^T
	=
	[D_{\rm{S}}]^{-1}
	\begin{bmatrix}
		\Delta\wAAobs \ \ \Delta\wABobs \ \ \Delta\wBBobs
	\end{bmatrix}^T
\label{eq: estimator delta}
}
Now defining $C_{\wsabest{}{}}(\thetak)$ as the covariance matrix for the \ttt{Decontaminated} estimators $\wsabest{\alpha}{\beta}(\thetak)$, we have
\eq{
	C_{\wsabest{}{}}(\thetak)
	=
	\ev{
	\begin{bmatrix}
		\Delta\wAAest \ \ \Delta\wABest \ \ \Delta\wBBest
	\end{bmatrix}^T
	\begin{bmatrix}
		\Delta\wAAest \ \ \Delta\wABest \ \ \Delta\wBBest
	\end{bmatrix}
	}
\label{eq: direct: cov definition}
}
Using Equation~\ref{eq: estimator delta} and its transpose, we then have
\eqs{
	C_{\wsabest{}{}}(\thetak)
	&=
	\ev{
	[D_{\rm{S}}]^{-1}
	\begin{bmatrix}
		\Delta\wAAobs \ \ \Delta\wABobs \ \ \Delta\wBBobs
	\end{bmatrix}^T
	\begin{bmatrix}
		\Delta\wAAobs \ \ \Delta\wABobs \ \ \Delta\wBBobs
	\end{bmatrix}
	 \bsqbr{[D_{\rm{S}}]^{-1} }^T
	 }
	 \\%
	 &=
	[D_{\rm{S}}]^{-1}
	\ev{
	\begin{bmatrix}
		\Delta\wAAobs \ \ \Delta\wABobs \ \ \Delta\wBBobs
	\end{bmatrix}^T
	\begin{bmatrix}
		\Delta\wAAobs \ \ \Delta\wABobs \ \ \Delta\wBBobs
	\end{bmatrix}
	}
	 \bsqbr{[D_{\rm{S}}]^{-1} }^T
	 \\%
	 &=
	[D_{\rm{S}}]^{-1}
	C_{\wsabobs{}{}}(\thetak)
	\bsqbr{[D_{\rm{S}}]^{-1} }^T
\label{eq: direct: cov}
}
where $C_{\wsabobs{}{}}$ is covariance matrix for the observed correlations, $\wsabobs{\alpha}{\beta}$. Note that the second equality is valid only under the assumption that $[D_{\rm{S}}]$ is constant.

Both $C_{\wsabobs{}{}}(\thetak)$ and $C_{\wsabest{}{}}(\thetak)$ can determined by bootstrap, as done for the example considered in Section~\ref{sec: results 3bin}, with the estimated covariance matrices presented in Figure~\ref{fig: 3bin full cov}. We note that $C_{\wsabest{}{}}(\thetak)$ may be calculated using $C_{\wsabobs{}{}}(\thetak)$ given Equation~\ref{eq: direct: cov}, assuming that $[D_{\rm{S}}]$  is constant across the bootstrapped samples. We also that one can construct covariance matrices for both $\wsabobs{}{}$ and $\wsabest{}{}$ spanning all $\theta$-bins via a block combination of the $\theta$-dependent matrices presented here; these larger matrices are only block diagonal to the extent that individual CFs are uncorrelated between neighboring $\theta$-bins. Finally, as a simple check of the expression in Equation~\ref{eq: direct: cov}, we note that if $C_{\wsabobs{}{}}(\thetak)$ is diagonal, i.e., there are no covariances in the observed correlations, Equation~\ref{eq: direct: cov} leads to the variance in the \ttt{Decontaminated} estimators as given by Equation~\ref{eq: direct variance}.

}
\replaced{
\section*{Decontamination: From Standard with Full Sample to Weighted\label{sec: dP: full sample to weighted}}
}
{
\section{Decontamination: From \ttt{Decontaminated} with Full Sample to Weighted\label{sec: dP: full sample to weighted}}
}
Here, we present the methodology to decontaminate the \replaced{weighted}{\ttt{Weighted}} correlation function introduced in Equation~\ref{eq: marked}, using the formalism introduced in \ref{sec: dP: standard}. To develop intuition, we first extend the methodology in \ref{sec: dP: standard} to consider an unweighted full observed sample, followed by considering the weighted full sample.
\replaced{
\subsection*{Standard: Full Sample\label{sec: dP: full sample}}
}
{
\subsection{\ttt{Decontaminated}: Full Sample\label{sec: dP: full sample}}
}
We extend the treatment in \ref{sec: dP: standard} to consider an unweighted full sample. Then, the analog of Equation~\ref{eq: dp 2} is
\eqs{
	dP(\thetak)
	= \mathcal{N}_{\rm{tot_{obs}}} \bsqbr{1+w^{\rm{full}}(\thetak)} \frac{d\Omega}{V} \frac{d\Omega}{V}
	= \sum_{\gamma,\delta}  \mathcal{N}_{\rm{tot_{obs}}}^{\gamma\delta, \rm{true}} [ 1 + w_{\gamma,\delta}^{\rm{true}} (\thetak) ] \frac{d\Omega}{V} \frac{d\Omega}{V}
	\label{eq: dp 2 full}
}
Note that we have dropped the $\alpha, \beta$ markers since there is only one correlation that can be measured for the unweighted full sample. Expanding the sum, we have
\eqs{
	\mathcal{N}_{\rm{tot_{obs}}} \bsqbr{1+w^{\rm{full}}(\thetak)}
	&=  \mathcal{N}_{\rm{tot_{obs}}}^{11, \rm{true}} \bsqbr{1+w_{11}^{\rm{true}}(\thetak)}
		+
		 \mathcal{N}_{\rm{tot_{obs}}}^{12, \rm{true}} \bsqbr{1+w_{12}^{\rm{true}}(\thetak)}
		\\
		&\hspace*{2em}
		+
		 \mathcal{N}_{\rm{tot_{obs}}}^{21, \rm{true}} \bsqbr{1+w_{21}^{\rm{true}}(\thetak)}
		+
		 \mathcal{N}_{\rm{tot_{obs}}}^{22, \rm{true}} \bsqbr{1+w_{22}^{\rm{true}}(\thetak)}
}
Now if we assume that our classification probabilities are unbiased, we can write
\eqs{
	\sum_{i}^{ N_{\rm{tot_{obs}}}^{\gamma} } \sum_{j\neq i}^{ N_{\rm{tot_{obs}}}^{\delta} } \ttt{q}_{i}^{\gamma} \ttt{q}_{j}^{\delta}  &= \widehat{\mathcal{N}}_{\rm{tot_{obs}}}^{\gamma\delta, \rm{true} }
	\label{eq: Nest}
	}
Note that technically $N_{\rm{tot_{obs}}}^{\gamma} = N_{\rm{tot_{obs}}}^{\delta} = N_{\rm{tot_{obs}}} $ but we keep $\gamma, \delta$ tags just to keep track of samples when reducing to \replaced{standard}{\ttt{Decontaminated}}. Now, simplifying the equation above, we have
\eqs{
	 \mathcal{N}_{\rm{tot_{obs}}} \bsqbr{1+w^{\rm{full}}(\thetak)}
	&= \sum_{i}^{ N_{\rm{tot_{obs}}}^{1} } \sum_{j\neq i}^{ N_{\rm{tot_{obs}}}^{1} } \ttt{q}_{i}^{1} \ttt{q}_{j}^{1}  \bsqbr{1+w_{11}^{\rm{true}}(\thetak)}
		+
		\sum_{i}^{ N_{\rm{tot_{obs}}}^{1} } \sum_{j\neq i}^{ N_{\rm{tot_{obs}}}^{2} } \ttt{q}_{i}^{1} \ttt{q}_{j}^{2}  \bsqbr{1+w_{12}^{\rm{true}}(\thetak)}
		\\
		&\hspace*{2em}
		+
		\sum_{i}^{ N_{\rm{tot_{obs}}}^{2} } \sum_{j\neq i}^{ N_{\rm{tot_{obs}}}^{1} } \ttt{q}_{i}^{2} \ttt{q}_{j}^{1}   \bsqbr{1+w_{21}^{\rm{true}}(\thetak)}
		+
		\sum_{i}^{ N_{\rm{tot_{obs}}}^{2} } \sum_{j\neq i}^{ N_{\rm{tot_{obs}}}^{2} } \ttt{q}_{i}^{2} \ttt{q}_{j}^{2}  \bsqbr{1+w_{22}^{\rm{true}}(\thetak)}
	\label{eq: dP: full} 
}

We now check what happens when we reduce the above equation to \replaced{standard}{\ttt{Decontaminated}}, i.e., we consider not the full sample but the target subsamples, while all the probabilities are represented by their averages. Then, for $\alpha, \beta =1, 2$, Equation~\ref{eq: dP: full} becomes
\eqs{
	N_{1, \rm{obs} } N_{ 2, \rm{obs} } \bsqbr{1+w_{11}^{\rm{obs}}(\thetak)}
	&= \bbr{ \sum_{i}^{ N_{ 1, \rm{obs} } } \sum_{j\neq i}^{ N_{2, \rm{obs}} } \ttt{q}_{i}^{1} \ttt{q}_{j}^{1} } \bsqbr{1+w_{11}^{\rm{true}}(\thetak)}
		+
		\bbr{ \sum_{i}^{ N_{ 1, \rm{obs} } } \sum_{j\neq i}^{ N_{2, \rm{obs}} } \ttt{q}_{i}^{1} \ttt{q}_{j}^{2} } \bsqbr{1+w_{12}^{\rm{true}}(\thetak)}
		\\
		&\hspace*{2em}
		+
		\bbr{ \sum_{i}^{ N_{ 1, \rm{obs} } } \sum_{j\neq i}^{ N_{2, \rm{obs}} } \ttt{q}_{i}^{2} \ttt{q}_{j}^{1} }  \bsqbr{1+w_{21}^{\rm{true}}(\thetak)}
		+
		\bbr{ \sum_{i}^{ N_{ 1, \rm{obs} } } \sum_{j\neq i}^{ N_{2, \rm{obs} } } \ttt{q}_{i}^{2} \ttt{q}_{j}^{2} } \bsqbr{1+w_{22}^{\rm{true}}(\thetak)}
	\\
	&= \bbr{ \sum_{i}^{ N_{ 1, \rm{obs} } } \sum_{j}^{ N_{2, \rm{obs}} } \ttt{q}_{i}^{1} \ttt{q}_{j}^{1} } \bsqbr{1+w_{11}^{\rm{true}}(\thetak)}
		+
		\bbr{ \sum_{i}^{ N_{ 1, \rm{obs} } } \sum_{j}^{ N_{2, \rm{obs}} } \ttt{q}_{i}^{1} \ttt{q}_{j}^{2} } \bsqbr{1+w_{12}^{\rm{true}}(\thetak)}
		\\
		&\hspace*{2em}
		+
		\bbr{ \sum_{i}^{ N_{ 1, \rm{obs} } } \sum_{j}^{ N_{2, \rm{obs}} } \ttt{q}_{i}^{2} \ttt{q}_{j}^{1} }  \bsqbr{1+w_{21}^{\rm{true}}(\thetak)}
		+
		\bbr{ \sum_{i}^{ N_{ 1, \rm{obs} } } \sum_{j}^{ N_{2, \rm{obs} } } \ttt{q}_{i}^{2} \ttt{q}_{j}^{2} } \bsqbr{1+w_{22}^{\rm{true}}(\thetak)}
	\\
	&\hspace*{-3em}\xrightarrow[\text{}]{\text{simplify\ $q$s}}
		\bbr{ \sum_{i}^{ N_{ 1, \rm{obs} } } \sum_{j}^{ N_{2, \rm{obs}} } q_{i, 11} q_{j, 12} } \bsqbr{1+w_{11}^{\rm{true}}(\thetak)}
		+
		\bbr{ \sum_{i}^{ N_{ 1, \rm{obs} } } \sum_{j}^{ N_{2, \rm{obs}} } q_{i, 11} q_{j, 22} } \bsqbr{1+w_{12}^{\rm{true}}(\thetak)}
		\\
		&\hspace*{2em}
		+
		\bbr{ \sum_{i}^{ N_{ 1, \rm{obs} } } \sum_{j}^{ N_{2, \rm{obs}} } q_{i, 12} q_{j, 21} }  \bsqbr{1+w_{21}^{\rm{true}}(\thetak)}
		+
		\bbr{ \sum_{i}^{ N_{ 1, \rm{obs} } } \sum_{j}^{ N_{2, \rm{obs} } } q_{i, 12} q_{j, 22} } \bsqbr{1+w_{22}^{\rm{true}}(\thetak)}
	\\
	&\hspace*{-3em}\xrightarrow[\text{}]{\text{$q$s=$f$s}}
		\bbr{  f_{11}f_{21} \sum_{i}^{ N_{ 1, \rm{obs} } } \sum_{j}^{ N_{2, \rm{obs}} } } \bsqbr{1+w_{11}^{\rm{true}}(\thetak)}
		+
		\bbr{  f_{11} f_{22} \sum_{i}^{ N_{1, \rm{obs}}} \sum_{j}^{ N_{2, \rm{obs}} }  } \bsqbr{1+w_{12}^{\rm{true}}(\thetak)}
		\\
		&\hspace*{2em}
		+
		\bbr{  f_{12} f_{21} \sum_{i}^{ N_{1, \rm{obs}} } \sum_{j}^{ N_{2, \rm{obs}} } }  \bsqbr{1+w_{21}^{\rm{true}}(\thetak)}
		+
		\bbr{  f_{12}f_{22} \sum_{i}^{ N_{1, \rm{obs}} } \sum_{j}^{ N_{2, \rm{obs}} } } \bsqbr{1+w_{22}^{\rm{true}}(\thetak)}
	\\
	&= f_{11} f_{21} N_{1, \rm{obs} } N_{ 2, \rm{obs} }  \bsqbr{1+w_{11}^{\rm{true}}(\thetak)}
		+
		f_{11} f_{22} N_{1, \rm{obs} } N_{ 2, \rm{obs} }  \bsqbr{1+w_{12}^{\rm{true}}(\thetak)}
		\\
		&\hspace*{2em}
		+
		f_{12} f_{21} N_{1, \rm{obs} } N_{ 2, \rm{obs} }  \bsqbr{1+w_{21}^{\rm{true}}(\thetak)}
		+
		f_{12} f_{22} N_{1, \rm{obs} } N_{ 2, \rm{obs} }  \bsqbr{1+w_{22}^{\rm{true}}(\thetak)}
		\label{eq: full: 12 reduce}
}
\eq{
\Rightarrow
	\bsqbr{ 1+w_{12}^{\rm{obs}}(\thetak) }
	= f_{11} f_{21} \bsqbr{1+w_{11}^{\rm{true}}(\thetak)}
		+
		\bcbr{ f_{11} f_{22} + f_{12} f_{21} } \bsqbr{1+w_{12}^{\rm{true}}(\thetak)}
		+
		f_{12} f_{22} \bsqbr{1+w_{22}^{\rm{true}}(\thetak)}
}
which agrees with Equation~\ref{eq: dp: w12 reduce}. Similar results follow for $(\alpha, \beta) = (1,1) = (2, 2)$.
\replaced{
\subsection*{Weighted: Full Sample\label{sec: dP: weighted}}
}
{
\subsection{\ttt{Weighted}: Full Sample\label{sec: dP: weighted}}
}
We now extend the analysis above further for the weighted (biased) estimator:
\eqs{
	d\widetilde{P}_{\alpha\beta}(\thetak) = \widetilde{ \mathcal{N} }_{\rm{tot_{obs}}}^{\alpha\beta, \rm{obs}}  \bsqbr{1+\widetilde{w}_{\alpha\beta}(\thetak)} \frac{d\Omega}{V} \frac{d\Omega}{V}
	\label{eq: dp weighted 2}
}
where we introduce $\widetilde{\mathcal{N}}$ to account for the weighted pair counts which we define as
\eq{
	\widetilde{\mathcal{N}}_{\rm{tot_{obs}}}^{\alpha\beta, \rm{obs}} = \sum_{i}^{ N_{\rm{tot_{obs}}}^{\alpha} } \sum_{j\neq i}^{ N_{\rm{tot_{obs}}}^{\beta} } \ttt{w}_{ij}^{\alpha\beta}
	\label{eq: new N}
}
Now, when writing the analog of Equations~\ref{eq: dp 2} -\ref{eq: dp 2 full}, we need to account for pair weights, leading us to
\eqs{
	d\widetilde{P}_{\alpha\beta}(\thetak)
	=  \widetilde{\mathcal{N}}_{\rm{tot_{obs}}}^{\alpha\beta, \rm{obs}} \bsqbr{1+\widetilde{w}_{\alpha\beta}(\thetak)} \frac{d\Omega}{V} \frac{d\Omega}{V}
	= \sum_{\gamma,\delta}  \widetilde{\mathcal{N}}_{\rm{tot_{obs}}}^{\gamma\delta, \rm{true}} [ 1 + w_{\gamma,\delta}^{\rm{true}} (\thetak) ] \frac{d\Omega}{V} \frac{d\Omega}{V}
	\label{eq: dp 2 full weighted}
}
where we have the analog of Equation~\ref{eq: Nest}:
\eqs{
	\sum_{i}^{ N_{\rm{tot_{obs}}}^{\alpha} } \sum_{j\neq i}^{ N_{\rm{tot_{obs}}}^{\beta} } \ttt{w}_{ij}^{\alpha\beta} \ttt{q}_{i}^{\alpha} \ttt{q}_{j}^{\beta}  &= \widehat{\widetilde{\mathcal{N}}}_{\rm{tot_{obs}}}^{\alpha\beta, \rm{true} }
	\label{eq: Nest 2}
}
Now, expanding the sum in Equation~\ref{eq: dp 2 full weighted}, we have
\eqs{
	\widetilde{\mathcal{N}}_{\rm{tot_{obs}}}^{\alpha\beta, \rm{obs}} \bsqbr{1+\widetilde{w}_{\alpha\beta}(\thetak)}
	&= \widetilde{\mathcal{N}}_{\rm{tot_{obs}}}^{11, \rm{true}} \bsqbr{1+w_{11}^{\rm{true}}(\thetak)}
		+
		\widetilde{\mathcal{N}}_{\rm{tot_{obs}}}^{12, \rm{true}} \bsqbr{1+w_{12}^{\rm{true}}(\thetak)}
		\\
		&\hspace*{2em}
		+
		\widetilde{\mathcal{N}}_{\rm{tot_{obs}}}^{21, \rm{true}} \bsqbr{1+w_{21}^{\rm{true}}(\thetak)}
		+
		\widetilde{\mathcal{N}}_{\rm{tot_{obs}}}^{22, \rm{true}}\bsqbr{1+w_{22}^{\rm{true}}(\thetak)}
}
Substituting Equation~\ref{eq: Nest} to estimate the true counts, we have
\eqs{
	\bbr{ \sum_{i}^{ N_{\rm{tot_{obs}}}^{\alpha} } \sum_{j\neq i}^{ N_{\rm{tot_{obs}}}^{\beta} } \ttt{w}_{ij}^{\alpha\beta} } \bsqbr{1+\widetilde{w}_{\alpha\beta}^{\rm{full}}(\thetak)}
	&= \bbr{ \sum_{i}^{ N_{\rm{tot_{obs}}}^{\alpha} } \sum_{j\neq i}^{ N_{\rm{tot_{obs}}}^{\beta} } \ttt{w}_{ij}^{\alpha\beta} \ttt{q}_{i}^{1} \ttt{q}_{j}^{1} } \bsqbr{1+w_{11}^{\rm{true}}(\thetak)}
		+
		\bbr{ \sum_{i}^{ N_{\rm{tot_{obs}}}^{\alpha} } \sum_{j\neq i}^{ N_{\rm{tot_{obs}}}^{\beta} } \ttt{w}_{ij}^{\alpha\beta} \ttt{q}_{i}^{1} \ttt{q}_{j}^{2} } \bsqbr{1+w_{12}^{\rm{true}}(\thetak)}
		\\
		&\hspace*{0em}
		+
		\bbr{ \sum_{i}^{ N_{\rm{tot_{obs}}}^{\alpha} } \sum_{j\neq i}^{ N_{\rm{tot_{obs}}}^{\beta} } \ttt{w}_{ij}^{\alpha\beta} \ttt{q}_{i}^{2} \ttt{q}_{j}^{1} }  \bsqbr{1+w_{21}^{\rm{true}}(\thetak)}
		+
		\bbr{ \sum_{i}^{ N_{\rm{tot_{obs}}}^{\alpha} } \sum_{j\neq i}^{ N_{\rm{tot_{obs}}}^{\beta} } \ttt{w}_{ij}^{\alpha\beta} \ttt{q}_{i}^{2} \ttt{q}_{j}^{2} } \bsqbr{1+w_{22}^{\rm{true}}(\thetak)}
	\label{eq: dP: weighted full} 
}
Note that, this equation reduces to \replaced{standard}{\ttt{Decontaminated}} as in Equation~ \ref{eq: full: 12 reduce} when weights are set to 1 for target subsample and 0 for the rest; and we basically have theta-independent decontamination.

\replaced{
\section*{Weighted Estimator: Variance and Practical Notes\label{sec: weighted variance + notes}}
}
{
\section{\ttt{Weighted} Estimator: Variance and Practical Notes\label{sec: weighted variance + notes}}
}

\replaced{
\subsection*{Weighted Estimator: Variance\label{sec: pweighted variance}}
}
{
\subsection{\ttt{Weighted} Estimator: Variance\label{sec: pweighted variance}}
}
Here, we follow the procedure in \LS\ to estimate the variance of the \replaced{weighted estimator}{\ttt{Weighted}} estimator introduced in Equation~\ref{eq: marked}, filling in additional details while accounting for the weights in the data-data pair counts. While the details may be of value to the interested reader, we note that the derivation is lengthy, culminating in the analytical expression for the variance in \ref{sec: var plug in}. Specifically, we write the pair counts, i.e., the unnormalized $\overline{DD}$, $\overline{RR}$ histograms in terms of the fluctuations about their means, i.e., we have
\eqs{
	(\overline{DD})_{\alpha \beta}(\thetak)&= \ev{(\overline{DD})_{\alpha \beta}(\thetak)}(1+\eta(\thetak)) \\
	(\overline{RR})(\thetak) &= \ev{(\overline{RR})(\thetak)}(1+\gamma(\thetak))
\label{eq: fluctuations}
}
where we use the overline to distinguish the \textit{unnormalized} histograms from the normalized ones (denoted with a tilde). Here, $\eta$ and $\gamma$ are the fluctuations in the histograms about their means, which follows
\eqs{
	\ev{\eta(\thetak)} = \ev{\gamma(\thetak)}= 0
\label{eq: avgs}
}
and hence, we have
\eqs{
\sigma_\eta^2(\thetak) &= \ev{\eta^2(\thetak)}-\cancelto{0}{\ev{\eta(\thetak)}^2}= \ev{\eta^2(\thetak)} \\
\sigma_\gamma^2(\thetak) &= \ev{\gamma^2(\thetak)}-\cancelto{0}{\ev{\gamma(\thetak)}^2}= \ev{\gamma^2(\thetak)} \\
\rm{cov}(\eta, \gamma)(\thetak) &= \ev{\eta(\thetak)\gamma(\thetak)}-\cancelto{0}{\ev{\eta(\thetak)}}\cancelto{0}{\ev{\gamma(\thetak)}}= 0
}
where $\ev{\eta(\thetak)\gamma(\thetak)}=0$ since the data and random catalogs are not correlated. Note that $\eta$ here is the same as $\alpha$ in \LS; we choose the former given that the latter letter is already in use here. Then, given Equation~\ref{eq: marked} and Equation~\ref{eq: fluctuations}, we have
\eqs{
1+\widetilde{w}_{\alpha \beta}(\thetak)
	=  \frac{(\widetilde{DD})_{\alpha \beta}(\thetak)}{RR(\thetak)}
	=  \frac{
		(\overline{DD})_{\alpha \beta}(\thetak)
		}{
		\sum_{j\neq i}^{\Ntot} \pw{ij}
		}
		\frac{ \NR (\NR-1)/2 }{ (\overline{RR})(\thetak) }
	= \frac{ \NR (\NR-1)}{ 2 \sum_{j\neq i}^{\Ntot} \pw{ij} }
	\frac{
		\ev{(\overline{DD})_{\alpha \beta}(\thetak)}(1+\eta(\thetak))
	}{
		\ev{(\overline{RR})(\thetak)}(1+\gamma(\thetak))
	}
}
where we have collapsed the double sums for brevity, and have defined 
\eq{
	RR(\thetak) = \frac{\sum_i^{\NR} \sum_{j> i}^{\NR} \heavisides{ij} }{\sum_i^{\NR} \sum_{j> i}^{\NR}}
				= \frac{\sum_i^{\NR} \sum_{j> i}^{\NR} \heavisides{ij} }{\NR (\NR-1)/2}
}
\eqs{
\Rightarrow
1+\ev{\widetilde{w}_{\alpha \beta}(\thetak)}
	&=  \ev{
		\frac{ \NR (\NR-1)}{ 2\sum_{j\neq i}^{\Ntot} \pw{ij} }
		\frac{
		\ev{(\overline{DD})_{\alpha \beta}(\thetak)}(1+\eta(\thetak))
		}{
		\ev{(\overline{RR})}(\thetak)(1+\gamma(\thetak))
		}
	}
	\\%
	&= 	\frac{ \NR (\NR-1) }{ 2 }
		\frac{ \ev{(\overline{DD})_{\alpha \beta}(\thetak)} }{ \ev{(\overline{RR})(\thetak)} }
		\ev{ \frac{ 1 }{ \sum_{j\neq i}^{\Ntot} \pw{ij} } }
		\ev{
		\frac{ (1+\eta(\thetak)) }{ (1+\gamma(\thetak)) }
		}
	\\%
	&\approx
		\frac{ \NR (\NR-1) }{ 2 \sum_{j\neq i}^{\Ntot} \pw{ij} }
		\frac{ \ev{(\overline{DD})_{\alpha \beta}(\thetak)} }{ \ev{(\overline{RR})(\thetak)} }
		\ev{(1+\eta(\thetak))(1-\gamma(\thetak)+\gamma^2(\thetak))}
	\\%
	&=	\frac{ \NR (\NR-1) }{ 2 \sum_{j\neq i}^{\Ntot} \pw{ij} }
		\frac{ \ev{(\overline{DD})_{\alpha \beta}(\thetak)} }{ \ev{(\overline{RR})(\thetak)} }
		\ev{ 1-\gamma(\thetak)+\gamma^2(\thetak) + \eta(\thetak) - \eta(\thetak)\gamma(\thetak) + \eta(\thetak)\gamma^2(\thetak) }
}
where we only keep the terms up to the second order in fluctuations. Note that the second equality is justified since the weights for individual galaxies are fixed across the different realizations. Now, we calculate the variance of the estimator as
\eqs{
\mathrm{var} \bsqbr{\widetilde{w}_{\alpha \beta}}(\thetak)
=
\sigma_{\widetilde{w}_{\alpha \beta}}^2(\thetak)
	&=  \mathrm{var}
		\bsqbr{
		\frac{ \NR (\NR-1) }{ 2 \sum_{j\neq i}^{\Ntot} \pw{ij} }
		\frac{ \ev{(\overline{DD})_{\alpha \beta}(\thetak)} }{ \ev{(\overline{RR})(\thetak)} }
		\bsqbr{ 1-\gamma(\thetak)+\gamma^2(\thetak) + \eta(\thetak) - \eta(\thetak)\gamma(\thetak) + \eta(\thetak)\gamma^2(\thetak) }
		}
	\\%
	&\approx
		\bsqbr{
		\frac{ \NR (\NR-1) }{ 2 \sum_{j\neq i}^{\Ntot} \pw{ij} }
		\frac{ \ev{(\overline{DD})_{\alpha \beta}(\thetak)} }{ \ev{(\overline{RR})(\thetak)} }
		}^2
		\mathrm{var}
		\bsqbr{ 1-\gamma(\thetak) + \eta(\thetak) }
	\\%
	&=
		\bsqbr{
		\frac{ \NR (\NR-1) }{ 2 \sum_{j\neq i}^{\Ntot} \pw{ij} }
		\frac{ \ev{(\overline{DD})_{\alpha \beta}(\thetak)} }{ \ev{(\overline{RR})(\thetak)} }
		}^2
		\bsqbr{\sigma_\gamma^2(\thetak) + \sigma_\eta^2(\thetak) - 2 \cancelto{0}{\rm{cov}(\eta(\thetak), \gamma(\thetak))}}
	\\%
	&=
		\bsqbr{
		\frac{ \NR (\NR-1) }{ 2 \sum_{j\neq i}^{\Ntot} \pw{ij} }
		\frac{ \ev{(\overline{DD})_{\alpha \beta}(\thetak)} }{ \ev{(\overline{RR})(\thetak)} }
		}^2
		\bsqbr{ \ev{\gamma^2(\thetak)} + \ev{\eta^2(\thetak)} }
\label{eq: w var}
}
where, again, we only keep the terms up to the second order in fluctuations. Here, as derived from Equation~\ref{eq: fluctuations}, we have the second moments of the fluctuations defined as
\eqs{
	\ev{\eta^2(\thetak)} &= \frac{ \ev{(\overline{DD})_{\alpha \beta}(\thetak) \cdot (\overline{DD})_{\alpha \beta}(\thetak)}-\ev{(\overline{DD})_{\alpha \beta}(\thetak)}^2
					}{
					\ev{(\overline{DD})_{\alpha \beta}(\thetak)}^2
					}
\label{eq: alpha2 intro}
}\vspace*{-1em}
\eqs{
	\ev{\gamma^2(\thetak)} &= \frac{ \ev{(\overline{RR})(\thetak)\cdot (\overline{RR})(\thetak)}-\ev{(\overline{RR})(\thetak)}^2
					}{
					\ev{(\overline{RR})(\thetak)}^2
					}
\label{eq: gamma2 intro}
}
In order to evaluate the variance, we calculate the second moments of the fluctuations using the first and second moments of the pair counts. Specifically, we only need $\ev{(\overline{RR}) (\theta_k)}$, $\ev{(\overline{DD})_{\alpha \beta} (\theta_k)}$, and $\ev{(\overline{DD})_{\alpha \beta} \cdot (\overline{DD})_{\alpha \beta} (\theta_k)}$; we do not need the second moment of the random pair counts, since $\ev{\gamma^2}$ is simply the variance of the random data and hence the variance of the Poisson distribution.

\subsubsection{Pair Counts: First and Second Moments\label{sec: moments}}
As in Section 2 in \LS, we consider counts in cells in order to write out the first and second moments of the pair counts. We calculate first moment of random pairs in \ref{sec: first moments: RR}; random pairs are uncorrelated in the limit of large $\NR$ and hence present a simpler case. Then, we calculate the first moment of correlated data pairs in \ref{sec: first moments: correlated DD}, followed by the second moment for the correlated data pairs in \ref{sec: second moments: correlated DD}.

\subsubsection{Random Pairs: First Moment \label{sec: first moments: RR}}
Here, we consider $\NR$ points distributed randomly over the survey area, which we divide into $K$ cells.  The probability of finding the $i$th random point in any cell is the continuum probability, $\ev{\rho_j}= \NR/K$, in the limit of large enough $K$ that we essentially have either zero or one point in each cell. This follows that the number of random pairs is
\eqs{
\ev{(\overline{RR}) (\theta_k)}
	&= \ev{\sum_{j<i}^K
		\rho_i \rho_j \heavisides{ij}
		}
	= \frac{1}{2} \sum_{i\neq j}^K \ev{ \rho_i \rho_j } \heavisides{ij}
\label{eq: RR intro}
}
where we have borrowed the notation introduced in Equation~\ref{eq: heaviside short} to express the heavisides.  Now, the probability of finding two random points in two cells, chosen without replacement, is
\eq{
\ev{\rho_i \rho_j}= \frac{\NR(\NR-1)}{K(K-1)}
\label{eq: rho_ij}
}
and, similar to \LS\ Equation 10, we have
\eq{
	\sum_{i\neq j}^K \heavisides{ij} = K(K-1) G_p(\thetak)
\label{eq: Gp intro}
}
where $G_p(\thetak)$ is the probability of finding two random points at separations $\thetak \pm d\thetak/2$. Hence $\sum_{i\neq j}^K \heavisides{ij}$ is just the total number of random points with separations between $\theta_{\mathrm{min}, k}$, $\theta_{\mathrm{max}, k}$ as we have $K(K-1)$ cells.  Substituting Equations~\ref{eq: rho_ij}-\ref{eq: Gp intro} into Equation~\ref{eq: RR intro}, we have
\eqs{
\ev{(\overline{RR}) (\theta_k)}
	&= \frac{1}{2} \frac{\NR(\NR-1)}{K(K-1)} \bsqbr{K(K-1) G_p(\thetak)}
	= \frac{\NR(\NR-1)}{2} G_p(\thetak)
\label{eq: RR}
}

\subsubsection{Data Pairs: First Moment \label{sec: first moments: correlated DD}}
Here, we have $\Ntot$ points distributed randomly over the survey area. As in \ref{sec: first moments: RR}, the probability of finding a galaxy in any cell is $\ev{\nu}= \Ntot/K$, in the limit of large enough $K$ that we essentially have either no galaxy or one galaxy in each cell. Furthermore, we assign the pair weight to the cells in which the pair falls. This follows, given Equation~\ref{eq: marked dd}, that
\eqs{
\ev{(\overline{DD})_{\alpha \beta} (\theta_k)}
	&= C_\Omega \ev{
		\sum_{i\neq j}^K
		\pw{ij} \nu_i \nu_j \heavisides{ij}
		}
	= C_\Omega \sum_{i\neq j}^K \ev{ \pw{ij} } \ev{\nu_i \nu_j} \heavisides{ij}
\label{eq: DD intro}
}
where $C_\Omega$ is a normalization constant to ensure that we recover the correct number of pairs, $\sum_{i\neq j}^{\Ntot} \pw{ij}$, when integrating over all angles. Here, the pair weights are assumed to be uncorrelated with the probability of finding galaxies in a particular pair of cells, allowing us to separate their expectation values in the second equality; this assumption is valid since we are assigning pair weights based upon galaxy properties rather than their locations. Now, since data pairs are generally correlated, we must account for the correlation explicitly when considering the probabilities of finding a pair of galaxies in any two cells, chosen without replacement. That is, we have the probability of finding two galaxies in two cells separated by $\thetak$, chosen without replacement, as
\eq{
\ev{\nu_i \nu_j} = \frac{\Ntot(\Ntot-1)}{K(K-1)}\bsqbr{1+w_{\alpha\beta}(\thetak)} = \frac{\Ntot(\Ntot-1)}{K(K-1)}\bsqbr{1+w_{\alpha\beta}(\thetak)}
\label{eq: nu_ij_2}
}
Therefore, using Equations~\ref{eq: Gp intro} and \ref{eq: nu_ij_2}, Equation~\ref{eq: DD intro} becomes
\eqs{
\ev{(\overline{DD})_{\alpha \beta} (\theta_k)}
	&= C_\Omega \ev{ \pw{ij} }|_{i\neq j} \frac{\Ntot(\Ntot-1)}{K(K-1)}\bsqbr{1+w_{\alpha\beta}(\thetak)} \bsqbr{ K(K-1) G_p(\thetak) }
	\\%
	&= C_\Omega \bsqbr{\frac{ \sum_{i\neq j}^{\Ntot} \pw{ij} }{ \Ntot(\Ntot-1) }} \bsqbr{1+w_{\alpha\beta}(\thetak)}  G_p(\thetak) \Ntot(\Ntot-1)
	\\%
	&= C_\Omega \bsqbr{1+w_{\alpha\beta}(\thetak)}  G_p(\thetak) \sum_{i\neq j}^{\Ntot} \pw{ij}
\label{eq: DD intro 2}
}
Now, before finding the normalization constant, we define $w_\Omega$ as the mean of $w_{\alpha\beta}(\thetak)$ over the sampling geometry, i.e.,
\eq{
	w_\Omega= \int_\Omega G_p(\thetak) w_{\alpha\beta}(\thetak) d\Omega
\label{eq: w omega}
}
with $G_p(\thetak)$ normalized to unity, i.e.,
\eq{
	\int_\Omega G_p(\thetak) d\Omega = 1
\label{eq: Gp norm}
}
Therefore, we have
\eqs{
	\int_\Omega \ev{(\overline{DD})_{\alpha \beta} (\theta_k)} d\Omega
	&= \sum\limits_{i\neq j}^{\Ntot} \pw{ij}
\\%
\Rightarrow
	\int_\Omega C_\Omega G_p(\thetak) \bsqbr{1+w_{\alpha\beta}(\thetak)} \sum_{i\neq j}^{\Ntot} \pw{ij}
	&= \sum\limits_{i\neq j}^{\Ntot} \pw{ij}
\\%
\Rightarrow
	C_\Omega&= \frac{1}{1+w_\Omega}
	\label{eq: c omega}
}
where we make use of Equation~\ref{eq: Gp norm}. Therefore, Equation~\ref{eq: DD intro 2} becomes
\eqs{
\ev{(\overline{DD})_{\alpha \beta} (\theta_k)}
	&= G_p(\thetak) \bsqbr{ \frac{1+w_{\alpha\beta}(\thetak)}{1+w_\Omega} } \sum_{i\neq j}^{\Ntot} \pw{ij}
\label{eq: DD correlated}
}

\subsubsection{Data-Data Pairs \label{sec: second moments: correlated DD}}
As in \LS, using counts in cells, the second moment is defined as
\eqs{
\ev{(\overline{DD})_{\alpha \beta} \cdot (\overline{DD})_{\alpha \beta} (\theta_k)}
	&= \ev{
		\sum_{j\neq i}^K \pw{ij} \nu_i \nu_j \heavisides{ij}
		\sum_{l\neq m}^K \pw{ml}\nu_m \nu_l \heavisides{ml}
		}
	\\%
	&=
		\ev{
		\sum_{j\neq i}^K \pw{ij} \nu_i \nu_j \heavisides{ij}
		\sum_{l\neq  m}^K \pw{ml}\nu_m \nu_l \heavisides{ml}
		}
	\\%
	&=
		\sum_{j\neq i}^K \sum_{l\neq  m}^K
		\ev{ \nu_i \nu_j \nu_m \nu_l }
		\ev{ \pw{ij} \pw{ml}}
		\heavisides{ij} \heavisides{ml}
\label{eq: DD-DD intro}
}

Now, there are three cases to consider, each of which needs to be normalized to give the right total weight from each case (as done in \ref{sec: first moments: correlated DD}):
\enumm{
\item No indices overlap: there are $K(K-1)(K-2)(K-3)$ cases of the sort since we choose each of the four cells without replacement. Since the data pairs are correlated, the probability of finding each of the four galaxies in the four cells, chosen without replacement, is given by
	\eq{
		\ev{ \nu_i \nu_j \nu_m \nu_l } = \frac{\Ntot(\Ntot-1)(\Ntot-2)(\Ntot-3)}{K(K-1)(K-2)(K-3)} \bsqbr{ 1 +  w_{ij}(\thetak) + w_{im}(\thetak) + w_{il}(\thetak) + w_{jm}(\thetak) + w_{jl}(\thetak) + w_{ml}(\thetak)}
	}
Here, since pairs $i,j$ and $m, l$ are separated by $\thetak \pm d\thetak/2$, $ w_{ij}(\thetak) = w_{ml}(\thetak) = w_{\alpha\beta}(\thetak)$ while the rest of the correlations can be approximated as $w_\Omega$. Therefore,
	\eq{
		\ev{ \nu_i \nu_j \nu_m \nu_l } = \frac{\Ntot(\Ntot-1)(\Ntot-2)(\Ntot-3)}{K(K-1)(K-2)(K-3)} \bsqbr{ 1 +  2w_{\alpha\beta}(\thetak) + 4w_\Omega}
	}
Also, as in \LS, we introduce $G_q(\thetak)$ as the probability of finding quadrilaterals, i.e., pairs $i,j$ and $m, l$ separated by $\thetak \pm d\thetak/2$. Then, the total number of quadrilaterals is
	\eq{
		\sum_{\mathrm{unique}\{{i, j, l, m}\}}^K \heavisides{ij} \heavisides{ml}
		= K(K-1)(K-2)(K-3) G_q(\thetak), \ \ \ i\neq j, m\neq l
	}
	Note that as in Equation~\ref{eq: Gp norm}, $G_q(\thetak)$ is also normalized to unity, i.e.,
	\eq{
		\int_\Omega G_q(\thetak) d\Omega = 1
	\label{eq: Gq norm}
	}
	Therefore, the contribution to the second moment of the pair counts by the quadrilaterals is given by
	\eqs{
		\ev{(\overline{DD})_{\alpha \beta} \cdot (\overline{DD})_{\alpha \beta} (\theta_k)}_{\mathrm{quad}}
		&=
		C_{\mathrm{quad}}
		\sum_{j\neq i \neq l\neq  m}^K
		\ev{ \nu_i \nu_j \nu_m \nu_l }
		\ev{ \pw{ij} \pw{ml}}
		\heavisides{ij} \heavisides{ml}
		\\%
		&\hspace*{-12em}=
		C_{\mathrm{quad}}
		\Ntot(\Ntot-1)(\Ntot-2)(\Ntot-3) \bsqbr{ 1 +  2w_{\alpha\beta}(\thetak) + 4w_\Omega} G_q(\thetak) \ev{ \pw{ij} \pw{ml}}_{ i\neq j\neq m \neq l }
		\\%
		&\hspace*{-12em}=
		C_{\mathrm{quad}}
		\Ntot(\Ntot-1)(\Ntot-2)(\Ntot-3) \bsqbr{ 1 +  2w_{\alpha\beta}(\thetak) + 4w_\Omega} G_q(\thetak)
		\bsqbr{
		\frac{
			\sum_{i\neq j\neq m \neq l}^{\Ntot} \pw{ij} \pw{ml}
			}{
			\Ntot(\Ntot-1)(\Ntot-2)(\Ntot-3)
			}
		}
		\\%
		&\hspace*{-12em}=
		C_{\mathrm{quad}}
		\bsqbr{ 1 +  2w_{\alpha\beta}(\thetak) + 4w_\Omega} G_q(\thetak)
		\sum_{i\neq j\neq m \neq l}^{\Ntot} \pw{ij} \pw{ml}
	}
	where $C_{\mathrm{quad}}$ is the normalization constant so that we get the correct weight for the quadrilaterals when integrating over all angles, i.e.,
	\eqs{
		\int \ev{(\overline{DD})_{\alpha \beta} \cdot (\overline{DD})_{\alpha \beta} (\theta_k)}_{\mathrm{quad}} d\Omega
		&= 
		\sum_{i\neq j\neq m \neq l}^{\Ntot} \pw{ij} \pw{ml}
		\\%
		\Rightarrow
		\int \bcbr{ 
			C_{\mathrm{quad}}
			\bsqbr{ 1 +  2w_{\alpha\beta}(\thetak) + 4w_\Omega} G_q(\thetak)
			} d\Omega &= 1
		\\%
		\Rightarrow
		C_{\mathrm{quad}}			
			&= \frac{1}{1 +  2 \int  w_{\alpha\beta}(\thetak) G_q(\thetak) d\Omega + 4  w_\Omega }
			= \frac{1}{1 + 2 w_{\Omega, q} + 4  w_\Omega }
	}
	where we have used Equation~\ref{eq: Gq norm} and have defined a new mean:
	\eq{
		w_{\Omega, q} \equiv \int  w_{\alpha\beta}(\thetak) G_q(\thetak) d\Omega
		\label{eq: w omega q}
	}
	Therefore,
		\eqs{
		\ev{(\overline{DD})_{\alpha \beta} \cdot (\overline{DD})_{\alpha \beta} (\theta_k)}_{\mathrm{quad}}
		&=
		\bsqbr{ \frac{1 +  2w_{\alpha\beta}(\thetak) + 4w_\Omega}{1 + 2 w_{\Omega, q} +  4  w_\Omega } }
			G_q(\thetak)
			\sum_{i\neq j\neq m \neq l}^{\Ntot} \pw{ij} \pw{ml}
		\label{eq: quad contribution}
		}
\item One of the indices is repeated: there are $K(K-1)(K-2)$ cases of the sort, since we choose only three cells without replacement, i.e., we choose two cells for the first $(\overline{DD})$ and one for the second $(\overline{DD})$. Note that we do not have to account for $m, l$ swap since we consider the two cases explicitly when calculating $\ev{\nu_i \nu_j  \nu_m \nu_l}$ (needed since the swap carries different meaning for the pair weights). As for the probabilities of finding the data points in the chosen cells, we have
	 \eqs{
		\ev{\nu_i \nu_j  \nu_m \nu_l}|_{i=m}&= \ev{\nu_i^2 \nu_j \nu_l} = \ev{\nu_i \nu_j \nu_l} \\
									&= \frac{\Ntot(\Ntot-1)(\Ntot-2)}{K(K-1)(K-2)} \bsqbr{ 1 +  w_{ij}(\thetak) + w_{il}(\thetak) + w_{jl}(\thetak) }  \\
									&= \frac{\Ntot(\Ntot-1)(\Ntot-2)}{K(K-1)(K-2)} \bsqbr{ 1 +  3w_{\alpha\beta}(\thetak) } \\
		\ev{\nu_i \nu_j  \nu_m \nu_l}|_{i=l}&= \ev{\nu_i^2 \nu_l \nu_m} = \ev{\nu_i \nu_l \nu_m} \\
									&=  \frac{\Ntot(\Ntot-1)(\Ntot-2)}{K(K-1)(K-2)} \bsqbr{ 1 +  w_{il}(\thetak) + w_{im}(\thetak) + w_{lm}(\thetak) } \\
									&= \frac{\Ntot(\Ntot-1)(\Ntot-2)}{K(K-1)(K-2)} \bsqbr{ 1 +  3w_{\alpha\beta}(\thetak) } \\
		\ev{\nu_i \nu_j  \nu_m \nu_l}|_{j=m}&= \ev{\nu_i \nu_j^2 \nu_l} = \ev{\nu_i \nu_j \nu_l} \\
									&= \frac{\Ntot(\Ntot-1)(\Ntot-2)}{K(K-1)(K-2)} \bsqbr{ 1 +  w_{ij}(\thetak) + w_{il}(\thetak) + w_{jl}(\thetak) } \\
									&= \frac{\Ntot(\Ntot-1)(\Ntot-2)}{K(K-1)(K-2)} \bsqbr{ 1 + 3 w_{\alpha\beta}(\thetak)} \\
		\ev{\nu_i \nu_j  \nu_m \nu_l}|_{j=l}&= \ev{\nu_i \nu_j^2 \nu_m} = \ev{\nu_i \nu_j \nu_m} \\
									&= \frac{\Ntot(\Ntot-1)(\Ntot-2)}{K(K-1)(K-2)} \bsqbr{ 1 +  w_{ij}(\thetak) + w_{im}(\thetak) + w_{jm}(\thetak) } \\
									&= \frac{\Ntot(\Ntot-1)(\Ntot-2)}{K(K-1)(K-2)} \bsqbr{ 1 +  3w_{\alpha\beta}(\thetak) } \\
		}
	where we note that $\ev{\nu} = \ev{\nu^2} = \Ntot/K$ since we are working in the large-$K$ regime where there is only 0 or 1 galaxy in each cell. Also, as in \LS, we introduce $G_t(\thetak)$ as the probability of finding triangles, i.e., two galaxies within $\thetak \pm d\thetak/2$ of a given galaxy. Then, the total number of triangles is
	\eq{
		\sum_{\mathrm{unique}\{{i, j, m\}; l=i}}^K \heavisides{ij} \heavisides{ml}
		= K(K-1)(K-2) G_t(\thetak), \ \ \ i\neq j, m\neq i
	}
	while $G_t(\thetak)$ it is also normalized to unity:
		\eq{
			\int_\Omega G_t(\thetak) d\Omega = 1
		\label{eq: Gt norm}
		}
	Therefore, the contribution to the second moment of the pair counts by the triangles is given by
	\eqs{
		\ev{(\overline{DD})_{\alpha \beta} \cdot (\overline{DD})_{\alpha \beta} (\theta_k)}_{\mathrm{tri}}
		&=
		C_{\mathrm{tri}}
		\Ntot(\Ntot-1)(\Ntot-2) G_t(\thetak) \bsqbr{ 1 +  3w_{\alpha\beta}(\thetak) }
		\times
		\\%
		&
		\bcbr{
		\ev{ \pw{ij} \pw{ml}}_{ i=m \neq j\neq l }
		+
		\ev{ \pw{ij} \pw{ml}}_{ i=l \neq j\neq m }
		+
		\ev{ \pw{ij} \pw{ml}}_{ i\neq j=m \neq l }
		+
		\ev{ \pw{ij} \pw{ml}}_{ i\neq j=l \neq m }
		}
		\\%
		&\hspace*{-5em}=
		C_{\mathrm{tri}}
		G_t(\thetak) \bsqbr{ 1 + 3 w_{\alpha\beta}(\thetak) }
		\sum_{i\neq j \neq l}^{\Ntot}
		\bcbr{ \pw{ij} \pw{il} + \pw{ij} \pw{li} + \pw{ij} \pw{jl} + \pw{ij} \pw{lj} }
	}
	where $C_{\mathrm{tri}}$ is the normalization constant so that we get the correct weight for the triangles when integrating over all angles, i.e.,
	\eqs{
		\int \ev{(\overline{DD})_{\alpha \beta} \cdot (\overline{DD})_{\alpha \beta} (\theta_k)}_{\mathrm{tri}} d\Omega
		&= 
		\sum_{i\neq j \neq l}^{\Ntot} \bcbr{ \pw{ij} \pw{il} + \pw{ij} \pw{li} + \pw{ij} \pw{jl} + \pw{ij} \pw{lj} }
		\\%
		\Rightarrow
		\int \bcbr{ 
			C_{\mathrm{tri}}
			\bsqbr{ 1 +  3w_{\alpha\beta}(\thetak) } G_t(\thetak)
			} d\Omega &= 1
		\\%
		\Rightarrow
		C_{\mathrm{tri}}			
			&= \frac{1}{1 +  3 \int  w_{\alpha\beta}(\thetak) G_t(\thetak) d\Omega + 3w_\Omega  }
			= \frac{1}{1 +  3w_{\Omega, t}  }
	}
	where we have used Equation~\ref{eq: Gt norm} and have defined a new mean:
	\eq{
		w_{\Omega, t} \equiv \int  w_{\alpha\beta}(\thetak) G_t(\thetak) d\Omega
		\label{eq: w omega t}
	}
	Therefore,
		\eqs{
		\ev{(\overline{DD})_{\alpha \beta} \cdot (\overline{DD})_{\alpha \beta} (\theta_k)}_{\mathrm{tri}}
		&=
		\bsqbr{ \frac{1 +  3w_{\alpha\beta}(\thetak)}{1 + 3w_{\Omega,t} } }
			G_t(\thetak)
			\sum_{i\neq j \neq l}^{\Ntot}
			\bcbr{ \pw{ij} \pw{il} + \pw{ij} \pw{li} + \pw{ij} \pw{jl} + \pw{ij} \pw{lj} }
		\label{eq: triangle contribution}
		}
\item Two of the indices overlap: there are  $K(K-1)$ cases, since we choose only two cells. This follows that the probability of finding two galaxies in the chosen cells is 
	\eqs{
		\ev{\nu_i \nu_j  \nu_m \nu_l}_{i=m, j=l}
			&= \ev{\nu_i \nu_j  \nu_i \nu_j}
			= \ev{\nu_i^2 \nu_j^2}
			= \ev{\nu_i \nu_j}
			= \frac{\Ntot(\Ntot-1)}{K(K-1)} \bsqbr{1+w_{\alpha\beta}(\thetak)}
		\\%
		\ev{\nu_i \nu_j  \nu_m \nu_l}_{i=l, j=m}
			&= \ev{\nu_i \nu_j  \nu_j \nu_i}
			= \ev{\nu_i^2 \nu_j^2}
			= \ev{\nu_i \nu_j}
			= \frac{\Ntot(\Ntot-1)}{K(K-1)} \bsqbr{1+w_{\alpha\beta}(\thetak)}
		}
	Here, Equation~\ref{eq: Gp intro} applies, giving us the contribution to the second moment of the pair counts by the pairs as
	\eqs{
		\ev{(\overline{DD})_{\alpha \beta} \cdot (\overline{DD})_{\alpha \beta} (\theta_k)}_{\mathrm{pairs}}
		&=
		C_{\mathrm{pairs}}
		\Ntot(\Ntot-1) G_p(\thetak) \bsqbr{1+w_{\alpha\beta}(\thetak)}
		\bcbr{
		\ev{ \pw{ij} \pw{ml}}_{ i=m \neq j=l }
		+
		\ev{ \pw{ij} \pw{ml}}_{ i=l \neq j=m }
		}
		\\%
		&=
		C_{\mathrm{pairs}}
		\Ntot(\Ntot-1) G_p(\thetak) \bsqbr{1+w_{\alpha\beta}(\thetak)}
		\bcbr{
		\ev{ \pw{ij} \pw{ij} }_{ i \neq j }
		+
		\ev{ \pw{ij} \pw{ji} }_{ i \neq j }
		}
		\\%
		&=
		C_{\mathrm{pairs}}
		G_p(\thetak) \bsqbr{1+w_{\alpha\beta}(\thetak)}		
		\sum_{i\neq j}^{\Ntot} \bcbr{ \pw{ij} \pw{ij} + \pw{ij} \pw{ji} }
	}
	where $C_{\mathrm{pairs}}$ is the normalization constant so that we get the correct weight for the pairs when integrating over all angles, i.e.,
	\eqs{
		\int \ev{(\overline{DD})_{\alpha \beta} \cdot (\overline{DD})_{\alpha \beta} (\theta_k)}_{\mathrm{pairs}} d\Omega
		&= 
		\sum_{i\neq j}^{\Ntot} \bcbr{ \pw{ij} \pw{ij} + \pw{ij} \pw{ji} }
		\\%
		\Rightarrow
		\int \bcbr{ 
			C_{\mathrm{pairs}}
			\bsqbr{ 1 +  w_{\alpha\beta}(\thetak) } G_p(\thetak)
			} d\Omega &= 1
		\\%
		\Rightarrow
		C_{\mathrm{pairs}}			
			&= \frac{1}{1 +  w_\Omega }
	}
	where we have used Equation~\ref{eq: Gp norm}; this results matches with Equation~\ref{eq: c omega} as it should. Therefore,
		\eqs{
		\ev{(\overline{DD})_{\alpha \beta} \cdot (\overline{DD})_{\alpha \beta} (\theta_k)}_{\mathrm{pairs}}
		=
		G_p(\thetak) \bsqbr{ \frac{1+w_{\alpha\beta}(\thetak)}	{1 +  w_\Omega} }
		\sum_{i\neq j}^{\Ntot} \bcbr{ \pw{ij} \pw{ij} + \pw{ij} \pw{ji} }
		\label{eq: pairs contribution}
		}
}
Combining the three cases, i.e., Equations~\ref{eq: quad contribution}, \ref{eq: triangle contribution} and \ref{eq: pairs contribution}, Equation~\ref{eq: DD-DD intro} becomes
\eqs{
\ev{(\overline{DD})_{\alpha \beta} \cdot (\overline{DD})_{\alpha \beta} (\theta_k)}
	&= 
	\sum_{j\neq i}^K \sum_{l\neq  m}^K
		\ev{ \nu_i \nu_j \nu_m \nu_l }
		\ev{ \pw{ij} \pw{ml}}
		\heavisides{ij} \heavisides{ml}
	\\%
	&=
	\bsqbr{ \frac{1 +  2w_{\alpha\beta}(\thetak) + 4w_\Omega}{1 + 2 w_{\Omega, q} +  4  w_\Omega } }
		G_p(\thetak)^2
		\sum_{i\neq j\neq m \neq l}^{\Ntot} \pw{ij} \pw{ml}
		\\%
		&\hspace*{3em}
	+
	\bsqbr{ \frac{1 +  3w_{\alpha\beta}(\thetak)}{1 + 3w_{\Omega,t}  } }
		G_t(\thetak)
		\sum_{i\neq j \neq l}^{\Ntot}
		\bcbr{ \pw{ij} \pw{il} + \pw{ij} \pw{li} + \pw{ij} \pw{jl} + \pw{ij} \pw{lj} }
		\\%
		&\hspace*{5em}
	+
	G_p(\thetak) \bsqbr{ \frac{1+w_{\alpha\beta}(\thetak)}	{1 +  w_\Omega} }
		\sum_{i\neq j}^{\Ntot} \bcbr{ \pw{ij} \pw{ij} + \pw{ij} \pw{ji} }
\label{eq: DD-DD}
}
where we have used the result $G_q(\thetak) = G_p^2(\thetak)$ from \LS, valid in the large-$K$ limit.

\subsubsection{Fluctuations\label{sec: fluctuations}}
Now, substituting Equations~\ref{eq: DD correlated}, \ref{eq: DD-DD} in Equation~\ref{eq: alpha2 intro}, we have
\eqs{
	\ev{\eta^2(\thetak)}
	&= \frac{
		\splitfrac{
		\bsqbr{ \frac{1 +  2w_{\alpha\beta}(\thetak) + 4w_\Omega}{1 + 2 w_{\Omega, q} +  4  w_\Omega } }
		G_p(\thetak)^2
		\sum_{i\neq j\neq m \neq l}^{\Ntot} \pw{ij} \pw{ml}
		}{
		\splitfrac{
		+
		\bsqbr{ \frac{1 +  3w_{\alpha\beta}(\thetak)}{1 + 3w_{\Omega,t}  }  }
		G_t(\thetak)
		\sum_{i\neq j \neq l}^{\Ntot}
		\bcbr{ \pw{ij} \pw{il} + \pw{ij} \pw{li} + \pw{ij} \pw{jl} + \pw{ij} \pw{lj} }
			}{
		+
		G_p(\thetak) \bsqbr{ \frac{1+w_{\alpha\beta}(\thetak)}{1 +  w_\Omega} }
		\sum_{i\neq j}^{\Ntot} \bcbr{ \pw{ij} \pw{ij} + \pw{ij} \pw{ji} }
			}
		}
	}{
		\bbr{ G_p(\thetak) \bsqbr{ \frac{1+w_{\alpha\beta}(\thetak)}{1+w_\Omega} } \sum_{i\neq j}^{\Ntot} \pw{ij}} ^2
	}
	-1
	\\%
	&= \frac{
		\splitfrac{
		\bsqbr{ \frac{1 +  2w_{\alpha\beta}(\thetak) + 4w_\Omega}{1 + 2 w_{\Omega, q} +  4  w_\Omega } }
		\sum_{i\neq j\neq m \neq l}^{\Ntot} \pw{ij} \pw{ml}
		}{
		\splitfrac{
		+
		\bsqbr{ \frac{1 +  3w_{\alpha\beta}(\thetak)}{1 + 3w_{\Omega,t}  }  }
		\frac{G_t(\thetak) }{ G_p^2(\thetak) }
		\sum_{i\neq j \neq l}^{\Ntot}
		\bcbr{ \pw{ij} \pw{il} + \pw{ij} \pw{li} + \pw{ij} \pw{jl} + \pw{ij} \pw{lj} }
			}{
		+
		\frac{1}{G_p(\thetak)}
		\bsqbr{ \frac{1+w_{\alpha\beta}(\thetak)}{1 +  w_\Omega} }
		\sum_{i\neq j}^{\Ntot} \bcbr{ \pw{ij} \pw{ij} + \pw{ij} \pw{ji} }
			}
		}
	}{
		\bbr{\bsqbr{ \frac{1+w_{\alpha\beta}(\thetak)}{1+w_\Omega} } \sum_{i\neq j}^{\Ntot} \pw{ij}} ^2
	}
	-1
\label{eq: alpha2}
}
As for $\ev{\gamma^2(\thetak)}$, given Equation~\ref{eq: RR}, it takes the form
\eq{
	\ev{\gamma^2(\thetak)}= \frac{2}{\NR(\NR-1)G_p(\thetak)}
	\label{eq: gamma2}
}

\subsubsection{Variance\label{sec: var plug in}}
We now go back to Equation~\ref{eq: w var}, and attempt to evaluate it. First, substituting Equations~\ref{eq: DD correlated} and \ref{eq: RR}, we have
\eqs{
\sigma_{\widetilde{w}_{\alpha \beta}}^2(\thetak)
	&= \bsqbr{
		\frac{ \NR (\NR-1) }{ 2 \sum_{j\neq i}^{\Ntot} \pw{ij} }
		\frac{
			G_p(\thetak) \bsqbr{ \frac{1+w_{\alpha\beta}(\thetak)}{1+w_\Omega} } \sum_{i\neq j}^{\Ntot} \pw{ij}
		}{
			\frac{\NR(\NR-1)}{2} G_p(\thetak)
		}
		}^2
		\bsqbr{ \ev{\gamma^2(\thetak)} + \ev{\eta^2(\thetak)} }
	\\%
	&= \bsqbr{ \frac{1+w_{\alpha\beta}(\thetak)}{1 +  w_\Omega} }^2
		\bsqbr{ \ev{\gamma^2(\thetak)} + \ev{\eta^2(\thetak)} }
}
Now, in the limit of large $\NR$, i.e., $\ev{\gamma^2} \rightarrow 0$, we have
\eqs{
\sigma_{\widetilde{w}_{\alpha \beta}}^2(\thetak)
	&\xrightarrow[\mathrm{large}\ \NR]{} \bsqbr{ \frac{1+w_{\alpha\beta}(\thetak)}{1 +  w_\Omega} }^2 \ev{\eta^2(\thetak)}
}
where $\ev{\eta^2(\thetak)}$ is given by Equation~\ref{eq: alpha2}. The expression can be simplified: we first look at leading order term, i.e., the quadrilateral contribution:
\eqs{
\sigma_{\widetilde{w}_{\alpha \beta}}^2(\thetak)
	\xrightarrow[\mathrm{order}]{\mathrm{leading}}
	& \frac{
		\bsqbr{ \frac{1 +  2w_{\alpha\beta}(\thetak) + 4w_\Omega}{1 + 2 w_{\Omega, q} +  4  w_\Omega } }
		\sum_{i\neq j\neq m \neq l}^{\Ntot} \pw{ij} \pw{ml}
		}{
		\bbr{\bsqbr{ \frac{1+w_{\alpha\beta}(\thetak)}{1+w_\Omega} } \sum_{i\neq j}^{\Ntot} \pw{ij}} ^2
		}
	-1
}
Then, in the limit of weak correlations as then $1 << w_{\alpha\beta}(\thetak) \sim w_\Omega < w_{\Omega,t} < w_{\Omega,q} $, we have
\eqs{
\sigma_{\widetilde{w}_{\alpha \beta}}^2(\thetak)
	\xrightarrow[\mathrm{correlations}]{\mathrm{weak}}
	& \frac{
		\sum_{i\neq j\neq m \neq l}^{\Ntot} \pw{ij} \pw{ml}
		}{
		\bbr{ \sum_{i\neq j}^{\Ntot} \pw{ij}} ^2
		}
	-1
	\label{eq: var approx}
}
where we note that $\pw{ij}=\ttt{w}_{ji}^{\beta\alpha}$.

Now, in order to get the analytical expression for the variance of the unbiased estimator, \added{i.e., the \ttt{Decontaminated Weighted} estimator}, we must consider not only the variance of each of the biased correlations but also the covariances. As an example, based on Equation~\ref{eq: pweighted} which is valid for when there are two galaxy types in our observed sample, we essentially have the unbiased estimator for the $AA$ auto-correlation function as
\eqs{
	\wAAest = C_{AA}(\thetak) \wAAestobs + C_{AB}(\thetak) \wABestobs + C_{BB}(\thetak) \wBBestobs
}
where $C_{AA}(\thetak), C_{AB}(\thetak), C_{BB}(\thetak)$ are the elements of the first row of the inverse matrix in Equation~\ref{eq: pweighted}. Given the dependency of all terms and factors on the pair weights, we have the variance of the unbiased estimator as
\eqs{
	\sigma_{\widehat{w}_{AA}}^2(\thetak) &= C_{AA}^2(\thetak) \sigma_{\widetilde{w}_{AA}}^2(\thetak)
									+ C_{AB}^2(\thetak) \sigma_{\widetilde{w}_{AB}}^2(\thetak)
									+ C_{BB}^2(\thetak) \sigma_{\widetilde{w}_{BB}}^2(\thetak)
								\\%
								&\hspace*{2em}
									-2 \mathrm{cov} \bsqbr{ C_{AA}(\thetak), \wAAestobs}
									-2 \mathrm{cov} \bsqbr{ C_{AB}(\thetak), \wABestobs}
									-2 \mathrm{cov} \bsqbr{ C_{BB}(\thetak), \wBBestobs}
								\\%
								&\hspace*{4em}
									-2 \wAAestobs \wABestobs \mathrm{cov} \bsqbr{ C_{AA}(\thetak),  C_{AB}(\thetak)}
									-2 \wAAestobs \wBBestobs \mathrm{cov} \bsqbr{ C_{AA}(\thetak),  C_{BB}(\thetak)}
								\\%
								&\hspace*{6em}
									-2 \wABestobs \wBBestobs \mathrm{cov} \bsqbr{ C_{AB}(\thetak),  C_{BB}(\thetak)}
}
This expression is unwieldy to evaluate for the general case, even if when we use the leading-order, weak-correlation approximation as in Equation~\ref{eq: var approx}. Therefore, we resort to numerical estimation of the variance.

\replaced{
\subsection*{Weighted Estimator: Practical Notes\label{eq: weighted notes}}
}{
\subsection{\ttt{Weighted} Estimator: Practical Notes\label{eq: weighted notes}}
}
\subsubsection{Weighted Data-Data Pair Counts\label{eq: weighted notes 1}}
Here, we note some points that are important when it comes to implementing the \replaced{weighted estimator}{\ttt{Weighted}} estimator proposed in Equation~\ref{eq: marked}. Specifically considering Equation~\ref{eq: marked dd} for the auto correlation, we have
\eq{
	(\widetilde{DD})_{AA}(\thetak) = \frac{ \sum_i^{\Ntot} \sum_{j\neq i}^{\Ntot}
									\ttt{w}_{ij}^{AA} \heavisides{ij}
								}{
								\sum_i^{\Ntot} \sum_{j\neq i}^{\Ntot} \ttt{w}_{ij}^{AA}
								}
	\label{eq: marked dd auto}
}
while for the cross, we have
\eq{
	(\widetilde{DD})_{AB}(\thetak) = \frac{ \sum_i^{\Ntot} \sum_{j\neq i}^{\Ntot}
									\ttt{w}_{ij}^{AB} \heavisides{ij}
								}{
								\sum_i^{\Ntot} \sum_{j\neq i}^{\Ntot} \ttt{w}_{ij}^{AB}
								}
	\label{eq: marked dd cross 1}
}
It might appear that $(\widetilde{DD})_{AB} \neq (\widetilde{DD})_{BA} $ since $\ttt{w}_{ij}^{AB} \neq \ttt{w}_{ij}^{BA}$ but we must realize that 
\eq{
	\ttt{w}_{ij}^{AB} = \ttt{w}_{ji}^{BA}
}
and since the sums are re-indexable, we have
\eqs{
	(\widetilde{DD})_{BA}(\thetak) &= \frac{ \sum_i^{\Ntot} \sum_{j\neq i}^{\Ntot}
									\ttt{w}_{ij}^{BA} \heavisides{ij}
								}{
								\sum_i^{\Ntot} \sum_{j\neq i}^{\Ntot} \ttt{w}_{ij}^{BA}
								}
							= \frac{ \sum_i^{\Ntot} \sum_{j\neq i}^{\Ntot}
									\ttt{w}_{ji}^{AB} \heavisides{ij}
								}{
								\sum_i^{\Ntot} \sum_{j\neq i}^{\Ntot} \ttt{w}_{ji}^{AB}
								}
							= (\widetilde{DD})_{AB}(\thetak)
	\label{eq: marked dd cross 2b}
}
Therefore, when implementing the weighted data-data histogram, we can work with either $\ttt{w}_{ij}^{\alpha\beta}$ or $\ttt{w}_{ij}^{\beta\alpha}$, even though $\ttt{w}_{ij}^{\alpha\beta} \neq \ttt{w}_{ij}^{\beta\alpha}$ when $\alpha\neq\beta$.

\subsubsection{Pair Weights\label{eq: weighted notes 2}}
While we have used simple pair weights in this work, i.e., $\pw{ij} = \ttt{q}_i^{\alpha} \ttt{q}_j^{\beta}$, the \replaced{weighted estimator}{\ttt{Weighted}} estimator presented in Equation~\ref{eq: marked} works with general pair weights. In the case where the pair weights are not separable (e.g., they account for a theta-dependence), we must circumvent the problem presented by the normalization of the data-data histogram in Equation~\ref{eq: marked dd}: it requires summing over all the pair weights -- a task that is computationally prohibitive when working with large datasets where standard correlation function algorithms focus on a specified range of separations to reduce compute time. We can address the challenge by \replaced{first}{two methods: 1) estimating the number of pairs and the average weights for the larger $\theta$-bins, and hence still being able to use the all-pairs normalization, and 2) introducing a new, exact normalization, which can be achieved by} considering Equation~\ref{eq: marked} with its full details, i.e.,
\eqs{
	\widetilde{w}_{\alpha\beta}^{\rm{obs}}(\thetak) + 1
		&= \frac{(\widetilde{DD})_{\alpha\beta}(\thetak)}{RR(\thetak)}
		= \frac{
			\sum_i^{\Ntot} \sum_{j\neq i}^{\Ntot} \pw{ij} \heavisides{ij}
			}{
			\sum_i^{\Ntot} \sum_{j\neq i}^{\Ntot} \pw{ij}
			}
		\frac{
			\sum_i^{\NR} \sum_{j\neq i}^{\NR}
			}{
			\sum_i^{\NR} \sum_{j\neq i}^{\NR} \heavisides{ij}
			}
		\\%
		&= \frac{
			\sum_i^{\Ntot} \sum_{j\neq i}^{\Ntot} \pw{ij} \heavisides{ij}
			}{
			\sum_i^{\NR} \sum_{j\neq i}^{\NR} \heavisides{ij}
			}
		\frac{
			\sum_i^{\NR} \sum_{j\neq i}^{\NR}
			}{
			\sum_i^{\Ntot} \sum_{j\neq i}^{\Ntot} \pw{ij}
			}
	\label{eq: marked expanded}
}
where the first fraction in the last line compares the data-data pair weight in bin $k$ with the random-random pairs in the same bins, while the second fraction normalizes the total data-data pair weight with the total random-random pair counts. Now, since \replaced{the}{exact} numerical calculation of the total data-data pair weight is prohibitive and affects only the overall normalization, we can normalize $both$ the total data-data pair weight and the total random pair counts in a less computationally challenging way, i.e.,
\eqs{
	\widetilde{w}_{\alpha\beta}^{\rm{obs}}(\thetak) + 1
		&= \frac{
			\sum_i^{\Ntot} \sum_{j\neq i}^{\Ntot} \pw{ij} \heavisides{ij}
			}{
			\sum_i^{\NR} \sum_{j\neq i}^{\NR} \heavisides{ij}
			}
		\frac{
			\sum_m^{N_{\rm{bin}}} \sum_i^{\NR} \sum_{j\neq i}^{\NR} \bar{\Theta}_{{ij}, m}
			}{
			\sum_m^{N_{\rm{bin}}} \sum_i^{\Ntot} \sum_{j\neq i}^{\Ntot} \pw{ij} \bar{\Theta}_{{ij}, m}
			}
	\label{eq: marked revised}
}
where have replaced the total counts over all possible scales to those in only the scales of interest.

\subsection{Direct Decontamination\label{sec: pweighted: optimize}}
Here we attempt to find weights that allow us to decontaminate $while$ estimating the correlations -- a step towards optimal weights. To achieve this, we consider Equation~\ref{eq: pweighted obs} which is reproduced here for convenience:
\eq{
	\begin{bmatrix}
		\ev{\wAAestobs} \\ \ev{\wABestobs} \\ \ev{\wBBestobs}
	\end{bmatrix}
	=
	\begin{bmatrix}
		\autofrac{A}{A}{A} & \crossfrac{A}{A}{A}{B} & \autofrac{A}{A}{B} \\
		\autofrac{A}{B}{A} & \crossfrac{A}{B}{A}{B}  &  \autofrac{A}{B}{B}  \\
		 \autofrac{B}{B}{A} & \crossfrac{B}{B}{A}{B}  &  \autofrac{B}{B}{B}  \\
	\end{bmatrix}
	\begin{bmatrix}
		\wAAtrue \\ \wABtrue \\ \wBBtrue
	\end{bmatrix}
\label{eq: pweighted obs rep}
}
In order to achieve our goal, we would like to find weights $\ttt{w}_{ij, \mathrm{opt}}^{\alpha\beta}$ such that we can write the above equation as
\eq{
	\begin{bmatrix}
		\ev{\wAAestobs} \\ \ev{\wABestobs} \\ \ev{\wBBestobs}
	\end{bmatrix}
	=
	\begin{bmatrix}
		\autofrac{A}{A}{A} & 0 & 0 \\
		0 & \crossfrac{A}{B}{A}{B}  &  0  \\
		 0 & 0  &  \autofrac{B}{B}{B}  \\
	\end{bmatrix}
	\begin{bmatrix}
		\wAAtrue \\ \wABtrue \\ \wBBtrue
	\end{bmatrix}
\label{eq: pweighted obs opt}
}
To consider a simple scenario, we first assume that the pair weights are a linear product of the weights of individual weights, i.e., $\ttt{w}_{ij, \mathrm{opt}}^{\alpha\beta} = \ttt{w}_{i, \mathrm{opt}}^{\alpha} \ttt{w}_{j, \mathrm{opt}}^{\beta}$, which follows that we only need to find $ \ttt{w}_{i, \mathrm{opt}}^{\alpha}$ and $ \ttt{w}_{i, \mathrm{opt}}^{\beta}$ (where we note $\alpha, \beta$ can be either $A$ or $B$). Then, we must have the non-diagonal terms in Equation~\ref{eq: pweighted obs rep} be zero, leading us to specific constraints on the pair weights. To demonstrate the method, we achieved the optimization by assuming a functional form for the optimized weights:
\eq{
	\ttt{w}_{i, \mathrm{opt}}^{\alpha} = \mu^\alpha + \nu^\alpha q_{i}^{\alpha}
	\label{eq: w func}
}
where $\mu, \nu$ are the optimization parameters and are allowed to be negative (which is what allows this method to mimic \replaced{the standard decontamination}{\ttt{Decontaminated}} by automatically subtracting off pairs in which one contributor is likely a contaminant). Using this method, we were able to decontaminate as effectively as \replaced{standard decontamination}{\ttt{Decontaminated}} for the 2-sample case, but without reducing the variance. We note that the equivalence between this direct decontamination with optimized weights and \replaced{standard decontamination}{\ttt{Decontaminated}} is not guaranteed for larger numbers of samples or for weights that are non-linear functions of probability, meriting further investigation as part of a larger investigation of optimizing the weights.

\section{Generalized Estimators\label{sec: general}}
\replaced{
\subsection*{Decontaminated Standard Estimator\label{sec: direct: general}}
}{
\subsection{\ttt{Decontaminated} Estimator\label{sec: direct: general}}
}
As an extension of our derivation for two samples in Section~\ref{sec: direct}, we now consider three samples, with galaxies of Types $A$, $B$, $C$ present in our sample. For instance, we have
\eqs{
	\wABobs &= \fAA\fBA\wAAtrue + \bcbr{ \fAA\fBB+\fAB\fBA } \wABtrue + \fAB\fBB\wBBtrue + \bcbr{ \fAB\fBC+\fAC\fBB } \wBCtrue \\
				&\hspace*{2em} + \fAC\fBC \wCCtrue + \bcbr{ \fAA\fBC+\fAC\fBA } \wCAtrue
}
Therefore, similar to the construction of Equation~\ref{eq: direct}, we have
\eq{
\begin{bmatrix}
	\wAAest \\ \wABest \\ \wBBest \\ \wBCest \\ \wCCest  \\ \wCAest
\end{bmatrix}
=
\begin{bmatrix}
    \sgen{A}{A}{A}{A}  & 2\sgen{A}{A}{A}{B} & \sgen{A}{B}{A}{B} & 2\sgen{A}{B}{A}{C} & \sgen{A}{C}{A}{C} & 2\sgen{A}{A}{A}{C} \\
    \sgen{A}{A}{B}{A} & \sgenadd{A}{A}{B}{B}  & \sgen{B}{B}{A}{B} & \sgenadd{A}{B}{B}{C}  & \sgen{A}{C}{B}{C} &   \sgenadd{A}{A}{B}{C} \\
    \sgen{B}{A}{B}{A} & 2\sgen{B}{B}{B}{A} & \sgen{B}{B}{B}{B} & 2\sgen{B}{B}{B}{C} & \sgen{B}{C}{B}{C} & 2\sgen{B}{A}{B}{C} \\
    \sgen{B}{A}{C}{B} & \sgenadd{B}{A}{C}{B} &  \sgen{B}{B}{C}{B} & \sgenadd{B}{B}{C}{C}  & \sgen{B}{C}{C}{C}& \sgenadd{B}{A}{B}{C} \\
    \sgen{C}{A}{C}{A} & 2\sgen{C}{A}{C}{B} & \sgen{C}{B}{C}{B} & 2\sgen{C}{B}{C}{C} & \sgen{C}{C}{C}{C} & 2\sgen{C}{A}{C}{C} \\
    \sgen{A}{A}{C}{A} & \sgenadd{A}{A}{C}{B}  & \sgen{A}{B}{C}{B} & \sgenadd{A}{B}{C}{C} & \sgen{A}{C}{C}{C} & \sgenadd{A}{A}{C}{C}
\end{bmatrix}^{-1}
\begin{bmatrix}
	\wAAobs \\ \wABobs \\ \wBBobs \\ \wBCobs \\ \wCCobs \\ \wCAobs
\end{bmatrix}
\label{eq: direct 3sample}
}
where we have defined the following for brevity:
\eqs{
	\varsigma_{mn}^{ij} &= f_{A_i A_j} f_{A_m A_n} = \varsigma_{ij}^{mn}
}
Extending the idea to $M$ samples, we can write the analog of the unbiased estimator for \replaced{standard decontamination}{\ttt{Decontamination}}, given by Equation~\ref{eq: direct}, as
\eq{
\scalemath{0.9}{
\begin{bmatrix}
	\wabest{1}{1} \\ \wabest{1}{2} \\ \vdots \\ \wabest{\gamma}{\gamma} \\ \wabest{\gamma}{(\gamma+1)} \\ \vdots \\ \wabest{M}{M}  \\ \wabest{M}{1}   \\
\end{bmatrix}
=
\begin{bmatrix}
   \sgen{1}{1}{1}{1}  & 2\sgen{1}{1}{1}{2} & \dots & \sgen{1}{\gamma}{1}{\gamma} & 2\sgen{1}{\gamma}{1}{(\gamma+1)} & \dots &  \sgen{1}{M}{1}{M} & 2\sgen{1}{1}{1}{M}\\
    \sgen{1}{1}{2}{1}  & \sgenadd{1}{1}{2}{2}  & \dots &  \sgen{1}{\gamma}{2}{\gamma}  & \sgenadd{1}{\gamma}{2}{(\gamma+1)} & \dots &   \sgen{1}{M}{2}{M}  & \sgenadd{1}{M}{2}{1} \\
    \vdots & \vdots & \dots & \vdots & \vdots & \dots & \vdots & \vdots \\
    \sgen{\gamma}{1}{\gamma}{1} & 2\sgen{\gamma}{1}{\gamma}{2} & \dots & \sgen{\gamma}{\gamma}{\gamma}{\gamma} & 2\sgen{\gamma}{\gamma}{\gamma}{(\gamma+1)} & \dots &  \sgen{\gamma}{M}{\gamma}{M} & 2 \sgen{\gamma}{M}{\gamma}{1}\\
    \sgen{\gamma}{1}{(\gamma+1)}{1}  & \sgenadd{\gamma}{1}{(\gamma+1)}{2}  & \dots &  \sgen{\gamma}{\gamma}{(\gamma+1)}{\gamma}  & \sgenadd{\gamma}{\gamma}{(\gamma+1)}{(\gamma+1)}  & \dots &   \sgen{\gamma}{M}{(\gamma+1)}{M}  & \sgenadd{\gamma}{M}{(\gamma+1)}{1}   \\
    \vdots & \vdots & \dots & \vdots & \vdots & \dots & \vdots & \vdots \\
    \sgen{M}{1}{M}{1}  & 2\sgen{M}{1}{M}{2} & \dots & \sgen{M}{\gamma}{M}{\gamma} & 2\sgen{M}{\gamma}{M}{(\gamma+1)} & \dots &  \sgen{M}{M}{M}{M} & 2 \sgen{M}{M}{M}{1}\\
    \sgen{M}{1}{1}{1}  & \sgenadd{M}{1}{1}{2} & \dots &  \sgen{M}{\gamma}{1}{\gamma}  & \sgenadd{M}{\gamma}{1}{(\gamma+1)} & \dots &   \sgen{M}{M}{1}{M}  & \sgen{M}{M}{1}{1}
\end{bmatrix}^{-1}
\begin{bmatrix}
	\wabobs{1}{1} \\ \wabobs{1}{2} \\ \vdots \\ \wabobs{\gamma}{\gamma} \\ \wabobs{\gamma}{(\gamma+1)} \\ \vdots \\ \wabobs{M}{M}  \\ \wabobs{M}{1}
\end{bmatrix}
\label{eq: direct general}
}
}

As for the 2-sample case, we can get the variance of the estimators for $M$ target samples as
\eq{
\scalemath{0.9}{
	\begin{bmatrix}
		\var{\wsabest{A_1}{A_1}} \var{\wsabest{A_1}{A_2}} \hdots \var{\wsabest{A_\gamma}{A_\gamma}} \var{\wsabest{A_\gamma}{A_{\gamma+1}}} \hdots \var{\wsabest{A_M}{A_M}}  \var{\wsabest{A_M}{A_1}}
	\end{bmatrix}^T
	=
	\replaced{[D_{\rm{S}}^{\rm{gen}}]^{-2}}{ \bcbr{ [D_{\rm{S}}^{\rm{gen}}]^{-1} }_{ij}^2 }
	\begin{bmatrix}
		\var{\wsabobs{A_1}{A_1}} \var{\wsabobs{A_1}{A_2}} \hdots \var{\wsabobs{A_\gamma}{A_\gamma}} \var{\wsabobs{A_\gamma}{A_{\gamma+1}}} \hdots \var{\wsabobs{A_M}{A_M}}  \var{\wsabobs{A_M}{A_1}}
	\end{bmatrix}^T
	}
}
where $[D_{\rm{S}}^{\rm{gen}}]$ is the square matrix in Equation~\ref{eq: direct general}\added{ and as in \ref{sec: direct: variance}, $\bcbr{ [D_{\rm{S}}^{\rm{gen}}]^{-1} }_{ij}^2$ denotes that matrix resulting from squaring each individual coefficient in the matrix $[D_{\rm{S}}^{\rm{gen}}]^{-1}$}. \added{The covariance matrix for the $M$-samples case follows the derivation in Equation~\ref{eq: direct: cov}, with all of its assumptions.}

\replaced{
\subsection*{Weighted Estimator\label{sec: pweighted: general}}
}{
\subsection{\ttt{Decontaminated Weighted} Estimator\label{sec: pweighted: general}}
}
Expanding our derivation for two samples to three samples, with galaxies of Types $A$, $B$, $C$ present in our sample, we have
\eq{
\hspace*{-3em}
\begin{bmatrix}
	\wabest{A}{A} \\ \wabest{A}{B} \\ \wabest{B}{B} \\ \wabest{B}{C} \\ \wabest{C}{C}  \\ \wabest{C}{A}   \\
\end{bmatrix}
=
\begin{bmatrix}
	\wgen{A}{A}{A}{A}  & 2\wgen{A}{A}{A}{B} & \wgen{A}{B}{A}{B} & 2\wgen{A}{B}{A}{C} & \wgen{A}{C}{A}{C} & 2\wgen{A}{A}{A}{C} \\
	\wgen{A}{A}{B}{A} & \wgenadd{A}{A}{B}{B}  & \wgen{B}{B}{A}{B} & \wgenadd{A}{B}{B}{C}  & \wgen{A}{C}{B}{C} &   \wgenadd{A}{A}{B}{C} \\
	\wgen{B}{A}{B}{A} & 2\wgen{B}{B}{B}{A} & \wgen{B}{B}{B}{B} & 2\wgen{B}{B}{B}{C} & \wgen{B}{C}{B}{C} & 2\wgen{B}{A}{B}{C} \\
	\wgen{B}{A}{C}{B} & \wgenadd{B}{A}{C}{B} &  \wgen{B}{B}{C}{B} & \wgenadd{B}{B}{C}{C}  & \wgen{B}{C}{C}{C}& \wgenadd{B}{A}{B}{C} \\
	\wgen{C}{A}{C}{A} & 2\wgen{C}{A}{C}{B} & \wgen{C}{B}{C}{B} & 2\wgen{C}{B}{C}{C} & \wgen{C}{C}{C}{C} & 2\wgen{C}{A}{C}{C} \\
	\wgen{A}{A}{C}{A} & \wgenadd{A}{A}{C}{B}  & \wgen{A}{B}{C}{B} & \wgenadd{A}{B}{C}{C} & \wgen{A}{C}{C}{C} & \wgenadd{A}{A}{C}{C}
\end{bmatrix}^{-1}
\begin{bmatrix}
	\wabestobs{A}{A} \\ \wabestobs{A}{B} \\ \wabestobs{B}{B} \\ \wabestobs{B}{C} \\ \wabestobs{C}{C}  \\ \wabestobs{C}{A}   \\
\end{bmatrix}
\label{eq: pweighted 3}
}
where we have defined the following for brevity:
\eqs{
	\varkappa_{mn}^{uv}
	&= \frac{
		\sum\limits_i^{\Ntot}\sum\limits_{j\neq i}^{\Ntot} \ttt{w}_{ij}^{A_u A_v} \ttt{q}_{i}^{A_m}\ttt{q}_{j}^{A_n}
		}{
		\sum\limits_i^{\Ntot}\sum\limits_{j\neq i}^{\Ntot} \ttt{w}_{ij}^{A_u A_v} }
}

Extending the idea to $M$ samples, we can write the analog of our unbiased estimator \added{for \ttt{Decontaminated Weighted}}, given by Equation~\ref{eq: pweighted}, as
\eq{
\scalemath{0.9}{
\begin{bmatrix}
	\wabest{1}{1} \\ \wabest{1}{2} \\ \vdots \\ \wabest{\gamma}{\gamma} \\ \wabest{\gamma}{(\gamma+1)} \\ \vdots \\ \wabest{M}{M}  \\ \wabest{M}{1}   \\
\end{bmatrix}
=
\begin{bmatrix}
	\wgen{1}{1}{1}{1}  & 2\wgen{1}{1}{1}{2} & \dots & \wgen{1}{\gamma}{1}{\gamma} & 2\wgen{1}{\gamma}{1}{(\gamma+1)} & \dots &  \wgen{1}{M}{1}{M} & 2\wgen{1}{1}{1}{M}\\
	\wgen{1}{1}{2}{1}  & \wgenadd{1}{1}{2}{2}  & \dots &  \wgen{1}{\gamma}{2}{\gamma}  & \wgenadd{1}{\gamma}{2}{(\gamma+1)} & \dots &   \wgen{1}{M}{2}{M}  & \wgenadd{1}{M}{2}{1} \\
	\vdots & \vdots & \dots & \vdots & \vdots & \dots & \vdots & \vdots \\
	\wgen{\gamma}{1}{\gamma}{1} & 2\wgen{\gamma}{1}{\gamma}{2} & \dots & \wgen{\gamma}{\gamma}{\gamma}{\gamma} & 2\wgen{\gamma}{\gamma}{\gamma}{(\gamma+1)} & \dots &  \wgen{\gamma}{M}{\gamma}{M} & 2 \wgen{\gamma}{M}{\gamma}{1}\\
	\wgen{\gamma}{1}{(\gamma+1)}{1}  & \wgenadd{\gamma}{1}{(\gamma+1)}{2}  & \dots &  \wgen{\gamma}{\gamma}{(\gamma+1)}{\gamma}  & \wgenadd{\gamma}{\gamma}{(\gamma+1)}{(\gamma+1)}  & \dots &   \wgen{\gamma}{M}{(\gamma+1)}{M}  & \wgenadd{\gamma}{M}{(\gamma+1)}{1}   \\
	\vdots & \vdots & \dots & \vdots & \vdots & \dots & \vdots & \vdots \\
	\wgen{M}{1}{M}{1}  & 2\wgen{M}{1}{M}{2} & \dots & \wgen{M}{\gamma}{M}{\gamma} & 2\wgen{M}{\gamma}{M}{(\gamma+1)} & \dots &  \wgen{M}{M}{M}{M} & 2 \wgen{M}{M}{M}{1}\\
	\wgen{M}{1}{1}{1}  & \wgenadd{M}{1}{1}{2} & \dots &  \wgen{M}{\gamma}{1}{\gamma}  & \wgenadd{M}{\gamma}{1}{(\gamma+1)} & \dots &   \wgen{M}{M}{1}{M}  & \wgen{M}{M}{1}{1}
\end{bmatrix}^{-1}
\begin{bmatrix}
	\wabestobs{1}{1} \\ \wabestobs{1}{2} \\ \vdots \\ \wabestobs{\gamma}{\gamma} \\ \wabestobs{\gamma}{(\gamma+1)} \\ \vdots \\ \wabestobs{M}{M}  \\ \wabestobs{M}{1}   \\
\end{bmatrix}
}
\label{eq: pweighted: general}
}

\end{document}